TECHNISCHE UNIVERSITÄT DRESDEN

**Faculty of Computer Science** Institute of Theoretical Computer Science, Chair of Automata Theory

# THE DISTRIBUTIVE, GRADED LATTICE OF $\mathcal{EL}$ CONCEPT DESCRIPTIONS AND ITS NEIGHBORHOOD RELATION (EXTENDED VERSION)

Francesco Kriegel

francesco.kriegel@tu-dresden.de

# ABSTRACT

For the description logic $\mathcal{EL}$, we consider the neighborhood relation which is induced by the subsumption order, and we show that the corresponding lattice of $\mathcal{EL}$ concept descriptions is distributive, modular, graded, and metric. In particular, this implies the existence of a rank function as well as the existence of a distance function.

# CONTENTS

# 1. INTRODUCTION

*Description Logics* (Baader, Horrocks, Lutz, and Sattler, 2017) are a family of well-founded languages for knowledge representation with a strong logical foundation as well as a widely explored hierarchy of decidability and complexity of common reasoning problems. The several reasoning tasks allow for an automatic deduction of implicit knowledge from given explicitly represented facts and axioms, and many reasoning algorithms have been developed. Description Logics are utilized in many different application domains, and in particular provide the logical underpinning of *Web Ontology Language (OWL)* (Hitzler, Krötzsch, and Rudolph, 2010) and its profiles.

$\mathcal{EL}$ is an example of a description logic with tractable reasoning problems, i.e., the usual inference problems can be decided in polynomial time, cf. Baader, Brandt, and Lutz in (Baader, Brandt, and Lutz, 2005). From a perspective of *lattice theory*, $\mathcal{EL}$ has not been deeply explored yet. Of course, it is apparent that the subsumption $\sqsubseteq$ with respect to some TBox $\mathcal{T}$ constitutes a quasi-order. Furthermore, in description logics supremums in the corresponding ordered set are usually called *least common subsumers*, and these exist in all cases if either no TBox is present, or if greatest fixed-point semantics are applied. Apart from that not much is known about the lattice of $\mathcal{EL}$ concept descriptions. In this document, we shall consider the neighborhood relation which is induced by the subsumption order, and we shall show that the lattice of $\mathcal{EL}$ concept descriptions is distributive, modular, graded, and metric. In particular, this implies the existence of a rank function as well as the existence of a distance function.

This report extends a previous publication (Kriegel, 2018b) by solving its remaining problems.

# 2. THE DESCRIPTION LOGIC $\mathcal{EL}$ AND SOME VARIANTS

In this section we shall introduce the syntax and semantics of the light-weight description logic $\mathcal{EL}$ Baader, Brandt, and Lutz, 2005; Baader, Horrocks, Lutz, and Sattler, 2017.

## 2.1. SYNTAX

Throughout the whole document assume that $\Sigma$ is a signature, i.e., $\Sigma = \Sigma_\mathsf{C} \uplus \Sigma_\mathsf{R}$ is a disjoint union of a set $\Sigma_\mathsf{C}$ of *concept names* and a set $\Sigma_\mathsf{R}$ of *role names*. An $\mathcal{EL}$ *concept description* over $\Sigma$ is a term that is constructed by means of the following inductive rule where $A \in \Sigma_\mathsf{C}$ and $r \in \Sigma_\mathsf{R}$.

$$C ::= \top \mid A \mid C \sqcap C \mid \exists\, r.\, C$$

The set of all $\mathcal{EL}$ concept descriptions over $\Sigma$ is denoted by $\mathcal{EL}(\Sigma)$. We call $\top$ the *top concept description*, $C \sqcap D$ the *conjunction* of $C$ and $D$, and $\exists\, r.\, C$ the *existential restriction* of $C$ with respect to $r$. Furthermore, we call concept names and existential restrictions *atomic*.

As syntactic sugar, we also allow using words of role names within existential restrictions: we define $\exists\, \epsilon.\, C \coloneqq C$ and $\exists\, rw.\, C \coloneqq \exists\, r.\, \exists\, w.\, C$ for any role name $r \in \Sigma_\mathsf{R}$ and for each non-empty role word $w \in \Sigma_\mathsf{R}^+$. Furthermore, if $\mathbf{C}$ is a finite set of $\mathcal{EL}$ concept descriptions, then the *conjunction* $\bigsqcap \mathbf{C}$ is, modulo equivalence, defined as $\top$ if $\mathbf{C}$ is empty and otherwise as the concept description $C_1 \sqcap (C_2 \sqcap (\ldots (C_{n-1} \sqcap C_n) \ldots))$ where $\{C_1, \ldots, C_n\}$ is an arbitrary enumeration of $\mathbf{C}$. Alternatively, we could also define $\bigsqcap \emptyset \coloneqq \top$ as well as $\bigsqcap \mathbf{C} \coloneqq C \sqcap \bigsqcap(\mathbf{C} \setminus \{C\})$ where $C$ is an arbitrary element of $\mathbf{C}$. As we shall see in the next section on the semantics, these two ambivalent definitions do not cause any problems and indeed always denote equivalent concept descriptions, since $(\mathcal{EL}(\Sigma), \sqcap, \top)$ is a commutative monoid in which each element is idempotent.

The *size* $||C||$ of an $\mathcal{EL}$ concept description $C$ is the number of nodes in its syntax tree, and we can recursively define it as follows.

$$||\top|| := 1$$
$$||A|| := 1$$
$$||C \sqcap D|| := ||C|| + 1 + ||D||$$
$$||\exists\, r.\, C|| := 1 + ||C||$$

The *role depth* $\mathsf{rd}(C)$ of some $\mathcal{EL}$ concept description $C$ is recursively defined as follows.

$$\mathsf{rd}(\top) := 0$$
$$\mathsf{rd}(A) := 0$$
$$\mathsf{rd}(C \sqcap D) := \mathsf{rd}(C) \vee \mathsf{rd}(D)$$
$$\mathsf{rd}(\exists\, r.\, C) := 1 + \mathsf{rd}(C)$$

For a role-depth bound $d \in \mathbb{N}$, we denote by $\mathcal{EL}(\Sigma)\restriction_d$ the set of all $\mathcal{EL}$ concept descriptions with a role depth not exceeding $d$. The set $\mathsf{Sub}(C)$ of all *subconcepts* of an $\mathcal{EL}^\bot$ concept description $C$ is recursively defined as follows.

$$\mathsf{Sub}(\top) := \{\top\}$$
$$\mathsf{Sub}(A) := \{A\}$$
$$\mathsf{Sub}(C \sqcap D) := \{C \sqcap D\} \cup \mathsf{Sub}(C) \cup \mathsf{Sub}(D)$$
$$\mathsf{Sub}(\exists\, r.\, C) := \{\exists\, r.\, C\} \cup \mathsf{Sub}(C)$$

A *concept inclusion* is an expression $C \sqsubseteq D$ where both the *premise* $C$ as well as the *conclusion* $D$ are concept descriptions. A *terminological box* (abbrv. *TBox*) is a finite set of concept inclusions. For a CI $C \sqsubseteq D$, we define its set of subconcepts as $\mathsf{Sub}(C \sqsubseteq D) := \mathsf{Sub}(C) \cup \mathsf{Sub}(D)$, and furthermore for a TBox $\mathcal{T}$, its set of subconcepts is defined as $\mathsf{Sub}(\mathcal{T}) := \bigcup\{\, \mathsf{Sub}(C \sqsubseteq D) \mid C \sqsubseteq D \in \mathcal{T}\,\}$.

## 2.2. SEMANTICS

An *interpretation* $\mathcal{I} := (\Delta^\mathcal{I}, \cdot^\mathcal{I})$ over $\Sigma$ consists of a non-empty set $\Delta^\mathcal{I}$ of *objects*, called the *domain*, and an *extension function* $\cdot^\mathcal{I}$ that maps concept names $A \in \Sigma_\mathsf{C}$ to subsets $A^\mathcal{I} \subseteq \Delta^\mathcal{I}$ and maps role names $r \in \Sigma_\mathsf{R}$ to binary relations $r^\mathcal{I} \subseteq \Delta^\mathcal{I} \times \Delta^\mathcal{I}$. Then, the extension function is canonically extended to all $\mathcal{EL}$ concept descriptions by the following definitions.

$$\top^\mathcal{I} := \Delta^\mathcal{I}$$
$$(C \sqcap D)^\mathcal{I} := C^\mathcal{I} \cap D^\mathcal{I}$$
$$(\exists\, r.\, C)^\mathcal{I} := \{\, \delta \in \Delta^\mathcal{I} \mid (\delta, \epsilon) \in r^\mathcal{I} \text{ for some } \epsilon \in C^\mathcal{I}\,\}$$

A concept inclusion $C \sqsubseteq D$ is *valid* in $\mathcal{I}$ if $C^\mathcal{I} \subseteq D^\mathcal{I}$. We then also refer to $\mathcal{I}$ as a *model* of $C \sqsubseteq D$, and denote this by $\mathcal{I} \models C \sqsubseteq D$. Furthermore, $\mathcal{I}$ is a *model* of a TBox $\mathcal{T}$, symbolized as $\mathcal{I} \models \mathcal{T}$, if each concept inclusion in $\mathcal{T}$ is valid in $\mathcal{I}$. The relation $\models$ is lifted to TBoxes as follows. A concept inclusion $C \sqsubseteq D$ is *entailed* by a TBox $\mathcal{T}$, denoted as $\mathcal{T} \models C \sqsubseteq D$, if each model of $\mathcal{T}$ is a model of $C \sqsubseteq D$ too.

We then also say that $C$ is *subsumed* by $D$ with respect to $\mathcal{T}$. A TBox $\mathcal{T}$ *entails* a TBox $\mathbf{U}$, symbolized as $\mathcal{T} \models \mathbf{U}$, if $\mathcal{T}$ entails each concept inclusion in $\mathbf{U}$, or equivalently if each model of $\mathcal{T}$ is also a model of $\mathbf{U}$.

Two $\mathcal{EL}$ concept descriptions $C$ and $D$ are *equivalent* with respect to $\mathcal{T}$, and we shall write $\mathcal{T} \models C \equiv D$, if $\mathcal{T} \models \{C \sqsubseteq D,\ D \sqsubseteq C\}$. As a further abbreviation, let $\mathcal{T} \models C \sqsubsetneq D$ if both $\mathcal{T} \models C \sqsubseteq D$ and $\mathcal{T} \not\models C \sqsupseteq D$, and we then say that $C$ is *strictly subsumed* by $D$ with respect to $\mathcal{T}$. We say that two $\mathcal{EL}$ concept descriptions $C$ and $D$ are *incomparable* with respect to $\mathcal{T}$, written $\mathcal{T} \models C \parallel D$, if $\mathcal{T} \not\models C \sqsubseteq D$ as well as $\mathcal{T} \not\models C \sqsupseteq D$ holds true. In the sequel of this document we may also write $C \leq_{\mathcal{Y}} D$ instead of $\mathcal{Y} \models C \leq D$ where $\mathcal{Y}$ is either an interpretation or a terminological box and $\leq$ is some suitable relation symbol, e.g., $\sqsubseteq$, $\sqsubsetneq$, $\equiv$, or $\parallel$.

## 2.3. COMPUTATIONAL COMPLEXITY

Reasoning in the description logic $\mathcal{EL}^\bot$ is *tractable*. More specifically, the *subsumption problem*, which is defined as follows, is decidable in deterministic polynomial time, cf. (Baader, Brandt, and Lutz, 2005; Baader, Lutz, and Brandt, 2008).

**Instance:** Let $\mathcal{T} \cup \{C \sqsubseteq D\}$ be an $\mathcal{EL}^\bot$ TBox.
**Question:** Is $C$ subsumed by $D$ w.r.t. $\mathcal{T}$?

Since the satisfiability problem in propositional Horn logic is **P**-complete and can be reduced to the subsumption problem for $\mathcal{EL}^\bot$, we conclude that the latter is **P**-complete as well.

## 2.4. REDUCED FORMS

It is not hard to find $\mathcal{EL}$ concept descriptions that are equivalent, i.e., have the same extension in *all* interpretations, but are not equal. It is therefore helpful for technical details to have a unique *normal form* for $\mathcal{EL}$ concept descriptions. According to Baader and Morawska, 2010; Küsters, 2001 an $\mathcal{EL}$ concept description $C$ can be transformed into a *reduced form* that is equivalent to $C$ by exhaustive application of the *reduction rule* $D \sqcap E \mapsto D$ whenever $\emptyset \models D \sqsubseteq E$ to the subconcepts of $C$ (modulo commutativity and associativity of $\sqcap$). It is immediately clear that each $\mathcal{EL}$ concept description $C$ is essentially a conjunction of atomic $\mathcal{EL}$ concept descriptions. In particular, if we define $\mathsf{Conj}(C)$ as the set of all atomic top-level conjuncts in $C$, then $C$ has the form $\bigsqcap \mathsf{Conj}(C)$ (modulo commutativity and associativity of $\sqcap$). Furthermore, for some role name $r \in \Sigma_\mathsf{R}$, we define the set of *$r$-successors* of $C$ as

$$\mathsf{Succ}(C, r) := \{\, D \mid \exists\, r.\, D \in \mathsf{Conj}(C)\,\}.$$

## 2.5. THE LATTICE OF CONCEPT DESCRIPTIONS

We do not refrain basic definition from order theory and lattice theory, and rather refer the interested reader to the following references: Birkhoff (1940), Davey and Priestley (2002), Ganter and Wille (1999), and Grätzer (2002).

It is readily verified that the *subsumption* $\sqsubseteq_\emptyset$ constitutes a quasi-order on $\mathcal{EL}(\Sigma)$. Hence, the quotient of $\mathcal{EL}(\Sigma)$ with respect to the induced *equivalence* $\equiv_\emptyset$ is an ordered set. In what follows we will not distinguish between the equivalence classes and their representatives. Furthermore, $\top$ is the greatest element, and the quotient set $\mathcal{EL}(\Sigma)/\emptyset$ is a lattice that we shall symbolize by $\boldsymbol{\mathcal{EL}}(\Sigma)$. It is easy to verify that the conjunction $\sqcap$ corresponds to the finitary *infimum* operation. In a description logic allowing for disjunctions $\sqcup$, it dually holds true that the disjunction $\sqcup$ corresponds to the finitary *supremum* operation. Unfortunately, this does not apply to our considered description logic $\mathcal{EL}$. As an obvious solution, we can simply define the notion of a *supremum* specifically tailored to the case of $\mathcal{EL}$ concept descriptions as follows. The *supremum* or *least common subsumer* (abbrv. *LCS*) of two $\mathcal{EL}$ concept descriptions $C$ and $D$ is an $\mathcal{EL}$ concept description $E$ with the following properties.

1. $C \sqsubseteq_\emptyset E$ and $D \sqsubseteq_\emptyset E$
2. For each $\mathcal{EL}$ concept description $F$, if $C \sqsubseteq_\emptyset F$ and $D \sqsubseteq_\emptyset F$, then $E \sqsubseteq_\emptyset F$.

Since all least common subsumers of $C$ and $D$ are unique up to equivalence, we may denote a representative of the corresponding equivalence class by $C \vee D$. It is well known that LCS-s always exist in $\mathcal{EL}$; in particular, the least common subsumer $C \vee D$ can be computed, modulo equivalence, by means of the following recursive formula.

$$\begin{aligned} C \vee D = {} & \bigsqcap(\Sigma_\mathsf{C} \cap \mathsf{Conj}(C) \cap \mathsf{Conj}(D)) \\ & \sqcap \bigsqcap\{\, \exists\, r.\, (E \vee F) \mid r \in \Sigma_\mathsf{R},\ \exists\, r.\, E \in \mathsf{Conj}(C), \text{ and } \exists\, r.\, F \in \mathsf{Conj}(D) \,\} \end{aligned}$$

It is easy to see that the equivalence $\equiv_\emptyset$ is compatible with both $\sqcap$ and $\vee$. Of course, the definition of a LCS can be extended to an arbitrary number of arguments in the obvious way, and we shall then denote the LCS of the concept descriptions $C_t$, $t \in T$, by $\bigvee\{\, C_t \mid t \in T \,\}$. We say that two concept descriptions $C, D \in \mathcal{EL}(\Sigma)$ are *orthogonal* or *disjoint* w.r.t. $\emptyset$, written $\emptyset \models C \perp D$ or $C \perp_\emptyset D$, if it holds true that $C \vee D \equiv_\emptyset \top$.

Let $\mathcal{T}$ be an $\mathcal{EL}$ TBox over some signature $\Sigma$. For any $\mathcal{EL}$ concept description $C$ over $\Sigma$, we denote by $[C]_\mathcal{T}$ the equivalence class of $C$ with respect to $\equiv_\mathcal{T}$, that is, we set it as follows.

$$[C]_\mathcal{T} := \{\, D \mid D \in \mathcal{EL}(\Sigma) \text{ and } C \equiv_\mathcal{T} D \,\}$$

Let $\mathbf{C} \subseteq \mathcal{EL}(\Sigma)$ be a set of concept descriptions and let $\mathcal{T}$ be an $\mathcal{EL}$ TBox. The *quotient* of $\mathbf{C}$ w.r.t. $\mathcal{T}$ consists of all equivalence classes w.r.t. $\equiv_\mathcal{T}$ that have representatives in $\mathbf{C}$, that is, we set

$$\mathbf{C}/\mathcal{T} := \{\, [C]_\mathcal{T} \mid C \in \mathbf{C} \,\}.$$

Furthermore, $\mathsf{Min}_\mathcal{T}(\mathbf{C})$ is the set of concept descriptions from $\mathbf{C}$ that are most specific w.r.t. $\mathcal{T}$; analogously, $\mathsf{Max}_\mathcal{T}(\mathbf{C})$ contains those concept descriptions from $\mathbf{C}$ which are most general w.r.t. $\mathcal{T}$. Formally, we define the following.

$$\begin{aligned} \mathsf{Min}_\mathcal{T}(\mathbf{C}) &:= \{\, C \mid C \in \mathbf{C} \text{ and there does not exist some } D \in \mathbf{C} \text{ such that } D \sqsubsetneq_\mathcal{T} C \,\} \\ \mathsf{Max}_\mathcal{T}(\mathbf{C}) &:= \{\, C \mid C \in \mathbf{C} \text{ and there does not exist some } D \in \mathbf{C} \text{ such that } C \sqsubsetneq_\mathcal{T} D \,\} \end{aligned}$$

## 2.6. SIMULATIONS AND CANONICAL MODELS

A *pointed interpretation* is a pair $(\mathcal{I}, \delta)$ consisting of an interpretation $\mathcal{I}$ and an element $\delta \in \Delta^\mathcal{I}$. Now let $(\mathcal{I}, \delta)$ and $(\mathcal{J}, \epsilon)$ be two pointed interpretations, and assume that $\Gamma \subseteq \Sigma$. A $\Gamma$-*simulation* from $(\mathcal{I}, \delta)$ to $(\mathcal{J}, \epsilon)$ is a relation $\mathfrak{S} \subseteq \Delta^\mathcal{I} \times \Delta^\mathcal{J}$ that satisfies $(\delta, \epsilon) \in \mathfrak{S}$ as well as the following conditions for all pairs $(\zeta, \eta) \in \mathfrak{S}$.

1. For all concept names $A \in \Gamma_\mathsf{C}$, if $\zeta \in A^\mathcal{I}$, then $\eta \in A^\mathcal{J}$.

2. For all role names $r \in \Gamma_\mathsf{R}$, if there is an element $\theta \in \Delta^\mathcal{I}$ such that $(\zeta, \theta) \in r^\mathcal{I}$, then there is an element $\iota \in \Delta^\mathcal{J}$ such that $(\eta, \iota) \in r^\mathcal{J}$ and $(\theta, \iota) \in \mathfrak{S}$.

We then also write $\mathfrak{S}\colon (\mathcal{I}, \delta) \precsim_\Gamma (\mathcal{J}, \epsilon)$, and to express the mere existence of a $\Gamma$-simulation from $(\mathcal{I}, \delta)$ to $(\mathcal{J}, \epsilon)$ we may write $(\mathcal{I}, \delta) \precsim_\Gamma (\mathcal{J}, \epsilon)$. Furthermore, if $\Gamma = \Sigma$, then we speak of *simulations* instead of $\Gamma$-simulations, and we leave out the subscript $\Gamma$, i.e., we use the symbol $\precsim$ instead of $\precsim_\Gamma$.

Assume that $(\mathcal{I}, \delta) \precsim (\mathcal{J}, \epsilon)$. It is easily verified that for all $\mathcal{EL}$ concept description $C$, it holds true that $\delta \in C^\mathcal{I}$ only if $\epsilon \in C^\mathcal{J}$. Further important notions and statements related to simulations are cited from Lutz and Wolter (2010) in the following.

**(Lutz and Wolter, 2010, Definition 11).** Let $\mathcal{T}$ be an $\mathcal{EL}$ TBox, and $C$ be an $\mathcal{EL}$ concept description. The *canonical model* $\mathcal{I}_{C,\mathcal{T}}$ of $\mathcal{T}$ and $C$ consists of the following components.

$$\Delta^{\mathcal{I}_{C,\mathcal{T}}} := \{C\} \cup \{\, D \mid \exists\, r \in \Sigma_\mathsf{R}\colon \exists\, r.\, D \in \mathsf{Sub}(\mathcal{T}) \cup \mathsf{Sub}(C) \,\}$$

$$\cdot^{\mathcal{I}_{C,\mathcal{T}}}\colon \begin{cases} A \mapsto \{\, D \mid D \sqsubseteq_\mathcal{T} A \,\} & \text{for any } A \in \Sigma_\mathsf{C} \\ r \mapsto \left\{ (D, E) \;\middle|\; \begin{array}{l} D \sqsubseteq_\mathcal{T} \exists\, r.\, E \text{ and } \exists\, r.\, E \in \mathsf{Sub}(\mathcal{T}), \\ \text{or } \exists\, r.\, E \in \mathsf{Conj}(D) \end{array} \right\} & \text{for any } r \in \Sigma_\mathsf{R} \end{cases}$$

Furthermore, we set $\mathcal{I}_C := \mathcal{I}_{C,\emptyset}$ for any $C \in \mathcal{EL}(\Sigma)$. △

**(Lutz and Wolter, 2010, Lemma 12).** *Let $\mathcal{T}$ be an $\mathcal{EL}$ TBox, and $C$ be an $\mathcal{EL}$ concept description. Then, the following statements hold true.*

1. $D \in D^{\mathcal{I}_{C,\mathcal{T}}}$ *for all* $D \in \Delta^{\mathcal{I}_{C,\mathcal{T}}}$

2. $\mathcal{I}_{C,\mathcal{T}} \models \mathcal{T}$

3. $(\mathcal{I}_{C,\mathcal{T}}, E) \precsim (\mathcal{I}_{D,\mathcal{T}}, E)$ *for all* $D \in \mathcal{EL}(\Sigma)$ *and all* $E \in \Delta^{\mathcal{I}_{C,\mathcal{T}}} \cap \Delta^{\mathcal{I}_{D,\mathcal{T}}}$ □

**(Lutz and Wolter, 2010, Lemma 13).** *Let $\mathcal{T}$ be an $\mathcal{EL}$ TBox, and $C$ be an $\mathcal{EL}$ concept description.*

1. *For all models $\mathcal{I}$ of $\mathcal{T}$ and all objects $\delta \in \Delta^\mathcal{I}$, the following statements are equivalent.*
   a) $\delta \in C^\mathcal{I}$
   b) $(\mathcal{I}_{C,\mathcal{T}}, C) \precsim (\mathcal{I}, \delta)$

2. *For all $\mathcal{EL}$ concept descriptions $D$, the following statements are equivalent.*
   a) $C \sqsubseteq_\mathcal{T} D$
   b) $C \in D^{\mathcal{I}_{C,\mathcal{T}}}$
   c) $(\mathcal{I}_{D,\mathcal{T}}, D) \precsim (\mathcal{I}_{C,\mathcal{T}}, C)$ □

As an immediate corollary, we get that the following two statements are equivalent for all $\mathcal{EL}$ concept descriptions $C$ and $D$, and thus yield a recursive procedure for checking subsumption with respect to $\emptyset$.

1. $C \sqsubseteq_\emptyset D$

2. $\mathsf{Conj}(D, \Sigma_\mathsf{C}) \subseteq \mathsf{Conj}(C, \Sigma_\mathsf{C})$, and for each existential restriction $\exists\, r.\, F \in \mathsf{Conj}(D)$, there is an existential restriction $\exists\, r.\, E \in \mathsf{Conj}(C)$ such that $E \sqsubseteq_\emptyset F$.

## 2.7. EDIT OPERATIONS

Let $(V, \prec)$ be tree-shaped directed graph with root vertex $v_0$ and such that all edges point away from $v_0$. Then the *induced partial order* $\leq$ on $V$ is defined as the reflexive transitive closure of $\prec$. Obviously, the root vertex $v_0$ is minimal with respect to $\leq$, and arbitrary, but not nullary, infima exist w.r.t. $\leq$. For a vertex $v \in V$, we define its *prime filter* as $\uparrow v := \{ w \in V \mid v \leq w \}$; the *filter* of a subset $U \subseteq V$ is defined as $\uparrow U := \bigcup \{ \uparrow u \mid u \in U \}$.

We say that an interpretation $\mathcal{I}$ is *tree-shaped* if the directed graph $(\Delta^{\mathcal{I}}, \bigcup\{ r^{\mathcal{I}} \mid r \in \Sigma_{\mathsf{R}} \})$ is tree-shaped. We shall denote the induced partial order as $\leq_{\mathcal{I}}$, or just as $\leq$ if it is clear from the context which interpretation is meant. Analogously, prime filters are symbolized as $\uparrow_{\mathcal{I}} \delta$, or simply as $\uparrow \delta$, for objects $\delta \in \Delta^{\mathcal{I}}$; and similarily for filters.

**Definitio 2.7.1.** We define the following *edit operations* on finite tree-shaped interpretations. For this purpose let $\mathcal{I}$ be such a finite tree-shaped interpretation with root $\delta$ and which is defined over a signature $\Sigma$.

1. For each $\epsilon \in \Delta^{\mathcal{I}}$ and each $A \in \Sigma_{\mathsf{C}}$, let $\mathsf{delete}(\mathcal{I}, \epsilon, A)$ be the interpretation with the following components.

$$\begin{aligned} \Delta^{\mathsf{delete}(\mathcal{I},\epsilon,A)} &:= \Delta^{\mathcal{I}} \\ A^{\mathsf{delete}(\mathcal{I},\epsilon,A)} &:= A^{\mathcal{I}} \setminus \{\epsilon\} \\ B^{\mathsf{delete}(\mathcal{I},\epsilon,A)} &:= B^{\mathcal{I}} && \text{for each } B \in \Sigma_{\mathsf{C}} \setminus \{A\} \\ r^{\mathsf{delete}(\mathcal{I},\epsilon,A)} &:= r^{\mathcal{I}} && \text{for each } r \in \Sigma_{\mathsf{R}} \end{aligned}$$

2. For each $\epsilon \in \Delta^{\mathcal{I}} \setminus \{\delta\}$ and each $n \in \mathbb{N}$, let $\mathsf{duplicate}(\mathcal{I}, \epsilon, n)$ be the interpretation with the following components.

$$\begin{aligned} \Delta^{\mathsf{duplicate}(\mathcal{I},\epsilon,n)} &:= (\Delta^{\mathcal{I}} \setminus \uparrow_{\mathcal{I}} \epsilon) \cup (\{1, \ldots, n\} \times \uparrow_{\mathcal{I}} \epsilon) \\ A^{\mathsf{duplicate}(\mathcal{I},\epsilon,n)} &:= (A^{\mathcal{I}} \setminus \uparrow_{\mathcal{I}} \epsilon) \cup (\{1, \ldots, n\} \times (A^{\mathcal{I}} \cap \uparrow_{\mathcal{I}} \epsilon)) \\ r^{\mathsf{duplicate}(\mathcal{I},\epsilon,n)} &:= (r^{\mathcal{I}} \setminus (\Delta^{\mathcal{I}} \times \uparrow_{\mathcal{I}} \epsilon)) \\ &\quad \cup \{ (\zeta, (i, \epsilon)) \mid (\zeta, \epsilon) \in r^{\mathcal{I}} \text{ and } i \in \{1, \ldots, n\} \} \\ &\quad \cup \{ ((i, \eta), (i, \theta)) \mid (\eta, \theta) \in r^{\mathcal{I}} \cap (\uparrow_{\mathcal{I}} \epsilon \times \uparrow_{\mathcal{I}} \epsilon) \text{ and } i \in \{1, \ldots, n\} \} \end{aligned}$$

We say that a finite tree-shaped interpretation $\mathcal{J}$ is *constructed* from $\mathcal{I}$ *with edit operations* if there is a finite sequence of edit operations starting with $\mathcal{I}$ and ending with $\mathcal{J}$. Accordingly, we say that an $\mathcal{EL}(\Sigma)$ concept description $D$ is *constructed* from an $\mathcal{EL}(\Sigma)$ concept description $C$ *with edit operations* if (an isomorphic copy of) the canonical model $\mathcal{I}_D$ is constructed from the canonical model $\mathcal{I}_C$ with edit operations. △

Note that we do not allow for applications of the $\mathsf{duplicate}$ operation to the root of an interpretation. This ensures that the result is always a tree-shaped finite interpretation too. Later in Section 3.1.2 we are going to ignore this current restriction; it is readily verified that the result of the $\mathsf{duplicate}$ operation applied to the root of a tree-shaped finite interpretation is a forest-shaped finite interpretation containing some copies of the input interpretation.

**Lemma 2.7.2.** *Let $C$ and $D$ be $\mathcal{EL}$ concept descriptions over some signature $\Sigma$. Then the following statements are equivalent.*

1. *$C$ is subsumed by $D$ with respect to the empty TBox.*

2. *$D$ is constructed from $C$ with edit operations.*

*Approbatio.* Let $C \sqsubseteq_\emptyset D$ and fix a homomorphism $\phi\colon (\mathcal{I}_D, D) \rightarrow (\mathcal{I}_C, C)$. Set $i := 1$, $\mathcal{I}_{C,1} := \mathcal{I}_C$, and $\phi_1 := \phi$. Apply the following rule exhaustively.

Find a minimal element $\delta_i \in \Delta^{\mathcal{I}_{C,i}}$ such that $|\phi_i^{-1}(\{\delta_i\})| \neq 1$, and set $n_i := |\phi_i^{-1}(\{\delta_i\})|$. Then set

$$\mathcal{I}_{C,i+1} := \mathsf{duplicate}(\mathcal{I}_{C,i}, \delta_i, n_i),$$

and define the mapping $\phi_{i+1}\colon \Delta^{\mathcal{I}_D} \to \Delta^{\mathcal{I}_{C,i+1}}$ as follows.

$$\begin{aligned}
\phi_{i+1}(\zeta) &:= \eta && \text{if } \phi_i(\zeta) = \eta \text{ and } \eta \in \Delta^{\mathcal{I}_{C,i}} \setminus \uparrow_{\mathcal{I}_{C,i}} \delta_i \\
\phi_{i+1}(\zeta_k) &:= (k, \delta_i) && \text{if } \phi_i^{-1}(\{\delta_i\}) = \{\zeta_1, \ldots, \zeta_{n_i}\} \text{ and } k \in \{1, \ldots, n_i\} \\
\phi_{i+1}(\zeta) &:= (k, \eta) && \text{if } \phi_i(\zeta) = \eta \text{ and } \eta \in \uparrow_{\mathcal{I}_{C,i}} \delta_i \setminus \{\delta_i\} \text{ and } \zeta_k \leq_{\mathcal{I}_{C,i}} \zeta \text{ for some } k \in \{1, \ldots, n_i\}
\end{aligned}$$

Finally, increment $i$.

We shall prove that each $\phi_i$ is a homomorphism from $(\mathcal{I}_D, D)$ to $(\mathcal{I}_{C,i}, C)$. Please note that the root node $C$ always satisfies that $\phi_i^{-1}(\{C\}) = \{D\}$, and is henceforth left untouched. Consequently, $\phi_i(D) = C$ holds true. Now consider an arbitrary element $\zeta \in \Delta^{\mathcal{I}_D}$; we proceed with a case distinction.

- Assume $\phi_i(\zeta) = \eta$ and $\eta \not\geq_{\mathcal{I}_{C,i}} \delta_i$, i.e., it holds true that $\phi_{i+1}(\zeta) = \eta$. Now if $\zeta \in A^{\mathcal{I}_D}$, then $\eta \in A^{\mathcal{I}_{C,i}}$, since $\phi_i$ is a homomorphism. Furthermore, $\eta \not\geq_{\mathcal{I}_{C,i}} \delta_i$ then implies that $\eta \in A^{\mathcal{I}_{C,i+1}}$ as well.

  Consider an $r$-successor $\zeta'$ of $\zeta$, i.e., an edge $(\zeta, \zeta') \in r^{\mathcal{I}_D}$. Since $\phi_i$ is a homomorphism, it follows that $(\phi_i(\zeta), \phi_i(\zeta')) \in r^{\mathcal{I}_{C,i}}$. We need to show that also $(\phi_{i+1}(\zeta), \phi_{i+1}(\zeta')) \in r^{\mathcal{I}_{C,i+1}}$ holds true. Let $\phi_i(\zeta') = \eta'$, i.e., $(\eta, \eta') \in r^{\mathcal{I}_{C,i}}$. Two cases are now possible: either $\eta' \not\geq_{\mathcal{I}_{C,i}} \delta_i$, or $\eta' = \delta_i$. In the first case it immediately follows that $\phi_{i+1}(\zeta') = \eta'$ and $(\eta, \eta') \in r^{\mathcal{I}_{C,i+1}}$, that is, $(\phi_{i+1}(\zeta), \phi_{i+1}(\zeta')) \in r^{\mathcal{I}_{C,i+1}}$. In the second case we know that $\phi_i^{-1}(\{\eta'\}) = \{\zeta_1, \ldots, \zeta_{n_i}\}$, and furthermore that $(\eta, (k, \eta')) \in r^{\mathcal{I}_{C,i+1}}$ as well as $\phi_{i+1}(\zeta_k) = (k, \eta')$ for all $k \in \{1, \ldots, n_i\}$. Since $\zeta' = \zeta_k$ for some $k$, we conclude that $\phi_{i+1}(\zeta') = (k, \eta')$ and thus $(\phi_{i+1}(\zeta), \phi_{i+1}(\zeta')) \in r^{\mathcal{I}_{C,i+1}}$.

- Let $\phi_i(\zeta) = \delta_i$, i.e., there exists a $k \in \{1, \ldots, n_i\}$ such that $\zeta = \zeta_k$ and $\phi_{i+1}(\zeta) = (k, \delta_i)$. If $\zeta \in A^{\mathcal{I}_D}$ for a concept name $A \in \Sigma_C$, then $\delta_i \in A^{\mathcal{I}_{C,i}}$, and consequently $(k, \delta_i) \in A^{\mathcal{I}_{C,i+1}}$.

  Assume that $(\zeta, \zeta') \in r^{\mathcal{I}_D}$, and hence $(\phi_i(\zeta), \phi_i(\zeta')) \in r^{\mathcal{I}_{C,i}}$. Since $\delta_i \leq_{\mathcal{I}_{C,i}} \phi_i(\zeta')$, we can immediately conclude that $((k, \phi_i(\zeta)), (k, \phi_i(\zeta'))) \in r^{\mathcal{I}_{C,i+1}}$, i.e., $(\phi_{i+1}(\zeta), \phi_{i+1}(\zeta')) \in r^{\mathcal{I}_{C,i+1}}$.

- Suppose that $\phi_i(\zeta) = \eta$, $\delta_i <_{\mathcal{I}_{C,i}} \eta$, and $\zeta_k \leq_{\mathcal{I}_D} \zeta$ for some $k \in \{1, \ldots, n_i\}$. By definition then $\phi_{i+1}(\zeta) = (k, \eta)$. If $\zeta \in A^{\mathcal{I}_D}$, then $\eta \in A^{\mathcal{I}_{C,i}}$ is immediate. The definition of $\mathcal{I}_{C,i+1}$ yields $(k, \eta) \in A^{\mathcal{I}_{C,i+1}}$.

  Let $(\zeta, \zeta') \in r^{\mathcal{I}_D}$, then we know that $(\phi_i(\zeta), \phi_i(\zeta')) \in r^{\mathcal{I}_{C,i}}$. Since $\delta_i <_{\mathcal{I}_{C,i}} \phi_i(\zeta) <_{\mathcal{I}_{C,i}} \phi_i(\zeta')$, we infer that $((k, \phi_i(\zeta)), (k, \phi_i(\zeta'))) \in r^{\mathcal{I}_{C,i+1}}$, i.e., $(\phi_{i+1}(\zeta), \phi_{i+1}(\zeta')) \in r^{\mathcal{I}_{C,i+1}}$.

If $n_i = 0$, then an application of the above mentioned rule would delete the whole subtree rooted at $\delta_i$. The following claim and its proof shows that this would not cause any problems and we may safely do so.

**Effatum 2.7.3.** *If $\delta \in \Delta^{\mathcal{I}_C} \setminus \mathsf{Ran}(\phi)$, then $\uparrow_{\mathcal{I}_C} \delta \cap \mathsf{Ran}(\phi) = \emptyset$.*

*Approbatio.* Assume that $\delta \leq_{\mathcal{I}_C} \epsilon$ and $\phi(\eta) = \epsilon$. We shall justify the existence of an element $\zeta$ such that $\phi(\zeta) = \delta$. We do this by induction on the length $\ell$ of *the* path from $\delta$ to $\epsilon$. If $\ell = 0$, then $\delta = \epsilon$ and henceforth we may choose $\zeta := \eta$. Now assume that $(\delta, \delta', \ldots, \epsilon)$ is a path of length $\ell + 1$. By induction hypothesis there exists an element $\zeta'$ with $\phi(\zeta') = \delta'$. Now let $\zeta$ be *the* unique parent of $\zeta'$, i.e., $(\zeta, \zeta') \in r^{\mathcal{I}_D}$ for some role name $r \in \Sigma_R$. Since $\zeta$ is reachable from the root node, $(\phi(\zeta), \phi(\zeta')) = (\phi(\zeta), \delta') \in r^{\mathcal{I}_C}$ must hold true. However, we also know that $(\delta, \delta') \in r^{\mathcal{I}_C}$, and we can thus infer that $\phi(\zeta) = \delta$. □

Let $\mathcal{I}_{C,*}$ be the interpretation and $\phi_*$ the homomorphism constructed by the last possible application of the above mentioned rule. Clearly, then $\phi_* \colon (\mathcal{I}_D, D) \precsim (\mathcal{I}_{C,*}, C)$ is a bijective homomorphism. The trees $\mathcal{I}_D$ and $\mathcal{I}_{C,*}$ can now only differ in the labels of vertices (and, of course, in the names of the vertices). To remove these differences, the following rule shall be applied exhaustively. Beforehand, set $i := 1$ and $\mathcal{I}_{C,*,1} := \mathcal{I}_{C,*}$.

> Find an element $\epsilon_i \in \Delta^{\mathcal{I}_{C,*,i}}$ and a concept name $A_i \in \Sigma_C$ such that $\epsilon_i \in A_i^{\mathcal{I}_{C,*,i}}$, but $\phi_*^{-1}(\epsilon_i) \notin A_i^{\mathcal{I}_D}$. Then set
>
> $$\mathcal{I}_{C,*,i+1} := \mathsf{delete}(\mathcal{I}_{C,*,i}, \epsilon_i, A_i),$$
>
> and increment $i$.

Denote by $\mathcal{I}_{C,*,*}$ the last constructed interpretation after which no further rule application is possible. It is readily verified that then $\phi_* \colon (\mathcal{I}_D, D) \precsim (\mathcal{I}_{C,*,*}, C)$ is an isomorphism, i.e., $\phi_*$ is a bijective homomorphism and its inverse $\phi_*^{-1}$ is a homomorphism too. Eventually, we have thus demonstrated that $D$ is constructed from $C$ with edit operations.

For proving the other direction, we show that applying an edit operation to a concept description $C$ always yields a subsumer of $C$. By a simple induction on the length of the sequence of edit operations the claim then follows.

- It is apparent that the identity is a homomorphism from $(\mathsf{delete}(\mathcal{I}, \epsilon, A), \delta)$ to $(\mathcal{I}, \delta)$ for arbitrary interpretations $\mathcal{I}$, objects $\delta, \epsilon \in \Delta^{\mathcal{I}}$, and concept names $A \in \Sigma_C$.
- Eventually, we define a homomorphism $\psi \colon (\mathsf{duplicate}(\mathcal{I}, \epsilon, n), \delta) \precsim (\mathcal{I}, \delta)$. Let $\zeta \in \Delta^{\mathcal{I}}$. If $\zeta \not\geq_{\mathcal{I}} \epsilon$, then $\psi(\zeta) := \zeta$. Otherwise, set $\psi(k, \zeta) := \zeta$ for all $k \in \{1, \ldots, n\}$. The proof that $\psi$ is indeed a homomorphism is obvious. □

## 2.8. GREATEST FIXED-POINT SEMANTICS

We cite two description logics introduced by Lutz, Piro, and Wolter (2010b) that are extensions of $\mathcal{EL}$ with greatest fixed-point semantics. According to (Lutz, Piro, and Wolter, 2010b, Theorem 10) there are polynomial time translations between both, and furthermore reasoning in these extensions remains **P**-complete, cf. (Lutz, Piro, and Wolter, 2010b, Theorem 12).

The description logic $\mathcal{EL}_{\mathsf{si}}$ extends $\mathcal{EL}$ by the concept constructor $\exists^{\mathsf{sim}}\,(\mathcal{I}, \delta)$ where $(\mathcal{I}, \delta)$ is a pointed interpretation such that $\mathcal{I}$ is finitely representable. The semantics of the additional concept constructor is defined as follows: for each interpretation $\mathcal{J}$ and any object $\epsilon \in \Delta^{\mathcal{J}}$, it holds true that $\epsilon \in (\exists^{\mathsf{sim}}\,(\mathcal{I}, \delta))^{\mathcal{J}}$ if $(\mathcal{I}, \delta) \precsim (\mathcal{J}, \epsilon)$. As shown in (Lutz, Piro, and Wolter, 2010b, Lemma 7), every $\mathcal{EL}_{\mathsf{si}}$ concept description is equivalent to a concept description of the form $\exists^{\mathsf{sim}}\,(\mathcal{I}, \delta)$, and furthermore, such an equivalent concept description can be constructed in linear time. Adding the bottom concept description $\bot$ yields the description logic $\mathcal{EL}^{\bot}_{\mathsf{si}}$.

Furthermore, Lutz, Piro, and Wolter (2010a, Definition 28) define the *nth characteristic concept description* $\mathsf{X}^n(\mathcal{I},\delta)$ of a pointed interpretation $(\mathcal{I},\delta)$ that has a finite active signature recursively as follows.

$$\mathsf{X}^0(\mathcal{I},\delta) := \bigsqcap\{\, A \mid A \in \Sigma_{\mathsf{C}} \text{ and } \delta \in A^{\mathcal{I}} \,\}$$
$$\mathsf{X}^{n+1}(\mathcal{I},\delta) := \mathsf{X}^0(\mathcal{I},\delta) \sqcap \bigsqcap\{\, \exists\, r.\, \mathsf{X}^n(\mathcal{I},\epsilon) \mid r \in \Sigma_{\mathsf{R}} \text{ and } (\delta,\epsilon) \in r^{\mathcal{I}} \,\}$$

For any finitely representable pointed interpretation $(\mathcal{I},\delta)$, the sequence $(\, \mathsf{X}^n(\mathcal{I},\delta) \mid n \in \mathbb{N}\,)$ converges to $\exists^{\mathsf{sim}}\,(\mathcal{I},\delta)$, that is, it holds true that

$$(\exists^{\mathsf{sim}}\,(\mathcal{I},\delta))^{\mathcal{J}} = \bigcap\{\, (\mathsf{X}^n(\mathcal{I},\delta))^{\mathcal{J}} \mid n \in \mathbb{N}\,\}$$

for every interpretation $\mathcal{J}$, and so we also call $\mathsf{X}^n(\mathcal{I},\delta)$ the *nth approximation* of $\exists^{\mathsf{sim}}\,(\mathcal{I},\delta)$. In general, we shall denote the $n$th approximation of an $\mathcal{EL}^{\bot}_{\mathsf{si}}$ concept description $C$ as $C\!\upharpoonright_n$ where we additionally need to define that $\bot\!\upharpoonright_n := \bot$ for each $n \in \mathbb{N}$. Clearly, if $C$ is an $\mathcal{EL}^{\bot}$ concept description with role depth $d$, then $C \equiv_{\emptyset} C\!\upharpoonright_n$ holds true for each $n \geq d$. Alternatively, we may call an $n$th approximation $C\!\upharpoonright_n$ also a *restriction* of $C$ to a role depth of $n$.

The description logic $\mathcal{EL}_{\mathsf{st}}$ extends $\mathcal{EL}$ by the concept constructor $\exists^{\mathsf{sim}}\,\Gamma.\,(\mathcal{T},C)$, where $\Gamma \subseteq \Sigma$ is a finite signature, $\mathcal{T}$ is a TBox, and $C$ is a concept description. More specifically, $\mathcal{EL}_{\mathsf{st}}$ concept descriptions, $\mathcal{EL}_{\mathsf{st}}$ concept inclusions, and $\mathcal{EL}_{\mathsf{st}}$ TBoxes are defined by simultaneous induction as follows.

1. Every $\mathcal{EL}$ concept description, $\mathcal{EL}$ concept inclusion, and $\mathcal{EL}$ TBox, is an $\mathcal{EL}_{\mathsf{st}}$ concept description, $\mathcal{EL}_{\mathsf{st}}$ concept inclusion, and $\mathcal{EL}_{\mathsf{st}}$ TBox, respectively;
2. if $\mathcal{T}$ is an $\mathcal{EL}_{\mathsf{st}}$ TBox, $C$ an $\mathcal{EL}_{\mathsf{st}}$ concept description, and $\Gamma \subseteq \Sigma$ a finite signature, then $\exists^{\mathsf{sim}}\,\Gamma.\,(\mathcal{T},C)$ is an $\mathcal{EL}_{\mathsf{st}}$ concept description;
3. if $C$ and $D$ are $\mathcal{EL}_{\mathsf{st}}$ concept descriptions, then $C \sqsubseteq D$ is an $\mathcal{EL}_{\mathsf{st}}$ concept inclusion;
4. an $\mathcal{EL}_{\mathsf{st}}$ TBox is a finite set of $\mathcal{EL}_{\mathsf{st}}$ concept inclusions.

The semantics of the additional concept constructor is defined as follows: let $\mathcal{I}$ be an interpretation, then $\delta \in (\exists^{\mathsf{sim}}\,\Gamma.\,(\mathcal{T},C))^{\mathcal{I}}$ if there exists a pointed interpretation $(\mathcal{J},\epsilon)$ such that $\mathcal{J}$ is a model of $\mathcal{T}$, $\epsilon \in C^{\mathcal{J}}$, and $(\mathcal{J},\epsilon) \precsim_{\Sigma\setminus\Gamma} (\mathcal{I},\delta)$. In case $\Gamma = \emptyset$ we may abbreviate $\exists^{\mathsf{sim}}\,\Gamma.\,(\mathcal{T},C)$ as $\exists^{\mathsf{sim}}\,(\mathcal{T},C)$. Adding the bottom concept description $\bot$ yields the description logic $\mathcal{EL}^{\bot}_{\mathsf{st}}$.

## 2.9. MOST SPECIFIC CONSEQUENCES

In (Kriegel, 2016a, 2018a), the author has introduced the notion of a most specific consequence as a new non-standard inference problem. Given a concept description $C$ and a TBox $\mathcal{T}$, the most specific consequence $C^{\mathcal{T}}$ is, informally speaking, a concept description which contains all information that follows from $C$ in $\mathcal{T}$. It has a number of interesting properties, e.g., we can reduce the problem of subsumption w.r.t. $\mathcal{T}$ to the problem of subsumption w.r.t. $\emptyset$ of the respective most specific consequences.

**(Kriegel, 2018a).** Consider a TBox $\mathcal{T}$ and an $\mathcal{EL}^{\bot}$ concept description $C$. Then an $\mathcal{EL}^{\bot}$ concept description $D$ is called a *most specific consequence* of $C$ with respect to $\mathcal{T}$ if it satisfies the following conditions.

1. $C \sqsubseteq_{\mathcal{T}} D$

2. For each $\mathcal{EL}^{\bot}$ concept description $E$, if $C \sqsubseteq_{\mathcal{T}} E$, then $D \sqsubseteq_{\emptyset} E$. △

It is readily verified that all most specific consequences of $C$ with respect to $\mathcal{T}$ are unique up to equivalence, and hence we shall denote *the* most specific consequence of $C$ with respect to $\mathcal{T}$ by $C^{\mathcal{T}}$—provided that it exists. Furthermore, it holds true that $C \sqsubseteq_{\mathcal{T}} C^{\mathcal{T}}$, and $C$ is of course a consequence of itself with respect to $\mathcal{T}$, i.e., $C^{\mathcal{T}} \sqsubseteq_{\emptyset} C$. Consequently, $C$ and its most specific consequence $C^{\mathcal{T}}$ are equivalent with respect to $\mathcal{T}$.

Most specific consequences need not exist in $\mathcal{EL}^{\bot}$. To see this, consider the exemplary TBox $\mathcal{T} := \{A \sqsubseteq \exists r. A\}$, and define $C_0 := A$ as well as $C_{n+1} := A \sqcap \exists r. C_n$ for all $n \in \mathbb{N}$. It is easy to verify that for each $n \in \mathbb{N}$, the concept description $C_n$ is a consequence of $A$ w.r.t. $\mathcal{T}$, and furthermore that $C_{n+1}$ is strictly more specific than $C_n$. Consequently, $A^{\mathcal{T}}$ does not exist in $\mathcal{EL}^{\bot}$. However, it always holds true that $C^{\mathcal{T}} \equiv_{\emptyset} \exists^{\mathsf{sim}}(\mathcal{I}_{C,\mathcal{T}}, C)$, i.e., most specific consequences exist in extensions of $\mathcal{EL}^{\bot}$ with greatest fixpoints.

**(Kriegel, 2018a).** *The mapping $\phi_{\mathcal{T}} \colon C \mapsto C^{\mathcal{T}}$ is a closure operator in the dual of $\boldsymbol{\mathcal{EL}}^{\bot}(\Sigma)$, i.e., for all $\mathcal{EL}^{\bot}$ concept descriptions $C$ and $D$, the following conditions are satisfied.*

1. $C^{\mathcal{T}} \sqsubseteq_{\emptyset} C$

2. $C^{\mathcal{T}} \equiv_{\emptyset} C^{\mathcal{T}\mathcal{T}}$

3. *$C \sqsubseteq_{\emptyset} D$ implies $C^{\mathcal{T}} \sqsubseteq_{\emptyset} D^{\mathcal{T}}$* □

**(Kriegel, 2018a).** *Let $\mathcal{T} \cup \{C \sqsubseteq D\}$ be an $\mathcal{EL}^{\bot}$ TBox. Then, the following statements are equivalent.*

1. $C \sqsubseteq_{\mathcal{T}} D$

2. $C^{\mathcal{T}} \sqsubseteq_{\emptyset} D$

3. $C^{\mathcal{T}} \sqsubseteq_{\emptyset} D^{\mathcal{T}}$

4. *$E^{\mathcal{T}} \sqsubseteq_{\emptyset} C$ implies $E^{\mathcal{T}} \sqsubseteq_{\emptyset} D$ for each $\mathcal{EL}^{\bot}$ concept description $E$.* □

## 2.10. MOST GENERAL DIFFERENCES

**Definitio 2.10.1.** Let $C, D \in \mathcal{EL}(\Sigma)$ be two concept descriptions such that $C \sqsubseteq_{\emptyset} D$. Then, some concept description $E \in \mathcal{EL}(\Sigma)$ is called *most general difference* (abbrv. MGD) of $C$ with respect to $D$ (or, alternatively, *complement of $D$ relative to $C$*) if it satisfies the following conditions.

1. $C \sqsubseteq_{\emptyset} E$

2. $C \equiv_{\emptyset} D \sqcap E$

3. $C \sqsubseteq_{\emptyset} F$ and $C \equiv_{\emptyset} D \sqcap F$ implies $F \sqsubseteq_{\emptyset} E$ for any $F \in \mathcal{EL}(\Sigma)$.

Furthermore, if $C \not\sqsubseteq_{\emptyset} D$, then a most general difference of $C$ w.r.t. $D$ is defined as a most general difference of $C$ w.r.t. $C \vee D$. △

It is an immediate consequence from the above definition that all most general differences of $C$ with respect to $D$ are equivalent. Thus, we shall denote *the* most general difference by $C \setminus D$ if it exists. Of course, in the extension $\mathcal{EL}^\bot$ of $\mathcal{EL}$ with the bottom concept description $\bot$ most general differences cannot exist, since $\bot \setminus C$ must be equivalent to the negation $\neg C$, which is a concept description that cannot be expressed in $\mathcal{EL}^\bot$ if $\bot \not\equiv_\emptyset C \not\equiv_\emptyset \top$. If we consider the exemplary concept descriptions $C := \exists\, r.\,(A \sqcap B)$ and $D := \exists\, r.\,A \sqcap \exists\, r.\,B$, then we see that $C \setminus D \equiv_\emptyset C$ holds true. We continue our investigations by considering the question whether such most general differences always exist.

**Definitio 2.10.2.** For two concept descriptions $C, D \in \mathcal{EL}^\bot(\Sigma)$, the *syntactic difference* of $C$ with respect to $D$ is defined as the following concept description.

$$C \mathbin{\setminus\!\!\setminus} D := \bigsqcap \mathsf{Conj}(C) \setminus \{\, E \mid D \sqsubseteq_\emptyset E \,\} \qquad \triangle$$

**Lemma 2.10.3.** *Most general differences always exist in $\mathcal{EL}$ and can be computed in deterministic polynomial time. In particular, $C \setminus D \equiv_\emptyset C \mathbin{\setminus\!\!\setminus} D$ holds true for any two concept descriptions $C, D \in \mathcal{EL}(\Sigma)$ satisfying $C \sqsubseteq_\emptyset D$.*

*Approbatio.* It is obvious that $C \sqsubseteq_\emptyset C \mathbin{\setminus\!\!\setminus} D$. This fact together with the precondition $C \sqsubseteq_\emptyset D$ implies that $C \sqsubseteq_\emptyset D \sqcap (C \mathbin{\setminus\!\!\setminus} D)$ is satisfied as well. Now fix some $X \in \mathsf{Conj}(C)$. In case $D \sqsubseteq_\emptyset X$ it immediately follows that $D \sqcap (C \mathbin{\setminus\!\!\setminus} D) \sqsubseteq_\emptyset X$. Otherwise if $D \not\sqsubseteq_\emptyset X$, then $X \in \mathsf{Conj}(C \mathbin{\setminus\!\!\setminus} D)$ holds true, which implies $D \sqcap (C \mathbin{\setminus\!\!\setminus} D) \sqsubseteq_\emptyset X$ as well. We conclude that $D \sqcap (C \mathbin{\setminus\!\!\setminus} D) \sqsubseteq_\emptyset C$ is satisfied.

Eventually, let $E \in \mathcal{EL}(\Sigma)$ such that $C \sqsubseteq_\emptyset E$ and $C \equiv_\emptyset D \sqcap E$. We shall show that $E \sqsubseteq_\emptyset C \mathbin{\setminus\!\!\setminus} D$. Fix some $Y \in \mathsf{Conj}(C \mathbin{\setminus\!\!\setminus} D)$, that is, $Y \in \mathsf{Conj}(C)$ such that $D \not\sqsubseteq_\emptyset Y$. Since $C \equiv_\emptyset D \sqcap E$, it follows that $D \sqcap E \sqsubseteq_\emptyset Y$ and so $D \not\sqsubseteq_\emptyset Y$ implies $E \sqsubseteq_\emptyset Y$. $\square$

**Lemma 2.10.4.** *The following statements hold true for any concept descriptions $C, D, E, F \in \mathcal{EL}(\Sigma)$.*

1. *$C \setminus D \sqsubseteq_\emptyset E \setminus F$ if $C \sqsubseteq_\emptyset E$ and $D \sqsupseteq_\emptyset F$*
2. *$C \setminus \top \equiv_\emptyset C$ or, more generally, $C \setminus D \equiv_\emptyset C$ if $C \perp_\emptyset D$*
3. *$(C \sqcap D) \setminus E \equiv_\emptyset (C \setminus E) \sqcap (D \setminus E)$*
4. *$(C \vee D) \setminus E \sqsupseteq_\emptyset (C \setminus E) \vee (D \setminus E)$*
5. *$(C \setminus D) \setminus E \equiv_\emptyset C \setminus (D \sqcap E)$*

*Approbatio.* Statements 1 and 2 easily follow from Lemma 2.10.3. Since $\mathsf{Conj}(C \sqcap D) = \mathsf{Conj}(C) \cup \mathsf{Conj}(D)$ holds true, Statement 3 follows from Lemma 2.10.3 as well. We shall now prove Statement 4.

It is well-known that $\mathsf{Conj}(C \vee D) = \{\, X \vee Y \mid X \in \mathsf{Conj}(C) \text{ and } Y \in \mathsf{Conj}(D) \,\}$ holds true. It then follows according to Lemma 2.10.3 that

$$\mathsf{Conj}((C \vee D) \setminus E) = \{\, X \vee Y \mid X \in \mathsf{Conj}(C) \text{ and } Y \in \mathsf{Conj}(D) \text{ such that } E \not\sqsubseteq_\emptyset X \vee Y \,\}$$

and likewise

$$\mathsf{Conj}((C \setminus E) \vee (D \setminus E)) = \{\, X \vee Y \mid X \in \mathsf{Conj}(C) \text{ and } Y \in \mathsf{Conj}(D) \text{ such that } E \not\sqsubseteq_\emptyset X \text{ and } E \not\sqsubseteq_\emptyset Y \,\}.$$

Clearly, we have that $\mathsf{Conj}((C \vee D) \setminus E) \subseteq \mathsf{Conj}((C \setminus E) \vee (D \setminus E))$, which yields the claim.

Statement 5 is again immediately clear due to Lemma 2.10.3. $\square$

Note that the converse direction of Statement 2 does not hold true: as a counterexample one can consider the concept descriptions $C \coloneqq \exists\, r.\, (A \sqcap B)$ and $D \coloneqq \exists\, r.\, A \sqcap \exists\, r.\, B$ again. Furthermore, a counterexample against the converse direction of Statement 4 is $C \coloneqq \exists\, r.\, (A \sqcap B_1)$, $D \coloneqq \exists\, r.\, (A \sqcap B_2)$, and $E \coloneqq \exists\, r.\, A$: it then holds true that $(C \vee D) \setminus E \equiv_\emptyset \top$ and $(C \setminus E) \vee (D \setminus E) \equiv_\emptyset E$.

We say that $C$ is *strongly not subsumed* by $D$, denoted as $\emptyset \models C \not\sqsubseteq\!\!\!\!/\; D$ or, alternatively, as $C \not\sqsubseteq\!\!\!\!/_\emptyset\; D$, if $C \not\sqsubseteq_\emptyset E$ for *each* $E \in \mathsf{Conj}(D)$. Note that, in contrast, it holds true that $C \not\sqsubseteq_\emptyset D$ if, and only if, there is *some* $E \in \mathsf{Conj}(D)$ such that $C \not\sqsubseteq_\emptyset E$.

**Lemma 2.10.5.** *Let $C, D, E \in \mathcal{EL}(\Sigma)$ be concept descriptions. If $C \sqsubseteq_\emptyset E$ and $D \not\sqsubseteq\!\!\!\!/_\emptyset\; E$, then $C \setminus D \sqsubseteq_\emptyset E$.*

*Approbatio.* Let $Z \in \mathsf{Conj}(E)$. Then, $C \sqsubseteq_\emptyset E$ implies there is some $X \in \mathsf{Conj}(C)$ such that $X \sqsubseteq_\emptyset Z$. Furthermore, from $D \not\sqsubseteq\!\!\!\!/_\emptyset\; E$ it follows that $D \not\sqsubseteq_\emptyset Z$. We infer that $D \not\sqsubseteq_\emptyset X$, that is, $X \in \mathsf{Conj}(C \setminus D)$ holds true as well. □

# 3. THE NEIGHBORHOOD PROBLEM FOR $\mathcal{EL}$ CONCEPT DESCRIPTIONS

In this section we consider the *neighborhood problem* for $\mathcal{EL}$. We have already seen that the set of $\mathcal{EL}$ concept descriptions constitutes a lattice. It is only natural to consider the question whether there exists a neighborhood relation which corresponds to the subsumption order. Remark that for an order relation $\leq$ on some set $P$ its *neighborhood relation* or *transitive reduction* is defined as

$$\prec := \lneq \setminus (\lneq \circ \lneq) = \{\, (p,q) \mid p \lneq q \text{ and there exists no } x \text{ such that } p \lneq x \lneq q \,\}.$$

Clearly, if $P$ is finite, then the transitive closure $\prec^+$ equals the irreflexive part $\lneq$. However, there are infinite ordered sets where this does not hold true; even worse, there are cases where $\prec^+$ is empty. Consider, for instance, the set $\mathbb{R}$ of real numbers with their usual ordering $\leq$. It is well-known that $\mathbb{R}$ is dense in itself, that is, for each pair $x \lneq y$, there is another real number $z$ such that $x \lneq z \lneq y$—thus, there are no neighboring real numbers. In general, we say that $\leq$ is *neighborhood generated* if $\prec^+ = \lneq$ is satisfied. Clearly, $\leq$ is a neighborhood generated order relation if, and only if, there is a finite path $p = x_0 \prec x_1 \prec \ldots \prec x_n = q$ for each pair $p \leq q$. An alternative formulation is the following. $\leq$ is not neighborhood generated if, and only if, there exists some pair $p \lneq q$ such that every finite path $p = x_0 \lneq x_1 \lneq \ldots \lneq x_n = q$ can be refined, that is, there is some index $i$ and an element $y$ such that $x_i \lneq y \lneq x_{i+1}$. Of course, if the order relation $\leq$ is *bounded*, i.e., for each element $p \in P$, there exists a finite upper bound on the lengths of $\lneq$-paths issuing from $p$, then $\leq$ is neighborhood generated.

Although boundedness of a poset $(P, \leq)$ is sufficient for neighborhood generatedness, it is not necessary. The following result of Ganter (2018) immediately implies that any unbounded poset can be order-embedded into some neighborhood generated poset, which then must be unbounded as well.

**(Ganter, 2018).** *Any poset $(P, \leq)$ is order-embeddable into some neighborhood generated poset.*

*Approbatio.* For some given poset $(P, \leq)$, we define another poset $(\leq, \sqsubseteq)$ where

$$(a,b) \sqsubseteq (c,d) \quad \text{if, and only if,} \quad (a,b) = (c,d) \text{ or } b \leq c.$$

As one quickly verifies, $\sqsubseteq$ is indeed reflexive, antisymmetric, and transitive. In the following, we shall denote the neighborhood relation of $\sqsubseteq$ by $\prec$.

We first show that $a \lneq b$ implies $(a,a) \prec (a,b)$. Let $a \lneq b$. Of course, we have that $(a,a) \sqsubsetneq (a,b)$. Now consider some pair $c \leq d$ such that $(a,a) \sqsubseteq (c,d) \sqsubsetneq (a,b)$. Then, it follows that $a \leq c \leq d \leq a$, which shows that $(c,d) = (a,a)$.

Analogously, we infer that $a \lneq b$ implies $(a,b) \prec (b,b)$. We conclude that $(a,a) \sqsubsetneq (b,b)$ always implies $(a,a) \prec (a,b) \prec (b,b)$.

Eventually, assume that $a \lneq b \lneq c \lneq d$, i.e., $(a,b) \sqsubsetneq (b,c) \sqsubsetneq (c,d)$ is satisfied. Applying the above yields that $(a,b) \prec (b,b) \prec (b,c) \prec (c,c) \prec (c,d)$. Consequently, we have that $\sqsubsetneq \subseteq \prec^+$. The converse inclusion is trivial. Thus, $(\leq, \sqsubseteq)$ is neighborhood generated.

It remains to show that there is an order-embedding of $(P, \leq)$ into $(\leq, \sqsubseteq)$. For this purpose, define the mapping $f \colon P \to \leq,\ p \mapsto (p,p)$. It is readily verified that $f$ is order-preserving as well as order-reflecting, which immediately implies that $f$ is injective as well. As a corollary, we obtain that $f$ is an order-embedding. □

In the sequel of this section, we shall address the neighborhood problem from different perspectives. We first consider the general problem of existence of neighbors, and then provide means for the computation of all upper neighbors and of all lower neighbors, respectively, in the cases where these exist. As it will turn out, neighbors only exist for all concept descriptions in the description logic $\mathcal{EL}$ without any TBox or in $\mathcal{EL}$ with respect to acyclic or cycle-restricted TBoxes. The presence of either a non-cycle-restricted TBox or of the bottom concept description $\bot$ prevents the existence of neighbors for some concept descriptions. Furthermore, the extensions of $\mathcal{EL}$ with greatest fixed-point semantics also allow for the construction of concept descriptions that do not possess neighbors. Eventually, a complexity analysis shows that deciding neighborhood in $\mathcal{EL}$ is in **P**, and that all upper neighbors of an $\mathcal{EL}$ concept description can be computed in deterministic polynomial time, while there exists some $\mathcal{EL}$ concept description that has exponentially many mutually distinct lower neighbors, and the sizes of reduced forms of lower neighbors are always linear.

**Definitio 3.1.** Consider a signature $\Sigma$, let $\mathcal{T}$ be a TBox over $\Sigma$, and further assume that $C$ and $D$ are concept descriptions over $\Sigma$. Then, $C$ is a *lower neighbor* or a *most general strict subsumee* of $D$ with respect to $\mathcal{T}$, denoted as $\mathcal{T} \models C \prec D$ or $C \prec_{\mathcal{T}} D$, if the following statements hold true.

1. $C \sqsubsetneq_{\mathcal{T}} D$
2. For each concept description $E$ over $\Sigma$, it holds true that $C \sqsubseteq_{\mathcal{T}} E \sqsubseteq_{\mathcal{T}} D$ implies $E \equiv_{\mathcal{T}} C$ or $E \equiv_{\mathcal{T}} D$.

Additionally, we then also say that $D$ is an *upper neighbor* or a *most specific strict subsumer* of $C$ with respect to $\mathcal{T}$, and we may also write $\mathcal{T} \models D \succ C$ or $D \succ_{\mathcal{T}} C$. △

Obviously, $\top$ does not have any upper neighbors, and dually $\bot$ does not have any lower neighbors.

We first observe that neighborhood of concept descriptions is not preserved by the concept constructors. It is easy to see that $A \sqcap B \prec_{\emptyset} A$. However, it holds true that $\exists\, r.\, (A \sqcap B) \sqsubsetneq_{\emptyset} \exists\, r.\, A \sqcap \exists\, r.\, B \sqsubsetneq_{\emptyset} \exists\, r.\, A$, which shows $\exists\, r.\, (A \sqcap B) \not\prec_{\emptyset} \exists\, r.\, A$. Furthermore, we have that $A \sqcap B \sqcap (A \sqcap B) \equiv_{\emptyset} A \sqcap (A \sqcap B)$, and consequently $A \sqcap B \sqcap (A \sqcap B) \not\prec_{\emptyset} A \sqcap (A \sqcap B)$. There are according counterexamples when neighborhood with respect to a non-empty TBox is considered.

It is easily verified that neighborhood with respect to the empty TBox $\emptyset$ does not coincide with neighborhood w.r.t. a non-empty TBox $\mathcal{T}$. For instance, $A \prec_{\emptyset} \top$ holds true, but $\{\top \sqsubseteq A\} \models A \equiv \top$.

For the converse direction, consider the counterexample where $\{A \sqsubseteq B,\ B \sqsubseteq A\} \models A \sqcap B \prec \top$ and $A \sqcap B \sqsubsetneq_\emptyset A \sqsubsetneq_\emptyset \top$.

## 3.1. THE EMPTY TBOX

Since Baader and Morawska (2010, Proof of Proposition 3.5) showed that $\sqsubseteq_\emptyset$ is bounded, we can immediately draw the following conclusion.

**Propositio 3.1.1.** *For any signature $\Sigma$, the subsumption relation $\sqsubseteq_\emptyset$ on $\mathcal{EL}(\Sigma)$ is neighborhood generated.* □

After this first promising result, we continue with describing the neighborhood relation $\prec_\emptyset$. As an immediate consequence of $\sqsubseteq_\emptyset$ being neighborhood generated, we can deduce that neighbors in arbitrary *directions* exist. More specifically, whenever $C \sqsubsetneq_\emptyset D$ holds true, there are $U$ and $L$ such that $C \prec_\emptyset U \sqsubseteq_\emptyset D$ as well as $C \sqsubseteq_\emptyset L \prec_\emptyset D$. We then also say that $U$ is an upper neighbor of $C$ *in direction* $D$ and, dually, that $L$ is some lower neighbor of $D$ *in direction* $C$.

**Lemma 3.1.2.** *Let $C$ and $D$ be $\mathcal{EL}^\bot$ concept descriptions over a signature $\Sigma$. Then $C \prec_\emptyset D$ holds true only if* $\mathsf{rd}(C) \in \{\mathsf{rd}(D),\ \mathsf{rd}(D)+1\}$.

*Approbatio.* Assume that $C$ is a lower neighbor of $D$ with respect to $\emptyset$. In particular, $C \sqsubseteq_\emptyset D$ follows, and so there is a simulation from the tree-shaped interpretation $(\mathcal{I}_D, D)$ to the tree-shaped interpretation $(\mathcal{I}_C, C)$. The mere existence of such a simulation yields that the depth of the tree $(\Delta^{\mathcal{I}_D}, \bigcup\{ r^{\mathcal{I}_D} \mid r \in \Sigma_\mathsf{R} \})$ is bounded by the depth of the tree $(\Delta^{\mathcal{I}_C}, \bigcup\{ r^{\mathcal{I}_C} \mid r \in \Sigma_\mathsf{R} \})$, that is, it must hold true that $\mathsf{rd}(D) \leq \mathsf{rd}(C)$.

Finally, assume that $\mathsf{rd}(C) > \mathsf{rd}(D) + 1$. Then $C \sqsubsetneq_\emptyset C\restriction_{\mathsf{rd}(D)+1} \sqsubsetneq_\emptyset D$. ↯ □

There is a well-known recursive characterization of $\sqsubseteq_\emptyset$ as follows: $C \sqsubseteq_\emptyset D$ if, and only if, $A \in \mathsf{Conj}(D)$ implies $A \in \mathsf{Conj}(C)$ for each concept name $A$, and for each $\exists\, r.\, F \in \mathsf{Conj}(D)$, there is some $\exists\, r.\, E \in \mathsf{Conj}(C)$ such that $E \sqsubseteq_\emptyset F$. With the help of that we can prove that there is the following necessary condition for neighboring concept descriptions.

### 3.1.1. A NECESSARY CONDITION

**Lemma 3.1.1.1.** *Let $C$ and $D$ be some reduced $\mathcal{EL}$ concept descriptions over a signature $\Sigma$. If $C \prec_\emptyset D$, then exactly one of the following statements holds true.*

1. *There is exactly one concept name $A \in \mathsf{Conj}(C)$ such that $C \equiv_\emptyset D \sqcap A$.*
2. *There is exactly one existential restriction $\exists\, r.\, E \in \mathsf{Conj}(C)$ such that $C \equiv_\emptyset D \sqcap \exists\, r.\, E$.*

*Approbatio.* Consider two reduced $\mathcal{EL}$ concept descriptions $C$ and $D$ over $\Sigma$ such that $C$ is a lower neighbor of $D$ with respect to $\emptyset$. It follows that $C \sqsubseteq_\emptyset D$, which means that $A \in \mathsf{Conj}(D)$ implies $A \in \mathsf{Conj}(C)$ for any concept name $A \in \Sigma_\mathsf{C}$ and further that, for each existential restriction $\exists\, r.\, F \in \mathsf{Conj}(D)$, there is some $\exists\, r.\, E \in \mathsf{Conj}(C)$ such that $E \sqsubseteq_\emptyset F$.

If there exist two distinct concept names $A, B \in \Sigma_C$ satisfying $\{A, B\} \subseteq \mathsf{Conj}(C) \setminus \mathsf{Conj}(D)$, then it would immediately follow that

$$C \sqsubseteq_\emptyset D \sqcap A \sqcap B \sqsubsetneq_\emptyset D \sqcap A \sqsubsetneq_\emptyset D,$$

which contradicts our assumption that $C \prec_\emptyset D$. ↯ Consequently, only one of the following two mutually exclusive cases can occur: either there is exactly one concept name $A \in \Sigma_C$ such that $\{A\} = (\mathsf{Conj}(C) \setminus \mathsf{Conj}(D)) \cap \Sigma_C$, or it holds true that $\mathsf{Conj}(C) \cap \Sigma_C = \mathsf{Conj}(D) \cap \Sigma_C$. We proceed with a case analysis.

1. Assume that $\{A\} = (\mathsf{Conj}(C) \setminus \mathsf{Conj}(D)) \cap \Sigma_C$ holds true for some concept name $A$. It follows that $C \sqsubseteq_\emptyset D \sqcap A \sqsubsetneq_\emptyset D$, and so $C \prec_\emptyset D$ implies $C \equiv_\emptyset D \sqcap A$.

2. Now let $\mathsf{Conj}(C) \cap \Sigma_C = \mathsf{Conj}(D) \cap \Sigma_C$. Since $C \not\sqsupseteq_\emptyset D$ holds true by assumption, there must exist some existential restriction $\exists\, r.\, E \in \mathsf{Conj}(C)$ such that $E \not\sqsupseteq_\emptyset F$ for any $\exists\, r.\, F \in \mathsf{Conj}(D)$. In particular, we have that $D \not\sqsubseteq_\emptyset \exists\, r.\, E$. Now suppose that there are two such existential restrictions $\exists\, r.\, E$ and $\exists\, s.\, F$ on the top-level conjunction of $C$. Since $C$ is assumed to be reduced, $\exists\, r.\, E$ and $\exists\, s.\, F$ are incomparable w.r.t. $\emptyset$. It follows that

$$C \sqsubseteq_\emptyset D \sqcap \exists\, r.\, E \sqcap \exists\, s.\, F \sqsubsetneq_\emptyset D \sqcap \exists\, r.\, E \sqsubsetneq_\emptyset D,$$

which obviously contradicts our assumption that $C \prec_\emptyset D$. As a consequence we obtain that there exists exactly one such existential restriction $\exists\, r.\, E \in \mathsf{Conj}(C)$ with $D \not\sqsubseteq_\emptyset \exists\, r.\, E$. It is now straight-forward to conclude that $C \sqsubseteq_\emptyset D \sqcap \exists\, r.\, E \sqsubsetneq_\emptyset D$ is satisfied, which together with the precondition $C \prec_\emptyset D$ implies that $C \equiv_\emptyset D \sqcap \exists\, r.\, E$. □

### 3.1.2. UPPER NEIGHBORHOOD

**Propositio 3.1.2.1.** *Let $C$ be a reduced $\mathcal{EL}$ concept description over some signature $\Sigma$, and recursively define*

$$\begin{aligned}\mathsf{Upper}(C) := \{ \textstyle\bigsqcap \mathsf{Conj}(C) \setminus \{A\} \mid A \in \mathsf{Conj}(C) \,\} \\ \cup \{ \textstyle\bigsqcap \mathsf{Conj}(C) \setminus \{\exists\, r.\, D\} \sqcap \bigsqcap \{ \exists\, r.\, E \mid E \in \mathsf{Upper}(D) \,\} \mid \exists\, r.\, D \in \mathsf{Conj}(C) \,\}.\end{aligned}$$

*Then $\mathsf{Upper}(C)$ contains, modulo equivalence, exactly all upper neighbors of $C$; more specifically, for each $\mathcal{EL}$ concept description $D$ over $\Sigma$, it holds true that*

$$C \prec_\emptyset D \quad \textit{if, and only if,} \quad \mathsf{Upper}(C) \ni D' \textit{ for some } D' \textit{ with } D \equiv_\emptyset D'.$$

*Approbatio.* We show the claim by induction on the role depth of $C$. The *induction base* where $\mathsf{rd}(C) = 0$ is obvious. For the *induction step* let now $\mathsf{rd}(C) > 0$.

*Soundness.* It is easily verified that, for any concept name $A \in \mathsf{Conj}(C)$, the concept description $\bigsqcap \mathsf{Conj}(C) \setminus \{A\}$ is an upper neighbor of $C$. Now fix some existential restriction $\exists\, r.\, E \in \mathsf{Conj}(C)$ and let

$$D := \textstyle\bigsqcap \mathsf{Conj}(C) \setminus \{\exists\, r.\, E\} \sqcap \bigsqcap \{ \exists\, r.\, F \mid F \in \mathsf{Upper}(E) \,\},$$

i.e., $D \in \mathsf{Upper}(C)$. We shall demonstrate that $C \prec_\emptyset D$.

1. It is easily verified that $C \sqsubseteq_\emptyset D$.

2. We proceed with proving that $C \not\sqsupseteq_\emptyset D$. In particular, we are going to show that there is no existential restriction $\exists\, r. F \in \mathsf{Conj}(D)$ such that $E \sqsupseteq_\emptyset F$. Assume that there was some such $F$. Since $C$ is reduced, we infer that $\exists\, r. F \notin \mathsf{Conj}(C)$, and hence $E \prec_\emptyset F$. ↯

3. Let $X$ be a concept description such that $C \sqsubsetneq_\emptyset X \sqsubseteq_\emptyset D$. We need to show that $X \sqsupseteq_\emptyset D$.

   a) According to the definition of $D$, it holds true that $\mathsf{Conj}(C) \cap \Sigma_\mathsf{C} = \mathsf{Conj}(D) \cap \Sigma_\mathsf{C}$. Furthermore, the precondition $C \sqsubsetneq_\emptyset X \sqsubseteq_\emptyset D$ implies that $\mathsf{Conj}(C) \cap \Sigma_\mathsf{C} \supseteq \mathsf{Conj}(X) \cap \Sigma_\mathsf{C} \supseteq \mathsf{Conj}(D) \cap \Sigma_\mathsf{C}$. We conclude that $A \in \mathsf{Conj}(D)$ implies $A \in \mathsf{Conj}(X)$ for any concept name $A \in \Sigma_\mathsf{C}$.

   b) Now consider an existential restriction $\exists\, s. Y \in \mathsf{Conj}(X)$. Then, $C \sqsubseteq_\emptyset X$ yields some $\exists\, s. G \in \mathsf{Conj}(C)$ satisfying $G \sqsubseteq_\emptyset Y$. If $s \neq r$ or $G \neq E$, then by definition of $D$ we have that $\exists\, s. G \in \mathsf{Conj}(D)$ as well.

      Eventually, we consider the case where $s = r$ and $G = E$. As $C \not\sqsupseteq_\emptyset X$, there exists some $\exists\, t. K \in \mathsf{Conj}(C)$ such that $K \not\sqsupseteq_\emptyset Z$ for all $\exists\, t. Z \in \mathsf{Conj}(X)$. If $t \neq r$ or $K \neq E$, then $\exists\, t. K \in \mathsf{Conj}(D)$ follows and immediately yields a contradiction to $X \sqsubseteq_\emptyset D$. As a corollary it follows that $E \sqsubsetneq_\emptyset Y$, and so there must be an upper neighbor $F$ of $E$ with $F \sqsubseteq_\emptyset Y$. The induction hypothesis ensures the existence of some concept description $F'$ with $F \equiv_\emptyset F'$ and $\exists\, r. F' \in \mathsf{Conj}(D)$.

Summing up, we have shown that $C \sqsubsetneq_\emptyset D$, and furthermore that, for each concept description $X$, it holds true that $C \sqsubsetneq_\emptyset X \sqsubseteq_\emptyset D$ implies $X \equiv_\emptyset D$. Hence, $C$ is a lower neighbor of $D$ with respect to $\emptyset$.

*Completeness.* Vice versa, consider a concept description $D$ such that $C \prec_\emptyset D$. Without loss of generality suppose that both $C$ and $D$ are reduced. We have to show that, up to equivalence, $\mathsf{Upper}(C) \ni D$. In accordance with Lemma 3.1.1.1 we shall only consider two cases. In the first case, if $C \equiv_\emptyset D \sqcap A$ for some unique concept name $A \in \mathsf{Conj}(C)$, the claim is trivial.

In the second case, there exists exactly one existential restriction $\exists\, r. E \in \mathsf{Conj}(C)$ such that $C \equiv_\emptyset D \sqcap \exists\, r. E$. We have already proven that $C \prec_\emptyset D'$ where

$$D' := \bigsqcap \mathsf{Conj}(C) \setminus \{\exists\, r. E\} \sqcap \bigsqcap \{\, \exists\, r. F \mid F \in \mathsf{Upper}(E)\,\}.$$

We proceed with demonstrating that $D' \sqsubseteq_\emptyset D$, which then immediately yields that $D' \equiv_\emptyset D$, and thus $D \in \mathsf{Upper}(C)$ modulo equivalence.

If $A \in \mathsf{Conj}(D)$, then it follows that $A \in \mathsf{Conj}(C)$ and further that $A \in \mathsf{Conj}(D')$. Now fix some existential restriction $\exists\, s. H \in \mathsf{Conj}(D)$. Then, there exists some $\exists\, s. G \in \mathsf{Conj}(C)$ satisfying $G \sqsubseteq_\emptyset H$. In case $s \neq r$ or $G \neq E$ we have that $\exists\, s. G \in \mathsf{Conj}(D')$ as well. Otherwise, $E \sqsubseteq_\emptyset H$ holds true. If $H \sqsubseteq_\emptyset E$ would hold true too, then it would follow that $D \sqsubseteq_\emptyset D \sqcap \exists\, r. E$, yielding $D \sqsubseteq_\emptyset C$—a contradiction to our assumption that $C \prec_\emptyset D$. So, we conclude that $E \sqsubsetneq_\emptyset H$. Thus, there exists an $F$ with $E \prec_\emptyset F \sqsubseteq_\emptyset H$. The induction hypothesis shows the existence of some $F' \in \mathsf{Upper}(E)$ with $F' \equiv_\emptyset F$, and then $\exists\, r. F' \in \mathsf{Conj}(D')$ is satisfied. □

For instance, consider the concept description $A \sqcap \exists\, r. B \sqcap \exists\, s. (A \sqcap B)$. It is in reduced form and has three upper neighbors, namely $\exists\, r. B \sqcap \exists\, s. (A \sqcap B)$, $A \sqcap \exists\, r. \top \sqcap \exists\, s. (A \sqcap B)$, and $A \sqcap \exists\, r. B \sqcap \exists\, s. A \sqcap \exists\, s. B$.

According to Propositio 3.1.2.1, each top-level conjunct $D$ of some concept description $C$ has exactly one upper neighbor $D^{\uparrow}$. If $C$ is reduced, then replacing $D$ with $D^{\uparrow}$ yields one upper neighbor of $C$, and (an equivalent concept description of) each upper neighbor of $C$ can be generated in this manner. We denote the concept description that is produced from $C$ by replacing $D$ with $D^{\uparrow}$ by $C^{\uparrow D}$. It then holds true that

$$\mathsf{Upper}(C) = \{\, C^{\uparrow D} \mid D \in \mathsf{Conj}(C) \,\}$$

modulo equivalence. Furthermore, there is a bijection between $\mathsf{Conj}(C)$ and $\mathsf{Upper}(C)$, cf. the next lemma.

**Lemma 3.1.2.2.** *Let $C$ be some reduced $\mathcal{EL}$ concept description. The mapping*

$$\begin{aligned} \upsilon_C \colon\ \mathsf{Conj}(C) &\to \mathsf{Upper}(C) \\ D &\mapsto C^{\uparrow D} \end{aligned}$$

*is bijective, that is, $|\mathsf{Conj}(C)| = |\mathsf{Upper}(C)|$ holds true.*

*Approbatio.* Apparently, $\upsilon_C$ is surjective. We proceed with demonstrating that it is injective as well. It is readily verified that removing one of the concept names on the top-level conjunction of $C$ yields one unique upper neighbor, that is, if $A, B \in \mathsf{Conj}(C)$ with $A \neq B$, then $C^{\uparrow A} = \bigsqcap \mathsf{Conj}(C) \setminus \{A\}$ and $C^{\uparrow B} = \bigsqcap \mathsf{Conj}(C) \setminus \{B\}$ are non-equivalent upper neighbors of $C$. Analogous statements obviously hold true for top-level conjuncts $A$ and $\exists\, r.\, D$, or $\exists\, r.\, D$ and $\exists\, s.\, E$ where $r \neq s$.

Eventually, assume that $\exists\, r.\, D$ and $\exists\, r.\, E$ are top-level conjuncts of $C$. Since we have assumed $C$ to be reduced, $D$ and $E$ are incomparable, i.e., it holds true that $D \not\sqsubseteq_\emptyset E$ as well as $E \not\sqsubseteq_\emptyset D$. These two conjuncts induce the following upper neighbors.

$$\begin{aligned} C^{\uparrow \exists\, r.\, D} &= \bigsqcap \mathsf{Conj}(C) \setminus \{\exists\, r.\, D\} \sqcap \bigsqcap \{\, \exists\, r.\, F \mid F \in \mathsf{Upper}(D) \,\} \\ C^{\uparrow \exists\, r.\, E} &= \bigsqcap \mathsf{Conj}(C) \setminus \{\exists\, r.\, E\} \sqcap \bigsqcap \{\, \exists\, r.\, F \mid F \in \mathsf{Upper}(E) \,\} \end{aligned}$$

For proving that $C^{\uparrow \exists\, r.\, D}$ and $C^{\uparrow \exists\, r.\, E}$ are incomparable, we assume the contrary, i.e., let $C^{\uparrow \exists\, r.\, D} \sqsubseteq_\emptyset C^{\uparrow \exists\, r.\, E}$. Since $\exists\, r.\, D \in \mathsf{Conj}(C^{\uparrow \exists\, r.\, E})$, there must exist some $\exists\, r.\, G \in \mathsf{Conj}(C^{\uparrow \exists\, r.\, D})$ satisfying $G \sqsubseteq_\emptyset D$. As $C$ is reduced, it cannot be the case that $\exists\, r.\, G \in \mathsf{Conj}(C) \setminus \{\exists\, r.\, D\}$; it can henceforth only happen that $\exists\, r.\, G \in \{\, \exists\, r.\, F \mid F \in \mathsf{Upper}(D) \,\}$, i.e., $G = F$ for some $F \in \mathsf{Upper}(D)$. Thus, we have that $F = G \sqsubseteq_\emptyset D \prec_\emptyset F$—a contradiction. ↯ □

**Lemma 3.1.2.3.** *Let $C \in \mathcal{EL}^\bot(\Sigma)$ be a concept description. For each set $\mathbf{D}$ containing only upper neighbors of $C$ and at least two incomparable upper neighbors of $C$, it holds true that $C \equiv_\emptyset \bigsqcap \mathbf{D}$.*

*Approbatio.* Without loss of generality let $C$ be reduced. Assume that $\mathbf{D}$ consists of upper neighbors of $C$ only and further contains two incomparable upper neighbors $D$ and $E$ of $C$. In particular, there must exist incomparable top-level conjuncts $X, Y \in \mathsf{Conj}(C)$ such that $D \equiv_\emptyset C^{\uparrow X}$ and $E \equiv_\emptyset C^{\uparrow Y}$. Obviously, it now follows that

$$C \sqsubseteq_\emptyset \bigsqcap \mathbf{D} \sqsubseteq_\emptyset D \sqcap E \equiv_\emptyset C^{\uparrow X} \sqcap C^{\uparrow Y} \equiv_\emptyset C.$$ □

### 3.1.3. LOWER NEIGHBORHOOD

#### A FIRST CHARACTERIZATION

**Propositio 3.1.3.1.** *For an $\mathcal{EL}$ concept description $C$ over some signature $\Sigma$, let*

$$\begin{aligned}\mathsf{Lower}(C) := \{\, C \sqcap A \mid A \in \Sigma_{\mathsf{C}} \text{ and } C \not\sqsubseteq_{\emptyset} A \,\} \\ \cup \{\, C \sqcap \exists\, r.\, D \mid r \in \Sigma_{\mathsf{R}},\ C \not\sqsubseteq_{\emptyset} \exists\, r.\, D, \text{ and } C \sqsubseteq_{\emptyset} \exists\, r.\, E \text{ for all } E \text{ with } D \prec_{\emptyset} E \,\}.\end{aligned}$$

*Then $\mathsf{Lower}(C)$ contains, modulo equivalence, exactly all lower neighbors of $C$; more specifically, for each $\mathcal{EL}$ concept description $D$ over $\Sigma$, it holds true that*

$$D \prec_{\emptyset} C \quad \text{if, and only if,} \quad D' \in \mathsf{Lower}(C) \text{ for some } D' \text{ with } D \equiv_{\emptyset} D'.$$

*Approbatio. Soundness.* We begin with proving *soundness*. Thus, fix some $L \in \mathsf{Lower}(C)$ and, without loss of generality, let $C$ be reduced. If $L = C \sqcap A$ for some concept name $A$ with $C \not\sqsubseteq_{\emptyset} A$, then it is apparent that $L$ is a lower neighbor of $C$. Henceforth, suppose $L = C \sqcap \exists\, r.\, D$ for some role name $r$ and a concept description $D$ which satisfies $C \not\sqsubseteq_{\emptyset} \exists\, r.\, D$ as well as $C \sqsubseteq_{\emptyset} \exists\, r.\, E$ for each upper neighbor $E$ of $D$. Then, it follows that $L \sqsubsetneq_{\emptyset} C$. Furthermore, $C$ is obviously equivalent to the concept description

$$C' := C \sqcap \bigsqcap \{\, \exists\, r.\, E \mid E \in \mathsf{Upper}(D) \,\},$$

and it is readily verified that $C' \in \mathsf{Upper}(L)$. Propositio 3.1.2.1 shows that $L \prec_{\emptyset} C'$ holds true, which yields $L \prec_{\emptyset} C$.

*Completeness.* We continue with showing *completeness*. For this purpose, consider a lower neighbor $L$ of $C$. Without loss of generality, assume that both $C$ and $L$ are reduced. According to Lemma 3.1.1.1, two mutually exclusive cases can occur. In the first case there exists a concept name $A$ such that $L \equiv_{\emptyset} C \sqcap A$. Clearly, $C \not\sqsubseteq_{\emptyset} A$ must hold true, as otherwise $L \equiv_{\emptyset} C \sqcap A \equiv_{\emptyset} C$. ↯ We conclude that $C \sqcap A \in \mathsf{Lower}(C)$. In the second case, there is exactly one existential restriction $\exists\, r.\, D \in \mathsf{Conj}(L)$ such that $L \equiv_{\emptyset} C \sqcap \exists\, r.\, D$. Since $L \prec_{\emptyset} C$ holds true, and Propositio 3.1.2.1 yields that

$$L \equiv_{\emptyset} C \sqcap \exists\, r.\, D \prec_{\emptyset} C \sqcap \bigsqcap \{\, \exists\, r.\, E \mid E \in \mathsf{Upper}(D) \,\} \sqsubseteq_{\emptyset} C,$$

it follows that $C \not\sqsubseteq_{\emptyset} \exists\, r.\, D$ as well as $C \equiv_{\emptyset} C \sqcap \bigsqcap \{\, \exists\, r.\, E \mid E \in \mathsf{Upper}(D) \,\}$, or equivalently, that $C \sqsubseteq_{\emptyset} \exists\, r.\, E$ for all $E$ with $D \prec_{\emptyset} E$. Summing up, we have shown that $C \sqcap \exists\, r.\, D \in \mathsf{Lower}(C)$. □

While the recursive characterization of Upper in Propositio 3.1.2.1 immediately yields a procedure for enumerating all upper neighbors of a given concept description, the situation is not that apparent for lower neighbors. We can, however, constitute a procedure for computing lower neighbors by means of Propositio 3.1.3.1. Let $C$ be an $\mathcal{EL}$ concept description over some signature $\Sigma$. Proceed as follows.

1. For each concept name $A \in \Sigma_{\mathsf{C}}$ with $C \not\sqsubseteq_{\emptyset} A$, output $C \sqcap A$ as a lower neighbor of $C$.
2. For each role name $r \in \Sigma_{\mathsf{R}}$, recursively proceed as follows.
   a) Let $D := \top$.
   b) While $C \sqsubseteq_{\emptyset} \exists\, r.\, D$, replace $D$ with a lower neighbor of $D$.
   c) If $C \sqsubseteq_{\emptyset} \exists\, r.\, E$ for all $E$ with $D \prec_{\emptyset} E$, then output $C \sqcap \exists\, r.\, D$ as a lower neighbor of $C$.

As we shall infer from the results in Section 4.5, the above algorithm always terminates but has non-elementary time complexity. Thus, we are going to develop a cheaper procedure for enumerating all lower neighbors of a given $\mathcal{EL}$ concept description in the next section. A complexity analysis shows that the proposed procedure needs only non-deterministic polynomial time or deterministic exponential time, and that there indeed exist $\mathcal{EL}$ concept descriptions with an exponential number of lower neighbors.

## A MORE EFFICIENT CHARACTERIZATION

According to Propositio 3.1.3.1, we can enumerate all lower neighbors of the form $C \sqcap A$ by simply iterating through the set of concept names while checking, for each such $A \in \Sigma_\mathsf{C}$, whether $C \not\sqsubseteq_\emptyset A$ or, equivalently, whether $A \notin \mathsf{Conj}(C)$ is satisfied and if so, then output $C \sqcap A$ as a lower neighbor of $C$. Clearly, this can be done in polynomial time with respect to the size of $C$ plus the size of $\Sigma$.

Let $C \in \mathcal{EL}(\Sigma)$ be some reduced concept description and consider a role name $r \in \Sigma_\mathsf{R}$. Then, for each subset $\mathbf{S} \subseteq \mathsf{Succ}(C, r)$, we define a mapping $\mathsf{Choices}_\mathbf{S} \colon \mathbf{S} \to \wp(\mathcal{EL}(\Sigma))$ as follows.

$$\mathsf{Choices}_\mathbf{S} \colon F \mapsto \{\, X \mid X \in \mathcal{EL}(\Sigma) \text{ such that } F \sqcap X \prec_\emptyset F \text{ and } F' \sqsubseteq_\emptyset X \text{ for each } F' \in \mathbf{S} \setminus \{F\} \,\}$$

According to Propositio 3.1.3.1, each such set $\mathsf{Choices}_\mathbf{S}(F)$ contains only atomic concept descriptions, i.e., concept descriptions that are either a concept name or some existential restriction. In the following, we consider choice functions in $\bigtimes \mathsf{Choices}_\mathbf{S} := \bigtimes\{\, \mathsf{Choices}_\mathbf{S}(F) \mid F \in \mathbf{S} \,\}$. We call some such choice function $\chi \in \bigtimes \mathsf{Choices}_\mathbf{S}$ *admissible* if $C \not\sqsubseteq_\emptyset \exists\, r. \bigsqcap \mathsf{Ran}(\chi)$.

**Lemma 3.1.3.2.** *A choice function $\chi \in \bigtimes \mathsf{Choices}_\mathbf{S}$ is admissible if, and only if, $\overline{F} \not\sqsubseteq_\emptyset \bigsqcap \mathsf{Ran}(\chi)$ for each $\overline{F} \in \mathsf{Succ}(C, r) \setminus \mathbf{S}$.*

*Approbatio.* Fix some $\chi \in \bigtimes \mathsf{Choices}_\mathbf{S}$. The *only if* direction is obvious. We continue with proving the *if* direction, for which it suffices to show that $D \not\sqsubseteq_\emptyset \bigsqcap \mathsf{Ran}(\chi)$ holds true for any $D \in \mathsf{Succ}(C, r)$. By assumption, this is satisfied for each $D \in \mathsf{Succ}(C, r) \setminus \mathbf{S}$. Furthermore, the above definition shows that $F \not\sqsubseteq_\emptyset \chi(F)$ for each $F \in \mathbf{S}$, which immediately implies that $F \not\sqsubseteq_\emptyset \bigsqcap \mathsf{Ran}(\chi)$ for any $F \in \mathbf{S}$, and we are done. □

**Propositio 3.1.3.3.** *A concept description $C \sqcap \exists\, r. D$ is a lower neighbor of $C$ if there is some subset $\mathbf{S} \subseteq \mathsf{Succ}(C, r)$ as well as an admissible choice function $\chi \in \bigtimes \mathsf{Choices}_\mathbf{S}$ such that $D \equiv_\emptyset \bigsqcap \mathsf{Ran}(\chi)$.*

*Approbatio.* Since $\chi$ is admissible, we have that $C \not\sqsubseteq_\emptyset \exists\, r. D$. Thus, in order to show that $C \sqcap \exists\, r. D$ is a lower neighbor of $C$ it remains to prove that $C \sqsubseteq_\emptyset \exists\, r. E$ for any upper neighbor $E$ of $D$, cf. Propositio 3.1.3.1.

We proceed with showing that all concept descriptions in $\mathsf{Ran}(\chi)$ are mutually incomparable, which implies that, modulo equivalence, the upper neighbors of $\bigsqcap \mathsf{Ran}(\chi)$ are exactly the concept descriptions $\bigsqcap \mathsf{Ran}(\chi)^{\uparrow \chi(F)}$ for $F \in \mathbf{S}$. If $F, F' \in \mathbf{S}$ are incomparable, then $X$ and $X'$ are incomparable as well for any $X \in \mathsf{Choices}_\mathbf{S}(F)$ and for any $X' \in \mathsf{Choices}_\mathbf{S}(F')$: otherwise it would hold true that $F' \sqsubseteq_\emptyset X \sqsubseteq_\emptyset X'$ or $F \sqsubseteq_\emptyset X' \sqsubseteq_\emptyset X$, which both yields a contradiction. ↯ Consequently, all top-level conjuncts in $\bigsqcap \mathsf{Ran}(\chi)$ must be mutually incomparable, and so there are bijections between $\mathbf{S}$, $\mathsf{Ran}(\chi)$, and $\mathsf{Upper}(\bigsqcap \mathsf{Ran}(\chi))$.

Fix some $F \in \mathbf{S}$, i.e., $\chi(F)$ is a top-level conjunct in $\bigsqcap \mathsf{Ran}(\chi)$ and $\bigsqcap \mathsf{Ran}(\chi)^{\uparrow \chi(F)}$ is an upper neighbor of $\bigsqcap \mathsf{Ran}(\chi)$. Then, we have that $F \sqcap \chi(F) \prec_\emptyset F$, and $F \sqsubseteq_\emptyset \chi(F')$ for each $F' \in \mathbf{S} \setminus \{F\}$. Furthermore, it holds true that $F \sqcap \chi(F) \sqsubseteq_\emptyset F \sqcap \chi(F)^\uparrow \sqsubseteq_\emptyset F$. Now assume that $F \sqcap \chi(F)^\uparrow \sqsubseteq_\emptyset F \sqcap \chi(F)$ would be satisfied, which would imply that $F \sqcap \chi(F)^{\uparrow} \sqsubseteq_{\emptyset} \chi(F)$. Since $\chi(F)^{\uparrow} \sqsubseteq_{\emptyset} \chi(F)$ cannot hold true, we would infer that $F \sqsubseteq_{\emptyset} \chi(F)$. However, this yields the contradiction $F \equiv_{\emptyset} F \sqcap \chi(F) \prec_{\emptyset} F$. ↯ We conclude that $F \sqcap \chi(F) \sqsubsetneq_{\emptyset} F \sqcap \chi(F)^{\uparrow} \sqsubseteq_{\emptyset} F$, which together with the precondition $F \sqcap \chi(F) \prec_{\emptyset} F$ implies that $F \sqcap \chi(F)^{\uparrow} \equiv_{\emptyset} F$. Clearly, this implies that $F \sqsubseteq_{\emptyset} \bigsqcap \mathsf{Ran}(\chi)^{\uparrow \chi(F)}$ and, thus, $C \sqsubseteq_{\emptyset} \exists\, r. \bigsqcap \mathsf{Ran}(\chi)^{\uparrow \chi(F)}$. □

**Lemma 3.1.3.4.** *Let $C \sqcap \exists\, r. D$ be a lower neighbor of $C$ where both $C$ and $D$ are reduced. Then, there is a mapping $\phi\colon \mathsf{Conj}(D) \to \mathsf{Succ}(C, r)$ with the following properties.*

1. *$\phi(X) \sqsubseteq_{\emptyset} D^{\uparrow X}$ for each $X \in \mathsf{Conj}(D)$*
2. *$\phi$ is injective*
3. *$\phi(X) \not\sqsubseteq_{\emptyset} X$ for any top-level conjunct $X \in \mathsf{Conj}(D)$*
4. *$\phi(Y) \sqsubseteq_{\emptyset} X$ for any two mutually distinct $X, Y \in \mathsf{Conj}(D)$*

*Approbatio.* Fix reduced concept descriptions $C, D \in \mathcal{EL}^{\bot}(\Sigma)$ and some role name $r \in \Sigma_{\mathsf{R}}$ such that $C \sqcap \exists\, r. D \prec_{\emptyset} C$. An application of Propositio 3.1.3.1 yields that $C \not\sqsubseteq_{\emptyset} \exists\, r. D$ and $C \sqsubseteq_{\emptyset} \exists\, r. E$ for each upper neighbor $E$ of $D$.

1. We start with defining such a mapping $\phi\colon \mathsf{Conj}(D) \to \mathsf{Succ}(C, r)$. Fix some top-level conjunct $X \in \mathsf{Conj}(D)$. Then, $D^{\uparrow X}$ is an upper neighbor of $D$. Since $C \sqsubseteq_{\emptyset} \exists\, r. D^{\uparrow X}$ is satisfied according to the preconditions, we conclude that there exists some successor $F_X \in \mathsf{Succ}(C, r)$ such that $F_X \sqsubseteq_{\emptyset} D^{\uparrow X}$. Thus, we can set $\phi(X) := F_X$.
2. We now show that $\phi$ is injective. Assume the contrary, i.e., there are two non-equivalent top-level conjuncts $X, Y \in \mathsf{Conj}(D)$ such that $\phi(X) = \phi(Y)$. It then holds true that $\phi(X) \sqsubseteq_{\emptyset} D^{\uparrow X} \sqcap D^{\uparrow Y}$. Now Lemma 3.1.2.3 implies that $D^{\uparrow X} \sqcap D^{\uparrow Y} \equiv_{\emptyset} D$, which contradicts the assumption that $C \not\sqsubseteq_{\emptyset} \exists\, r. D$. ↯
3. Assume to the contrary that $\phi(X) \sqsubseteq_{\emptyset} X$ is satisfied. Of course, it then immediately follows that $\phi(X) \sqsubseteq_{\emptyset} X \sqcap D^{\uparrow X} \equiv_{\emptyset} D$ would be satisfied, which contradicts the assumption that $C \not\sqsubseteq_{\emptyset} \exists\, r. D$. ↯
4. Let $X$ be a top-level conjunct of $D$. It then follows that $D^{\uparrow X}$ is some upper neighbor of $D$ and, for each upper neighbor $E$ of $D$ that is incomparable to $D^{\uparrow X}$, it holds true that $E \sqsubseteq_{\emptyset} X$, cf. Propositio 3.1.2.1. Fix a further top-level conjunct $Y \in \mathsf{Conj}(D)$ that is incomparable to $X$. Of course, $D^{\uparrow Y}$ is incomparable to $D^{\uparrow X}$, since $\upsilon_D\colon Z \mapsto D^{\uparrow Z}$ is a bijection between $\mathsf{Conj}(D)$ and $\mathsf{Upper}(D)$, cf. Lemma 3.1.2.2. We conclude that $\phi(Y) \sqsubseteq_{\emptyset} D^{\uparrow Y} \sqsubseteq_{\emptyset} X$. □

As a corollary we obtain that $|\mathsf{Conj}(D)| = |\mathsf{Upper}(D)| \leq |\mathsf{Succ}(C, r)|$ holds true.

**Propositio 3.1.3.5.** *A concept description $C \sqcap \exists\, r. D$ is a lower neighbor of $C$ only if there is some subset $\mathbf{S} \subseteq \mathsf{Succ}(C, r)$ as well as an admissible choice function $\chi \in \bigtimes \mathsf{Choices}_{\mathbf{S}}$ such that $D \equiv_{\emptyset} \bigsqcap \mathsf{Ran}(\chi)$.*

*Approbatio.* We know that there is some injective mapping $\phi\colon \mathsf{Conj}(D) \to \mathsf{Succ}(C, r)$ with all the properties stated in Lemma 3.1.3.4. Set $\mathbf{S} := \mathsf{Ran}(\phi)$, and define a mapping $\chi$ by $\chi(F) := X$ if $F = \phi(X)$.

We proceed with showing that $\chi(F) \in \mathsf{Choices}_{\mathbf{S}}(F)$ for each $F \in \mathbf{S}$, and for this purpose we have to show that $F \sqcap \chi(F) \prec_{\emptyset} F$ and $F' \sqsubseteq_{\emptyset} \chi(F)$ for each $F' \in \mathbf{S} \setminus \{F\}$. Fix some $F \in \mathbf{S}$. Then, $\chi(F) = X$ if, and only if, $F = \phi(X)$.

- We have that $\phi(X) \sqsubseteq_\emptyset D^{\uparrow X}$. In particular, this implies that $\phi(X) \sqsubseteq_\emptyset X^\uparrow$, that is, $F \sqsubseteq_\emptyset \chi(F)^\uparrow$. From $\phi(X) \not\sqsubseteq_\emptyset X$, we immediately infer that $F \not\sqsubseteq_\emptyset \chi(F)$. It follows that $F \sqcap \chi(F) \sqsubsetneq_\emptyset F$ and, since $(F \sqcap \chi(F))^{\uparrow\chi(F)} = F \sqcap \chi(F)^\uparrow \equiv_\emptyset F$ is an upper neighbor of $F \sqcap \chi(F)$, we conclude that $F \sqcap \chi(F)$ is a lower neighbor of $F$ as claimed.

- Furthermore, we have that $\phi(Y) \sqsubseteq_\emptyset X$ for each $Y \in \mathsf{Conj}(D) \setminus \{X\}$. If we now consider some $F' \in \mathbf{S} \setminus \{F\}$, then there is some $Y \in \mathsf{Conj}(D) \setminus \{X\}$ satisfying $\chi(F') = Y$ or, equivalently, $F' = \phi(Y)$. We conclude that $F' \sqsubseteq_\emptyset \chi(F)$.

Summing up, we have that $\chi$ is a choice function in $\bigtimes \mathsf{Choices}_{\mathbf{S}}$. Obviously, it holds true that $D \equiv_\emptyset \bigsqcap \mathsf{Ran}(\chi)$ and, thus, $\chi$ is admissible. □

**Corollarium 3.1.3.6.** *Let $C \in \mathcal{EL}(\Sigma)$ be some reduced concept description and define the following.*

$$\mathsf{Lower}^*(C) = \{\, C \sqcap A \mid A \in \Sigma_\mathsf{C} \text{ and } C \not\sqsubseteq_\emptyset A \,\} \cup \left\{ C \sqcap \exists r. \bigsqcap \mathsf{Ran}(\chi) \;\middle|\; \begin{array}{l} r \in \Sigma_\mathsf{R} \text{ and there exists some } \mathbf{S} \subseteq \mathsf{Succ}(C,r) \\ \text{such that } \chi \in \bigtimes \mathsf{Choices}^*_{\mathbf{S}} \text{ and } \chi \text{ is admissible} \end{array} \right\}$$

*Note that, for each subset $\mathbf{S} \subseteq \mathsf{Succ}(C,r)$, we define the mapping $\mathsf{Choices}^*_{\mathbf{S}} \colon \mathbf{S} \to \wp(\mathcal{EL}(\Sigma))$ slightly different from $\mathsf{Choices}_{\mathbf{S}}$, namely as follows.*

$$\mathsf{Choices}^*_{\mathbf{S}} \colon F \mapsto \{\, X \mid X \in \mathcal{EL}(\Sigma) \text{ such that } F \sqcap X \in \mathsf{Lower}^*(F) \text{ and } F' \sqsubseteq_\emptyset X \text{ for each } F' \in \mathbf{S} \setminus \{F\} \,\}$$

*Then $\mathsf{Lower}^*(C)$ contains, modulo equivalence, exactly all lower neighbors of $C$; more specifically, for each $\mathcal{EL}$ concept description $D$ over $\Sigma$, it holds true that*

$$D \prec_\emptyset C \quad \text{if, and only if,} \quad D' \in \mathsf{Lower}^*(C) \text{ for some } D' \text{ with } D \equiv_\emptyset D'.$$ □

**Lemma 3.1.3.7.** *For a fixed concept description $C$ as well as a fixed role name $r$, all admissible choice functions are incomparable with respect to $\subseteq$. In particular, if $\mathbf{S} \subsetneq \mathbf{T} \subseteq \mathsf{Succ}(C,r)$, then there does not exist admissible choice functions $\chi \in \bigtimes \mathsf{Choices}^*_{\mathbf{S}}$ and $\psi \in \bigtimes \mathsf{Choices}^*_{\mathbf{T}}$ such that $\chi \subseteq \psi$.*

*Approbatio.* Consider some $G \in \mathbf{T} \setminus \mathbf{S}$. Further assume that $\chi \in \mathsf{Choices}^*_{\mathbf{S}}$ is admissible, i.e., it follows that $G \not\sqsubseteq_\emptyset \bigsqcap \mathsf{Ran}(\chi)$, which shows that there exists some $F \in \mathbf{S}$ such that $G \not\sqsubseteq_\emptyset \chi(F)$. Consequently, we cannot extend $\chi$ to some (admissible) choice function $\psi$ in $\bigtimes \mathsf{Choices}_{\mathbf{T}}$. □

**Corollarium 3.1.3.8.** *For any reduced $\mathcal{EL}$ concept description $C$, it holds true that all lower neighbors in $\mathsf{Lower}^*(C)$ are mutually incomparable.* □

### 3.1.4. COMPUTATIONAL COMPLEXITY

Eventually, we finish our investigations of $\prec_\emptyset$ with analyzing the computational complexity of three problems related to the neighborhood of $\mathcal{EL}$ concept descriptions. In particular, we shall prove the following results.

- $\prec_\emptyset$ is in **P**.

- Upper can be computed in deterministic quadratic time. In particular, each upper neighbor in $\mathsf{Upper}(C)$ has a quadratic size, and $\mathsf{Upper}(C)$ has a linear cardinality.

- Lower$^*$ can be computed in deterministic exponential time. Furthermore, any lower neighbor in Lower$^*(C)$ has a quadratic size, and Lower$^*(C)$ has an exponential cardinality.
- There is a non-deterministic polynomial time procedure which on input $C$ has one (successful) computation path that returns a concept description equivalent to $L$ for any lower neighbor $L$ of $C$.

### ENUMERATING ALL UPPER NEIGHBORS

**Propositio 3.1.4.1.** *The mapping* Upper *can be computed in deterministic polynomial time. More specifically,* Upper$(C)$ *can be enumerated in deterministic quadratic time w.r.t.* $||C||$ *for each reduced* $\mathcal{EL}$ *concept description* $C$*.*

*Approbatio.* We could try to prove the claim by induction on the role depth of $C$. However, the straight-forward attempt to do so would only yield that Upper$(C)$ is computable in deterministic time $\mathcal{O}(||C||^{\mathsf{rd}(C)+2})$. Thus, we shall follow a more sophisticated approach.

For a finite set $\mathbf{C}$ of reduced $\mathcal{EL}$ concept descriptions, its size is defined by $||\mathbf{C}|| := \sum(\,||C|| \mid C \in \mathbf{C}\,)$; further let

$$\begin{aligned} \mathsf{Upper}(\mathbf{C})\colon \mathbf{C} &\to \wp(\mathcal{EL}(\Sigma)) \\ C &\mapsto \mathsf{Upper}(C), \end{aligned}$$

and the size of Upper$(\mathbf{C})$ is defined as $||\mathsf{Upper}(\mathbf{C})|| := \sum(\,||\mathsf{Upper}(C)|| \mid C \in \mathbf{C}\,)$. More generally, we shall show by induction on the maximal role depth $\mathsf{rd}(\mathbf{C}) := \bigvee\{\,\mathsf{rd}(C) \mid C \in \mathbf{C}\,\}$ that Upper$(\mathbf{C})$ can be computed in deterministic time $\mathcal{O}(||\mathbf{C}||^2)$, which implies that $||\mathsf{Upper}(\mathbf{C})|| \in \mathcal{O}(||\mathbf{C}||^2)$.

The induction base where $\mathsf{rd}(\mathbf{C}) = 0$ is obvious. For the induction step assume $\mathsf{rd}(\mathbf{C}) > 0$. For computing a single Upper$(C)$ we can proceed as follows. For each top-level conjunct of $C$, create a fresh copy of $C$. Clearly, the number of copies is bounded by $||C||$, and creating these copies hence takes time quadratic in $||C||$. From some of those copies one concept name is removed, which reduces the size of that copy, and one removal needs constant time. The sequence of these removal operations thus requires time linear in $||C||$. Furthermore, for some other copies, a top-level conjunct $\exists\, r.\, D$ is replaced by $\bigsqcap\{\,\exists\, r.\, E \mid E \in \mathsf{Upper}(D)\,\}$. Let $\mathsf{Succ}(C) := \bigcup\{\,\mathsf{Succ}(C,r) \mid r \in \Sigma_{\mathsf{R}}\,\} = \{\,D \mid \exists\, r.\, D \in \mathsf{Conj}(C)\,\}$. By induction hypothesis, the object Upper(Succ$(C)$) can be computed in deterministic time $\mathcal{O}(||\mathsf{Succ}(C)||^2)$ and has size $\mathcal{O}(||\mathsf{Succ}(C)||^2)$. It is apparent that $||\mathsf{Succ}(C)|| \leq ||C||$, and henceforth Upper(Succ$(C)$) can be computed in time $\mathcal{O}(||C||^2)$ and has size $\mathcal{O}(||C||^2)$. For each top-level conjunct $\exists\, r.\, D \in \mathsf{Conj}(C)$, we choose a distinct and so far untouched copy of $C$, remove $\exists\, r.\, D$, which takes time linear in $||C||$, and add $\exists\, r.\, E$ as new top-level conjunct for each $E \in \mathsf{Upper}(D)$, which takes constant time for finding Upper$(D)$ within Upper(Succ$(C)$) if Upper(Succ$(C)$) is computed as a function like above, and requires constant time for adding each $\exists\, r.\, E$ for $E \in \mathsf{Upper}(D)$ as a new top-level conjunct, since $E$ is already computed and we only need to link it to the copy we are editing. Since the number of top-level conjuncts of $C$ which are existential restrictions is bounded by $||C||$, and each replacement takes linear time in $||C||$, as the number of concept descriptions in each Upper$(D)$ is bounded by $|\mathsf{Conj}(C)| \leq ||C||$, we conclude that only quadratic time in $||C||$ is necessary for the replacement of the existential restrictions. Furthermore, the size of Upper$(C)$ is quadratic in $||C||$ too, since in the set of the in $||C||$ linearily many copies of $C$ we have removed some nodes and edges, and have added existential restrictions the fillers of which are from the in $||C||$ quadratically sized Upper(Succ$(C)$). Finally, if we consider the task of computing $\mathsf{Upper}(\mathbf{C})$, then we can compute, for each $C \in \mathbf{C}$, the set $\mathsf{Upper}(C)$ in time $\mathcal{O}(||C||^2)$ and collect the results in a function. Clearly, this takes $\mathcal{O}(\sum( ||C||^2 \mid C \in \mathbf{C} )) = \mathcal{O}(||\mathbf{C}||^2)$ time, and $||\mathsf{Upper}(\mathbf{C})||$ can similarily be bounded. □

### DECIDING NEIGHBORHOOD

**Lemma 3.1.4.2.** *It holds true that $\prec_\emptyset \in \mathbf{P}$. More specifically, we can decide in deterministic polynomial time w.r.t. $||C|| + ||D||$ whether $C$ is a lower neighbor of $D$ for any $\mathcal{EL}$ concept descriptions $C$ and $D$.*

*Approbatio.* We leave out picky details like the encoding of $\mathcal{EL}$ concept description, and recognizing correctly encoded $\mathcal{EL}$ concept descriptions. So, assume that $C$ and $D$ are $\mathcal{EL}$ concept descriptions. We want to show the existence of a procedure which, given $C$ and $D$ as input, decides in deterministic polynomial time whether $C$ is a lower neighbor of $D$ with respect to $\emptyset$. Such a procedure can, e.g., work as follows for input concept descriptions $C$ and $D$.

1. Reduce $C$.
2. Compute $\mathsf{Upper}(C)$.
3. Check whether there is some concept description $E \in \mathsf{Upper}(C)$ such that $D \equiv_\emptyset E$. If yes, accept $(C, D)$, and otherwise reject $(C, D)$.

Step 1 needs polynomial time in $||C||$. Step 2 also needs polynomial time in $||C||$, cf. Propositio 3.1.4.1. Since $\sqsubseteq_\emptyset \in \mathbf{P}$ holds true, $|\mathsf{Upper}(C)| \leq |\mathsf{Conj}(C)| \leq ||C||$ is satisfied, and $||E|| \leq ||\mathsf{Upper}(C)|| \in \mathcal{O}(||C||^2)$ for each $E \in \mathsf{Upper}(C)$, we infer that, for some $n$ that is the exponent for deciding $\sqsubseteq_\emptyset$, Step 3 requires deterministic time in $\mathcal{O}(||C|| \cdot (||C||^2 + ||D||)^n)$, which clearly is polynomial in $||C|| + ||D||$. □

### ENUMERATING ALL LOWER NEIGHBORS

**Lemma 3.1.4.3.** *Let $C$ be some reduced $\mathcal{EL}$ concept description over the signature $\Sigma$. Then, it holds true that*

$$|\mathsf{Lower}^*(C)| \leq |\Sigma| \cdot (|\Sigma| \cdot ||C|| \cdot 2^{||C||-1})^{\mathsf{rd}(C)}.$$

*Approbatio.* We show the claim by induction on the role depth of $C$. If $\mathsf{rd}(C) = 0$, then it holds true that $|\mathsf{Lower}^*(C)| \leq |\Sigma|$, simply because any lower neighbor in $\mathsf{Lower}^*(C)$ is either of the form $C \sqcap A$ for some concept name $A \in \Sigma_\mathsf{C}$ or of the form $C \sqcap \exists\, r. \top$ for a role name $r \in \Sigma_\mathsf{R}$.

Now assume that $\mathsf{rd}(C) > 0$. Then, we have the following.

$$|\mathsf{Lower}^*(C)| \leq |\Sigma_\mathsf{C}| + \sum_{r \in \Sigma_\mathsf{R}} \sum_{\mathbf{S} \subseteq \mathsf{Succ}(C,r)} |\bigtimes \mathsf{Choices}^*_\mathbf{S}|$$

Furthermore, we can an estimate upper bound for each $|\bigtimes \mathsf{Choices}^*_\mathbf{S}|$ as follows.

$$\begin{aligned} |\bigtimes \mathsf{Choices}^*_\mathbf{S}| &\leq |\mathbf{S}| \cdot \max_{F \in \mathbf{S}} |\mathsf{Choices}^*_\mathbf{S}(F)| \\ &\leq |\mathbf{S}| \cdot \max_{F \in \mathbf{S}} |\mathsf{Lower}^*(F)| \end{aligned}$$

Applying the induction hypothesis to any $F \in \mathbf{S}$ yields the following.

$$\begin{aligned} |\mathsf{Lower}^*(F)| &\leq |\Sigma| \cdot (|\Sigma| \cdot ||F|| \cdot 2^{||F||-1})^{\mathsf{rd}(F)} \\ &\leq |\Sigma| \cdot (|\Sigma| \cdot ||C|| \cdot 2^{||C||-1})^{\mathsf{rd}(C)-1} \end{aligned}$$

Summing up shows the following.

$$|\mathsf{Lower}^*(C)| \leq |\Sigma_\mathsf{C}| + \sum_{r\in\Sigma_\mathsf{R}} \sum_{\mathbf{S}\subseteq\mathsf{Succ}(C,r)} |\mathbf{S}| \cdot |\Sigma| \cdot (|\Sigma| \cdot ||C|| \cdot 2^{||C||-1})^{\mathsf{rd}(C)-1}$$

It is easy to verify that $\sum_{k=0}^{n} \binom{n}{k} \cdot k = n \cdot 2^{n-1}$, and so we can continue with the following.

$$\begin{aligned}\sum_{\mathbf{S}\subseteq\mathsf{Succ}(C,r)} |\mathbf{S}| &= \sum_{k=0}^{|\mathsf{Succ}(C,r)|} \binom{|\mathsf{Succ}(C,r)|}{k} \cdot k \\ &= |\mathsf{Succ}(C,r)| \cdot 2^{|\mathsf{Succ}(C,r)|-1} \\ &\leq ||C|| \cdot 2^{||C||-1}\end{aligned}$$

Putting the last two results together provides the following.

$$\begin{aligned}|\mathsf{Lower}^*(C)| &\leq |\Sigma_\mathsf{C}| + |\Sigma_\mathsf{R}| \cdot (||C|| \cdot 2^{||C||-1}) \cdot |\Sigma| \cdot (|\Sigma| \cdot ||C|| \cdot 2^{||C||-1})^{\mathsf{rd}(C)-1} \\ &= |\Sigma_\mathsf{C}| + |\Sigma_\mathsf{R}| \cdot (|\Sigma| \cdot ||C|| \cdot 2^{||C||-1})^{\mathsf{rd}(C)} \\ &\leq |\Sigma| \cdot (|\Sigma| \cdot ||C|| \cdot 2^{||C||-1})^{\mathsf{rd}(C)}\end{aligned}$$

□

**Propositio 3.1.4.4.** *Fix some reduced concept description $C \in \mathcal{EL}(\Sigma)$. Then, for each lower neighbor $D \in \mathsf{Lower}^*(C)$, it holds true that the size of $D$ is quadratic in the size of $C$.*

*Approbatio.* We show the claim by induction on the role depth of $C$—more specifically, we prove that any lower neighbor $D \in \mathsf{Lower}^*(C)$ satisfies $||D|| \leq (3 + \mathsf{rd}(C)) \cdot ||C|| + 1$.

Clearly, if $\mathsf{rd}(C) = 0$, then each $D \in \mathsf{Lower}(C)$ must be of the form $C \sqcap A$ for some concept name $A \in \Sigma_\mathsf{C}$ or of the form $C \sqcap \exists r.\top$ for some role name $r \in \Sigma_\mathsf{R}$. Obviously, this shows that $||D|| \leq ||C|| + 3 \leq 3 \cdot ||C|| + 1$.

Now assume that $\mathsf{rd}(C) > 0$. If $D$ is of the form $C \sqcap A$, we again have that $||D|| \leq 3 \cdot ||C|| + 1 \leq (3 + \mathsf{rd}(C)) \cdot ||C|| + 1$. Thus, we continue with the non-trivial case where $D$ has a form $C \sqcap \exists r. \bigsqcap \mathsf{Ran}(\chi)$ for some role name $r \in \Sigma_\mathsf{R}$ and a subset $\mathbf{S} \subseteq \mathsf{Succ}(C,r)$ such that $\chi \in \bigtimes \mathsf{Choices}^*_\mathbf{S}$ is an admissible choice function. It follows that $F \sqcap \chi(F) \in \mathsf{Lower}^*(F)$ for each $F \in \mathbf{S}$. An application of the induction hypothesis yields that $||F \sqcap \chi(F)|| \leq (3 + \mathsf{rd}(F)) \cdot ||F|| + 1$, and so we infer that $||\chi(F)|| \leq (2 + \mathsf{rd}(F)) \cdot ||F||$. Summing up, we have that

$$\begin{aligned}&||\bigsqcap \mathsf{Ran}(\chi)|| \\ &= |\mathsf{Ran}(\chi)| - 1 + \sum(\, ||\chi(F)|| \mid F \in \mathbf{S}\,) \\ &\leq ||C|| - 1 + \sum(\, (2 + \mathsf{rd}(F)) \cdot ||F|| \mid F \in \mathbf{S}\,) \\ &\leq ||C|| - 1 + \sum(\, (2 + \mathsf{rd}(C) - 1) \cdot ||F|| \mid F \in \mathbf{S}\,) \\ &\leq ||C|| - 1 + (2 + \mathsf{rd}(C) - 1) \cdot \sum(\, ||F|| \mid F \in \mathbf{S}\,) \\ &\leq ||C|| - 1 + (2 + \mathsf{rd}(C) - 1) \cdot ||C|| \\ &= (2 + \mathsf{rd}(C)) \cdot ||C|| - 1\end{aligned}$$

and hence $||C \sqcap \exists r. \bigsqcap \mathsf{Ran}(\chi)|| \leq (3 + \mathsf{rd}(C)) \cdot ||C|| + 1$ holds true. □

**Corollarium 3.1.4.5.** *For each reduced $\mathcal{EL}$ concept description $C$ over some signature $\Sigma$, it holds true that the size of an (efficient) encoding of $\mathsf{Lower}^*(C)$ is exponential in $||C|| + |\Sigma|$.* □

**Propositio 3.1.4.6.** *The mapping* Lower$^*$ *can be computed in deterministic exponential time. More specifically, for any reduced $C \in \mathcal{EL}(\Sigma)$, the set* Lower$^*(C)$ *is computable in deterministic exponential time with respect to $||C|| + |\Sigma|$.*

*Approbatio.* Using arguments from the proof of Lemma 3.1.4.3, the fact that subsumption in $\mathcal{EL}$ can be decided in polynomial time, and Propositio 3.1.4.4, we see that enumerating all admissible choice functions as required in Corollarium 3.1.3.6 takes at most exponential time with respect to $||C|| + |\Sigma|$. This shows the claim. □

As a further result regarding the computational complexity of computing lower neighbors, we have the following. While it shows a lower complexity for the problem of generating one lower neighbor of some given $\mathcal{EL}$ concept description, one can obviously not expect the proposed procedure to outperform algorithms that efficiently implement Corollarium 3.1.3.6. However, it would be not to hard to suitably adapt the deterministic manner of these algorithms to let them work in a non-deterministic fashion. That way, we can significantly decrease the number of failing computation paths.

**Propositio 3.1.4.7.** *For any $\mathcal{EL}$ concept description $C$, we can compute one lower neighbor of $C$ in non-deterministic polynomial time with respect to $||C|| + |\Sigma|$. More specifically, there is a non-deterministic polynomial time procedure such that, for any lower neighbor $L$ of $C$, it has a (successful) computation path that returns some concept description equivalent to $L$, when started on $C$ as input.*

*Approbatio.* The claim essentially is a consequence of Lemma 3.1.4.2 and Propositio 3.1.4.4, and the well-known fact that any $\mathcal{EL}$ concept description can be reduced in polynomial time. In particular, a suitable algorithm could work as follows on an input $C$.

1. Reduce $C$.
2. Guess some $\mathcal{EL}$ concept description $L$ such that $||L|| \leq (3 + \mathsf{rd}(C)) \cdot ||C|| + 1$ is satisfied.
3. Check whether $L$ is a lower neighbor of $C$. If yes, then return $L$; otherwise fail. □

The next lemma's aim is to show that the means of enumerating all lower neighbors from Corollarium 3.1.3.6 is optimal in terms of computational complexity. In particular, each efficient algorithmization of Corollarium 3.1.3.6 runs in exponential time, cf. the above proposition, and there is some example showing that $\mathcal{EL}$ concept descriptions can indeed have exponentially many lower neighbors, cf. the below lemma.

**Lemma 3.1.4.8.** *There is a sequence of signatures $\Sigma_n$ and concept descriptions $C_n \in \mathcal{EL}(\Sigma_n)$ such that, for any $n \in \mathbb{N}$, the set* Lower$^*(C_n)$ *of (representatives of) lower neighbors of $C_n$ has a cardinality that is exponential in the size of $\Sigma_n$ plus the size of $C_n$.*

*Approbatio.* We define a sequence of signatures $\Sigma_n$ and concept descriptions $C_n \in \mathcal{EL}(\Sigma_n)$ as follows. Fix some $n \in \mathbb{N}$ such that $n \geq 2$. Set $(\Sigma_n)_\mathsf{C} := \{ A_i, B_i \mid i \in \{1, \ldots, n\} \}$ and $(\Sigma_n)_\mathsf{R} := \{r\}$. Furthermore, let

$$C_n := \bigsqcap\{ \exists r. D_n^i \mid i \in \{1, \ldots, n\} \} \quad \text{where} \quad D_n^i := \bigsqcap\{ A_j, B_j \mid j \in \{1, \ldots, n\} \setminus \{i\} \}.$$

If we now set $\mathbf{S} := \mathsf{Succ}(C_n, r)$, then it obviously holds true that $\mathsf{Choices}^*_\mathbf{S}(D_n^i) = \{A_i, B_i\}$ for any index $i \in \{1, \ldots, n\}$. It is easy to verify that any choice function $\chi \in \bigtimes \mathsf{Choices}^*_\mathbf{S}$ is admissible and further that there are exponentially many such choice functions, i.e., $C_n$ has $\Omega(2^n)$ mutually incomparable lower neighbors while the size of $C_n$ is $\mathcal{O}(n^2)$ and the size of $\Sigma_n$ is $\mathcal{O}(n)$. □

### 3.1.5. APPLICATIONS

**Propositio 3.1.5.1.** *Let $\Xi \subseteq \mathcal{EL}(\Sigma)$ be a problem that is closed under subsumees, that is, $C \in \Xi$ and $C \sqsupseteq_\emptyset D$ implies $D \in \Xi$. We consider the problem* $\mathsf{Max}_\emptyset(\Xi)$*, which consists of all most general elements of* $\Xi$*.*

1. $\Xi \in \mathbf{C}$ *implies* $\mathsf{Max}_\emptyset(\Xi) \in \mathbf{P}^\mathbf{C}$ *for each complexity class* $\mathbf{C}$*.*
2. $\Xi \in \mathbf{P}$ *implies* $\mathsf{Max}_\emptyset(\Xi) \in \mathbf{P}$
3. $\Xi \in \mathbf{\Sigma}^\mathbf{P}_n$ *implies* $\mathsf{Max}_\emptyset(\Xi) \in \mathbf{\Delta}^\mathbf{P}_{n+1}$ *for any number* $n \in \mathbb{N}$*.*
4. $\Xi \in \mathbf{C}$ *implies* $\mathsf{Max}_\emptyset(\Xi) \in \mathbf{C}$ *for any complexity class* $\mathbf{C}$ *such that* $\mathbf{PSpace} \subseteq \mathbf{C}$*.*
5. $\Xi \in \mathbf{PSpace}$ *implies* $\mathsf{Max}_\emptyset(\Xi) \in \mathbf{PSpace}$

*Approbatio.* We only prove Statement 1; the others are then obtained as corollaries. In particular, for Statement 4 we need that $\mathbf{P}^\mathbf{C} \subseteq \mathbf{C}$ holds true for any complexity class $\mathbf{C}$ such that $\mathbf{PSpace} \subseteq \mathbf{C}$. In case $\mathbf{C} = \mathbf{PSpace}$ this follows from

$$\mathbf{PSpace} \subseteq \mathbf{P}^{\mathbf{PSpace}} \subseteq \mathbf{NP}^{\mathbf{PSpace}} \subseteq \mathbf{NPSpace} \subseteq \mathbf{PSpace},$$

cf. (Papadimitriou, 1994, Proof of Theorem 14.4). With similar arguments, we see that $\mathbf{P}^\mathbf{C} \subseteq \mathbf{C}$ holds true as well for the general case $\mathbf{PSpace} \subseteq \mathbf{C}$, since each polynomial time Turing machine with $\mathbf{C}$-oracle can be "recompiled" to a $\mathbf{C}$-Turing machine.

A deterministic procedure that decides $\mathsf{Max}_\emptyset(\Xi)$ could work as follows when given some $\mathcal{EL}$ concept description $C$ as input.

1. Check if $C \in \Xi$. If not, then reject $C$.
2. Enumerate all upper neighbors of $C$.
3. If there exists some upper neighbor $D$ of $C$ with $D \in \Xi$, then reject $C$; otherwise accept $C$.

According to Propositio 3.1.4.1, Step 2 requires polynomial time. We conclude that this procedure shows that $\mathsf{Max}_\emptyset(\Xi) \in \mathbf{P}^\mathbf{C}$ holds true. □

**Propositio 3.1.5.2.** *Let $\Xi \subseteq \mathcal{EL}(\Sigma)$ be some problem that is closed under subsumers, that is, $C \in \Xi$ and $C \sqsubseteq_\emptyset D$ implies $D \in \Xi$. We consider the problem* $\mathsf{Min}_\emptyset(\Xi)$*, which consists of all most specific elements of* $\Xi$*.*

1. $\Xi \in \mathbf{C}$ *implies* $\mathsf{Min}_\emptyset(\Xi) \in \mathbf{co}(\mathbf{NP}^\mathbf{C})$ *for each complexity class* $\mathbf{C}$*.*
2. $\Xi \in \mathbf{P}$ *implies* $\mathsf{Min}_\emptyset(\Xi) \in \mathbf{coNP}$
3. $\Xi \in \mathbf{\Sigma}^\mathbf{P}_n$ *implies* $\mathsf{Min}_\emptyset(\Xi) \in \mathbf{\Pi}^\mathbf{P}_{n+1}$ *for each number* $n \in \mathbb{N}$*.*
4. $\Xi \in \mathbf{C}$ *implies* $\mathsf{Min}_\emptyset(\Xi) \in \mathbf{coC}$ *for any complexity class* $\mathbf{C}$ *such that* $\mathbf{PSpace} \subseteq \mathbf{C}$*.*
5. $\Xi \in \mathbf{PSpace}$ *implies* $\mathsf{Min}_\emptyset(\Xi) \in \mathbf{PSpace}$

*Approbatio.* It is sufficient to show Statement 1, since the others are then obtained as immediate consequences. For Statement 4 we use the fact that $\mathbf{co}(\mathbf{NP}^{\mathbf{C}}) \subseteq \mathbf{C}$ is satisfied for each complexity class $\mathbf{C}$ satisfying $\mathbf{PSpace} \subseteq \mathbf{C}$. If $\mathbf{C} = \mathbf{PSpace}$, then this follows from $\mathbf{NP}^{\mathbf{PSpace}} \subseteq \mathbf{NPSpace} \subseteq \mathbf{PSpace}$, cf. (Papadimitriou, 1994, Proof of Theorem 14.4), since we can conclude that $\mathbf{co}(\mathbf{NP}^{\mathbf{PSpace}}) \subseteq \mathbf{coPSpace} = \mathbf{PSpace}$. More generally if $\mathbf{PSpace} \subseteq \mathbf{C}$, we can "recompile" any non-deterministic polynomial time Turing machine with $\mathbf{C}$-oracle to some deterministic polynomial space Turing machine with $\mathbf{C}$-oracle, which itself can be "recompiled" to a $\mathbf{C}$-Turing machine.

The following non-deterministic procedure decides the complement of $\mathsf{Min}_{\emptyset}(\Xi)$. Let $C$ be an $\mathcal{EL}$ concept description that is given as input.

1. Check whether $C \in \Xi$. If not, then accept $C$.
2. Guess some lower neighbor $D$ of $C$.
3. If $D \in \Xi$, then accept $C$; otherwise reject $C$.

Now Propositio 3.1.4.7 implies that the above is a procedure that needs non-deterministic polynomial time, and since it uses a $\mathbf{C}$-oracle to decide $\Xi$, we conclude that the complement of $\mathsf{Min}_{\emptyset}(\Xi)$ is in $\mathbf{NP}^{\mathbf{C}}$, which implies that $\mathsf{Min}_{\emptyset}(\Xi) \in \mathbf{co}(\mathbf{NP}^{\mathbf{C}})$. □

## 3.2. THE BOTTOM CONCEPT DESCRIPTION

Now consider the extension of $\mathcal{EL}$ with the *bottom concept description* $\bot$ the semantics of which is defined as $\bot^{\mathcal{I}} := \emptyset$ for any interpretation $\mathcal{I}$. Then $\sqsubseteq_{\emptyset}$ is not bounded and $\sqsupseteq_{\emptyset}$ is not well-founded, since the following infinite chain exists.

$$\bot \sqsubsetneq_{\emptyset} \ldots \sqsubsetneq_{\emptyset} \exists\, r^{n+1}.\top \sqsubsetneq_{\emptyset} \exists\, r^{n}.\top \sqsubsetneq_{\emptyset} \ldots \sqsubsetneq_{\emptyset} \exists\, r^{2}.\top \sqsubsetneq_{\emptyset} \exists\, r.\top \sqsubsetneq_{\emptyset} \top$$

Furthermore, $\sqsubseteq_{\emptyset}$ is not neighborhood generated, as $\bot$ does not have any upper neighbors. To see this, consider a concept description $C$ such that $\bot \sqsubsetneq_{\emptyset} C$; it then follows that $\bot \sqsubsetneq_{\emptyset} C \sqcap \exists\, r.C \sqsubsetneq_{\emptyset} C$. Anyway, $\bot$ is the only concept description that causes problems here: for each satisfiable $\mathcal{EL}^{\bot}$ concept description, that is, for any $C \in \mathcal{EL}^{\bot}(\Sigma)$ such that $C \not\equiv_{\emptyset} \bot$, we can enumerate all upper and lower neighbors with the same techniques as in Section 3.1. This is due to the fact that some $\mathcal{EL}^{\bot}$ concept description is satisfiable if, and only if, it does not contain $\bot$ as a subconcept.

## 3.3. GREATEST FIXED-POINT SEMANTICS

Unfortunately, the situation is also not rosy for extensions of $\mathcal{EL}$ with *greatest fixed-point semantics* Baader, 2003; Lutz, Piro, and Wolter, 2010b. It then also holds true that $\sqsubseteq_{\emptyset}$ is neither bounded nor neighborhood generated, and $\sqsupseteq_{\emptyset}$ is not well-founded. One culprit is a concept description which represents a cycle, for instance $\nu X.\exists\, r.X$, the extension of which is maximal w.r.t. the property of containing elements that have some other element in that extension as an $r$-successor.

The following infinite chain justifies that $\sqsubseteq_{\emptyset}$ is not bounded and further that $\sqsupseteq_{\emptyset}$ is not well-founded.

$$\nu X.\exists\, r.X \sqsubsetneq_{\emptyset} \ldots \sqsubsetneq_{\emptyset} \exists\, r^{n+1}.\top \sqsubsetneq_{\emptyset} \exists\, r^{n}.\top \sqsubsetneq_{\emptyset} \ldots \sqsubsetneq_{\emptyset} \exists\, r^{2}.\top \sqsubsetneq_{\emptyset} \exists\, r.\top \sqsubsetneq_{\emptyset} \top$$

**Lemma 3.3.1.** *There is some signature $\Sigma$ such that the subsumption relation $\sqsubseteq_\emptyset$ on $\mathcal{EL}_\nu(\Sigma)$ is not neighborhood generated.*

*Approbatio.* Consider the signature $\Sigma$ with $\Sigma_\mathsf{C} := \emptyset$ and $\Sigma_\mathsf{R} := \{r\}$. We will show that then $(\mathcal{EL}_\nu(\Sigma), \sqsubseteq_\emptyset)/\emptyset$ is isomorphic to the ordinal $(\omega + 1, \geq)$, which is apparently not neighborhood generated.

In (Lutz, Piro, and Wolter, 2010b), Lutz, Piro, and Wolter showed that all $\mathcal{EL}_\nu(\Sigma)$ concept descriptions are equivalent to an $\mathcal{EL}_{\mathsf{si}}(\Sigma)$ concept description, which has the form $\exists^{\mathsf{sim}}(\mathcal{I}, \delta)$ for a finite pointed interpretation $(\mathcal{I}, \delta)$ and the extension of which in an interpretation $\mathcal{J}$ is given as follows.

$$(\exists^{\mathsf{sim}}(\mathcal{I}, \delta))^\mathcal{J} := \{\, \epsilon \mid \epsilon \in \Delta^\mathcal{J} \text{ and } (\mathcal{I}, \delta) \precsim (\mathcal{J}, \epsilon) \,\}$$

As an immediate consequence of this definition we infer that $\exists^{\mathsf{sim}}(\mathcal{I}, \delta) \sqsubseteq_\emptyset \exists^{\mathsf{sim}}(\mathcal{J}, \epsilon)$ if, and only if, $(\mathcal{J}, \epsilon) \precsim (\mathcal{I}, \delta)$.

The finite interpretations over $\Sigma$ essentially are just finite directed graphs, in which the vertices have no labels and in which the edges are (virtually) labeled with $r$. We first show that the loop (the one-element cycle) is maximal with respect to the simulation order. In particular, we show that the following finite pointed interpretation $(\mathcal{I}_\omega, \delta_0)$ is maximal w.r.t. $\precsim$.

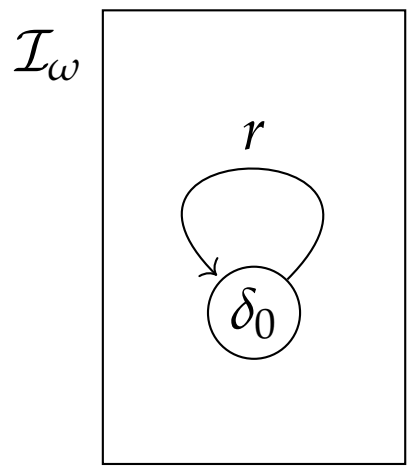

If $(\mathcal{J}, \epsilon)$ is an arbitrary pointed interpretation over $\Sigma$, then the binary relation $\sigma := \{\, (\eta, \delta_0) \mid \eta \in \Delta^\mathcal{J} \,\}$ apparently is a simulation from $(\mathcal{J}, \epsilon)$ to $(\mathcal{I}_\omega, \delta_0)$.

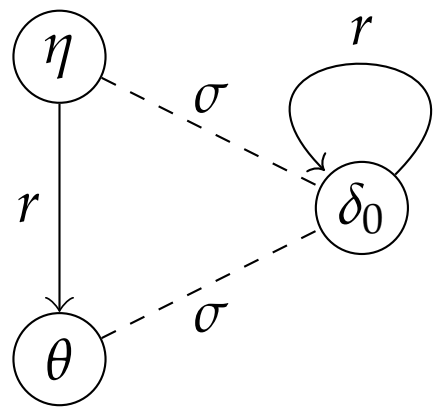

As an immediate corollary we obtain that $\exists^{\mathsf{sim}}(\mathcal{I}_\omega, \delta_0) \equiv \nu X.\exists r.X =: C_\omega$ is the smallest $\mathcal{EL}_\nu(\Sigma)$ concept description.

As next step we prove that each finite pointed interpretation $(\mathcal{J}, \epsilon)$ over $\Sigma$ which contains a cycle through $\epsilon$ is equi-similar to $(\mathcal{I}_\omega, \delta_0)$. It is only left to show the existence of a simulation from $(\mathcal{I}_\omega, \delta_0)$ to such $(\mathcal{J}, \epsilon)$. We essentially do this by connecting $\delta$ with each element in the cycle, i.e., if

$$\epsilon \; r^\mathcal{J} \; \epsilon_1 \; r^\mathcal{J} \; \epsilon_2 \; r^\mathcal{J} \; \ldots \; r^\mathcal{J} \; \epsilon_n \; r^\mathcal{J} \; \epsilon$$

is the cycle containing $\epsilon$, then the binary relation $\tau := \{(\delta_0, \epsilon)\} \cup \{\, (\delta_0, \epsilon_i) \mid i \in \{1, \ldots, n\} \,\}$ clearly is a simulation from $(\mathcal{I}_\omega, \delta_0)$ to $(\mathcal{J}, \epsilon)$.

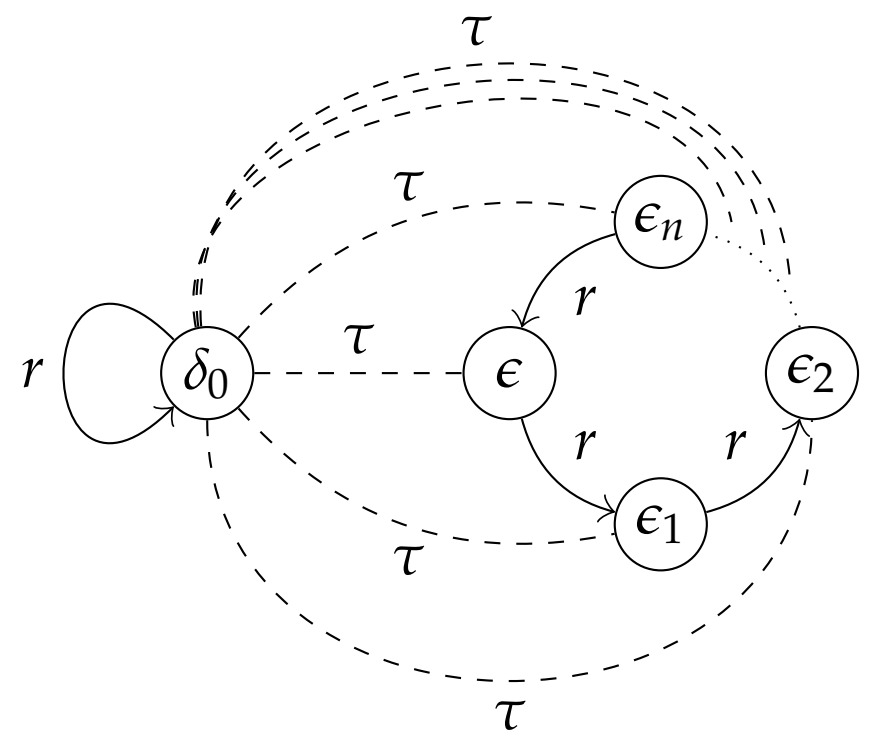

Continuing our investigations, we now consider a finite pointed interpretation $(\mathcal{J}, \epsilon)$ which does not contain a cycle through $\epsilon$, i.e., all paths starting with $\epsilon$ do not contain $\epsilon$ twice. It is readily verified that then there cannot exist a simulation from $(\mathcal{I}_\omega, \delta_0)$ to $(\mathcal{J}, \epsilon)$, simply because the loop in $(\mathcal{I}_\omega, \delta_0)$ cannot be simulated in $(\mathcal{J}, \epsilon)$. As a corollary, each of these finite pointed interpretations $(\mathcal{J}, \epsilon)$ is strictly smaller than $(\mathcal{I}_\omega, \delta_0)$ with respect to the simulation order $\precsim$. Furthermore, we observe that each path starting with $\epsilon$ is of finite length, and only finitely many mutually distinct paths starting with $\epsilon$ exist. If we now define $n$ as the maximal length of a path starting with $\epsilon$, we conclude that $(\mathcal{J}, \epsilon)$ and the finite pointed interpretation $(\mathcal{I}_n, \delta_0)$ are equi-similar where

$$\mathcal{I}_n \coloneqq (\{\delta_0, \delta_1, \ldots, \delta_n\}, \{r \mapsto \{\, (\delta_{i-1}, \delta_i) \mid i \in \{1, \ldots, n\} \,\}\}).$$

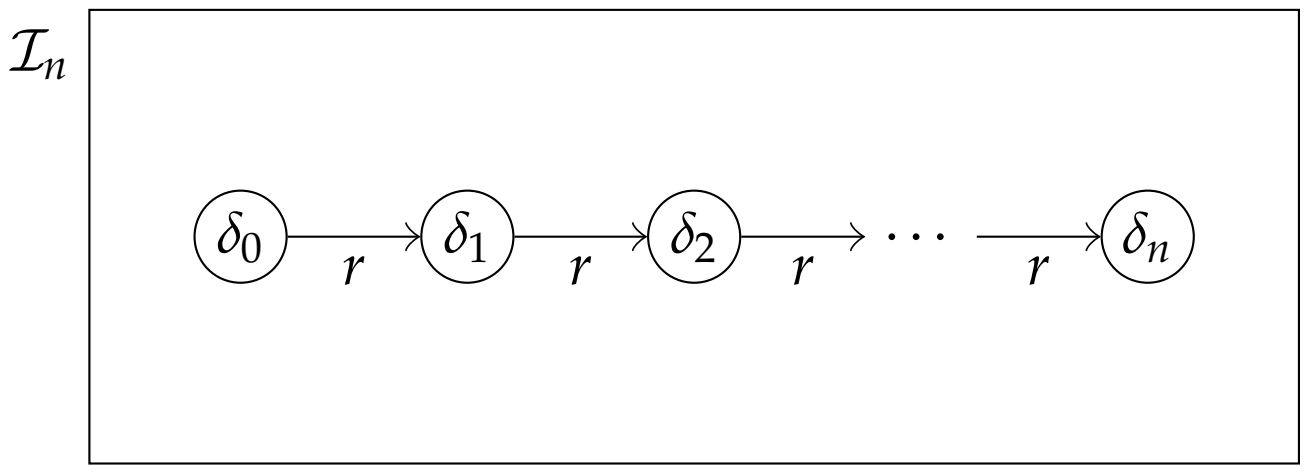

Clearly, it holds true that $\exists^{\mathsf{sim}}(\mathcal{I}_n, \delta_0) \equiv \exists\, r^n. \top \eqqcolon C_n$, and since $(\mathcal{I}_m, \delta_0) \precnsim (\mathcal{I}_n, \delta_0)$ is satisfied if $m < n$, it follows that $C_n \sqsubsetneq_\emptyset C_m$ whenever $m < n$.

In summary, we have shown that

$$\mathcal{EL}_\mathsf{v}(\Sigma)/\emptyset = \{\, [C_n]_\emptyset \mid n < \omega \,\} \cup \{[C_\omega]_\emptyset\},$$

and that these concept descriptions are ordered as follows.

$$C_\omega \sqsubsetneq_\emptyset \ldots \sqsubsetneq_\emptyset C_2 \sqsubsetneq_\emptyset C_1 \sqsubsetneq_\emptyset C_0$$

This immediately proves that $C_\omega$ does not have upper neighbors although it is subsumed by each $C_n$.

□

## 3.4. CYCLE-RESTRICTED TBOXES

According to Baader, Borgwardt, and Morawska (2012), a TBox $\mathcal{T}$ is called *cycle-restricted* if there does not exist a word $w \in \Sigma_\mathsf{R}^+$ and a concept description $C \in \mathcal{EL}(\Sigma)$ such that $C \sqsubseteq_\mathcal{T} \exists\, w. C$. Furthermore, deciding whether a TBox is cycle-restricted can be done in polynomial time. In (Kriegel, 2018a), the author has shown that most specific consequences with respect to cycle-restricted TBoxes always exist in $\mathcal{EL}$ (without greatest fixed-point semantics). Thus, we can utilize our results on neighborhood in $\mathcal{EL}$ without any TBox to constitute procedures for deciding neighborhood and for enumerating all neighbors in $\mathcal{EL}$ with respect to cycle-restricted TBoxes.

**(Baader, Borgwardt, and Morawska, 2012, Definition 2).** An $\mathcal{EL}$ TBox $\mathcal{T}$ is *cycle-restricted* if there is no $\mathcal{EL}$ concept description $C$ and no non-empty role word $w \in \Sigma_{\mathsf{R}}^{+}$ such that $C \sqsubseteq_{\mathcal{T}} \exists\, w.\, C$. △

**(Kriegel, 2018a).** *Let $\mathcal{T}$ be an $\mathcal{EL}$ TBox. Then, the following statements are equivalent.*

1. *$\mathcal{T}$ is cycle-restricted.*
2. *The canonical model $\mathcal{I}_{C,\mathcal{T}}$ is tree-shaped for every $\mathcal{EL}$ concept description $C$.*
3. *The most specific consequence $C^{\mathcal{T}}$ exists in $\mathcal{EL}$ for any $\mathcal{EL}$ concept description $C$.* □

**Lemma 3.4.1.** *For each cycle-restricted TBox $\mathcal{T}$, the subsumption relation $\sqsubseteq_{\mathcal{T}}$ is neighborhood generated.*

*Approbatio.* It is readily verified that $C \prec_{\mathcal{T}} D$ if, and only if, $C^{\mathcal{T}} \sqsubsetneq_{\emptyset} D^{\mathcal{T}}$ and there is no most specific consequence $E^{\mathcal{T}}$ such that $C^{\mathcal{T}} \sqsubsetneq_{\emptyset} E^{\mathcal{T}} \sqsubsetneq_{\emptyset} D^{\mathcal{T}}$. According to Section 3.4, all most specific consequences of $\mathcal{T}$ exist in $\mathcal{EL}$. Furthermore, we know that $\sqsubseteq_{\emptyset}$ is bounded, cf. (Baader and Morawska, 2010, Proof of Proposition 3.5). Of course, if we now restrict the subsumption relation $\sqsubseteq_{\emptyset}$ to the most specific consequences of $\mathcal{T}$, that is, if we consider the relation $\sqsubseteq_{\emptyset} \cap \mathsf{MSS}(\mathcal{T}) \times \mathsf{MSS}(\mathcal{T})$ where $\mathsf{MSS}(\mathcal{T}) :=$ $\{\, C^{\mathcal{T}} \mid C \in \mathcal{EL}(\Sigma) \,\}$, then this relation must also be bounded. Now since there exists an order isomorphism $[C]_{\mathcal{T}} \mapsto [C^{\mathcal{T}}]_{\emptyset}$ between $(\mathcal{EL}(\Sigma), \sqsubseteq_{\mathcal{T}})/\mathcal{T}$ and $(\mathsf{MSS}(\mathcal{T}), \sqsubseteq_{\emptyset} \cap \mathsf{MSS}(\mathcal{T}) \times \mathsf{MSS}(\mathcal{T}))/\emptyset$, we conclude that $\sqsubseteq_{\mathcal{T}}$ is bounded as well and is, thus, neighborhood generated. □

**Lemma 3.4.2.** *Fix some cycle-restricted $\mathcal{EL}$ TBox $\mathcal{T}$ as well as two $\mathcal{EL}$ concept descriptions $C$ and $D$. It then holds true that $C \prec_{\mathcal{T}} D$ if, and only if, $C \sqsubsetneq_{\mathcal{T}} D$ and $C \sqsubseteq_{\mathcal{T}} L$ implies $C \equiv_{\mathcal{T}} L$ for any $L \in \mathsf{Lower}^{*}(D^{\mathcal{T}})$. Furthermore, it holds true that $\prec_{\mathcal{T}} \in$ **coNP**, i.e., neighborhood of two $\mathcal{EL}$ concept descriptions is decidable in non-deterministic polynomial time w.r.t. $||C|| + ||D|| + ||\mathcal{T}|| + |\Sigma|$.*

*Approbatio.* We start with proving the *if* statement. Let $C \sqsubseteq_{\mathcal{T}} X \sqsubsetneq_{\mathcal{T}} D$, that is, $C^{\mathcal{T}} \sqsubseteq_{\emptyset} X^{\mathcal{T}} \sqsubsetneq_{\emptyset} D^{\mathcal{T}}$. Now there is some $L \in \mathsf{Lower}^{*}(D^{\mathcal{T}})$ such that $X^{\mathcal{T}} \sqsubseteq_{\emptyset} L$, and it follows that $X \sqsubseteq_{\mathcal{T}} L$. We conclude that $C \equiv_{\mathcal{T}} L$ holds true, which implies $C \equiv_{\mathcal{T}} X$.

We proceed with the *only if* direction. Assume $C \prec_{\mathcal{T}} D$, which immediately yields that $C \sqsubsetneq_{\mathcal{T}} D$, and further let $L \in \mathsf{Lower}^{*}(D^{\mathcal{T}})$ such that $C \sqsubseteq_{\mathcal{T}} L$. The very definition of a most specific consequence shows that $L \prec_{\emptyset} D^{\mathcal{T}}$ implies $L \sqsubsetneq_{\mathcal{T}} D$. Eventually, our assumption yields that $C \equiv_{\mathcal{T}} L$.

The complexity result can be obtained as a corollary of the following facts.

- Subsumption in $\mathcal{EL}$ can be decided in polynomial time.
- Most specific consequences w.r.t. cycle-restricted TBoxes always exist in $\mathcal{EL}$ and can be computed in polynomial time.
- Lower neighbors of an $\mathcal{EL}$ concept description can be guessed in polynomial time, cf. Propositio 3.1.4.7. □

**Lemma 3.4.3.** *Let $\mathcal{T}$ be a cycle-restricted $\mathcal{EL}$ TBox and $C$ an $\mathcal{EL}$ concept description. Then the set*

$$\mathsf{Lower}_{\mathcal{T}}(C) := \mathsf{Max}_{\mathcal{T}}(\mathsf{Lower}^*(C^{\mathcal{T}}))$$

*contains exactly all lower neighbors of $C$ with respect to $\mathcal{T}$ modulo equivalence and can further be computed in exponential time w.r.t. $||C|| + ||\mathcal{T}|| + |\Sigma|$.*

*Approbatio. Soundness.* Let $L \in \mathsf{Lower}_{\mathcal{T}}(C)$ and assume that $L \sqsubseteq_{\mathcal{T}} X \sqsubsetneq_{\mathcal{T}} C$. It then follows that $L^{\mathcal{T}} \sqsubseteq_{\emptyset} X^{\mathcal{T}} \sqsubsetneq_{\emptyset} C^{\mathcal{T}}$ and, thus, there is some $M$ such that $L^{\mathcal{T}} \sqsubseteq_{\emptyset} X^{\mathcal{T}} \sqsubseteq_{\emptyset} M \prec_{\emptyset} C^{\mathcal{T}}$. We conclude that $L \sqsubseteq_{\mathcal{T}} X \sqsubseteq_{\mathcal{T}} M$. As $L$ is $\sqsubseteq_{\mathcal{T}}$-maximal in $\mathsf{Lower}^*(C^{\mathcal{T}})$, we conclude that $L \equiv_{\mathcal{T}} M$, which shows that $X \equiv_{\mathcal{T}} L$, that is, $L \prec_{\mathcal{T}} C$.

*Completeness.* Vice versa, assume that $L \prec_{\mathcal{T}} C$. We infer that $L \sqsubsetneq_{\mathcal{T}} C$ and further that $L^{\mathcal{T}} \sqsubsetneq_{\emptyset} C^{\mathcal{T}}$. According to Corollarium 3.1.3.6, there exists some lower neighbor $M \in \mathsf{Lower}^*(C^{\mathcal{T}})$ satisfying $L^{\mathcal{T}} \sqsubseteq_{\emptyset} M \prec_{\emptyset} C^{\mathcal{T}}$. Thus, it follows that $L \equiv_{\mathcal{T}} L^{\mathcal{T}} \sqsubseteq_{\mathcal{T}} M \sqsubsetneq_{\mathcal{T}} C^{\mathcal{T}} \equiv_{\mathcal{T}} C$, which yields $L \equiv_{\mathcal{T}} M$. It remains to prove that $M$ is $\sqsubseteq_{\mathcal{T}}$-maximal. If $M \sqsubsetneq_{\mathcal{T}} N$ for some $N \in \mathsf{Lower}^*(C^{\mathcal{T}})$, then $M \sqsubsetneq_{\mathcal{T}} N \prec_{\emptyset} C^{\mathcal{T}}$ immediately implies the contradiction $M \sqsubsetneq_{\mathcal{T}} N \sqsubsetneq_{\mathcal{T}} C$.

*Complexity.* The complexity result can be obtained as a corollary of the following facts.

- Subsumption in $\mathcal{EL}$ can be decided in polynomial time.
- Most specific consequences w.r.t. cycle-restricted TBoxes always exist in $\mathcal{EL}$ and can be computed in polynomial time.
- (Representatives of) all lower neighbors of some $\mathcal{EL}$ concept description can be enumerated in exponential time, cf. Propositio 3.1.4.6. □

**Propositio 3.4.4.** *Fix some cycle-restricted $\mathcal{EL}$ TBox $\mathcal{T}$ and consider an $\mathcal{EL}$ concept description $C$. Then the set*

$$\mathsf{Upper}_{\mathcal{T}}(C) := \mathsf{Min}_{\emptyset}(\bigcup\{\, \mathsf{Upper}(X) \mid X \in \mathsf{Max}_{\emptyset}([C]_{\mathcal{T}})\,\})$$

*contains exactly all upper neighbors of $C$ with respect to $\mathcal{T}$ modulo equivalence.*

*Approbatio. Soundness.* Assume that $C \prec_{\mathcal{T}} D$ holds true. It follows that $C^{\mathcal{T}} \sqsubsetneq_{\emptyset} D^{\mathcal{T}}$. Now consider some concept description $E$ such that $C^{\mathcal{T}} \sqsubseteq_{\emptyset} E \sqsubseteq_{\emptyset} D^{\mathcal{T}}$. According to the properties of most specific consequences, we can infer that $C \sqsubseteq_{\mathcal{T}} E \sqsubseteq_{\mathcal{T}} D$, which yields that either $E \equiv_{\mathcal{T}} C$ or $E \equiv_{\mathcal{T}} D$. Formulated alternatively, we have that $E \equiv_{\mathcal{T}} C$ if, and only if $E \not\equiv_{\mathcal{T}} D$.

It is readily verified that some $E$ satisfying $C^{\mathcal{T}} \sqsubseteq_{\emptyset} E \sqsubseteq_{\emptyset} D^{\mathcal{T}}$ and $E \equiv_{\mathcal{T}} C$ exists, namely $E = C^{\mathcal{T}}$. We now fix some such $E$ that is most general (w.r.t. $\emptyset$) such that $C^{\mathcal{T}} \sqsubseteq_{\emptyset} E \sqsubseteq_{\emptyset} D^{\mathcal{T}}$ and $E \equiv_{\mathcal{T}} C$. Then, we immediately conclude that $E \not\equiv_{\mathcal{T}} D$ as well as $E \sqsubsetneq_{\emptyset} D^{\mathcal{T}}$, and furthermore we have that $C \not\equiv_{\mathcal{T}} F \equiv_{\mathcal{T}} D$ for any $F$ with $E \prec_{\emptyset} F \sqsubseteq_{\emptyset} D^{\mathcal{T}}$. In particular, at least one such upper neighbor $F$ of $E$ must exist and we infer that $F \equiv_{\emptyset} D^{\mathcal{T}}$ holds true. Summing up, we have shown that $D \equiv_{\mathcal{T}} F$ for some $F \in \bigcup\{\, \mathsf{Upper}(X) \mid X \in \mathsf{Max}_{\emptyset}([C]_{\mathcal{T}})\,\}$.

It remains to show that $F$ is most specific w.r.t. $\emptyset$. Assume the contrary, i.e., let $E' \in \mathsf{Max}_{\emptyset}([C]_{\mathcal{T}})$ and $F' \in \mathsf{Upper}(E')$ such that $F' \sqsubsetneq_{\emptyset} F$ and $D \equiv_{\mathcal{T}} F'$. It then follows that $D \sqsubseteq_{\mathcal{T}} F'$, which implies $D^{\mathcal{T}} \sqsubseteq_{\emptyset} F'$. Putting everything together yields the contradiction $D^{\mathcal{T}} \sqsubseteq_{\emptyset} F' \sqsubsetneq_{\emptyset} F \equiv_{\emptyset} D^{\mathcal{T}}$. ↯

*Completeness.* Vice versa, assume that there are two concept descriptions $X$ and $D$ such that $X \in \mathsf{Max}_\emptyset([C]_\mathcal{T})$, $D \in \mathsf{Upper}(X)$, and where $D$ is most specific with respect to these two properties, that is, there does not exist any $X' \in \mathsf{Max}_\emptyset([C]_\mathcal{T})$ and some $D' \in \mathsf{Upper}(X')$ with $D' \sqsubsetneq_\emptyset D$. We claim that then $C \prec_\mathcal{T} D$ holds true. Before we proceed with proving this, we show the following auxiliary claim.

**Lemma.** *For each $Y$ such that $C \sqsubseteq_\mathcal{T} Y \sqsubsetneq_\emptyset D$, we have $C \equiv_\mathcal{T} Y$.*

*Approbatio.* Let $C \sqsubsetneq_\mathcal{T} Y \sqsubsetneq_\emptyset D$. Then, there must exist some $Z \in \mathsf{Max}_\emptyset([C]_\mathcal{T})$ such that $C^\mathcal{T} \sqsubseteq_\emptyset Z \sqsubsetneq_\emptyset Y$. Thus, there is some $U \in \mathsf{Upper}(Z)$ such that $C^\mathcal{T} \sqsubseteq_\emptyset Z \prec_\emptyset U \sqsubseteq_\emptyset Y$. It follows that $U \sqsubsetneq_\emptyset D$ where $U \in \mathsf{Upper}(Z)$ and $Z \in \mathsf{Max}_\emptyset([C]_\mathcal{T})$. $\lightning$ □

From $D \succ_\emptyset X \sqsupseteq_\emptyset X^\mathcal{T} \equiv_\emptyset C^\mathcal{T}$ we infer that $C^\mathcal{T} \sqsubsetneq_\emptyset D$, which immediately implies that $C \sqsubseteq_\mathcal{T} D$. Apparently, $C \equiv_\mathcal{T} D$ would contradict the precondition that $X$ is most general in $[C]_\mathcal{T}$; we conclude that $C \sqsubsetneq_\mathcal{T} D$.

Furthermore, it holds true that $D \equiv_\emptyset D^\mathcal{T}$. To see this, assume the contrary, i.e., let $D^\mathcal{T} \sqsubsetneq_\emptyset D$. Since $C \sqsubseteq_\mathcal{T} D$ and $D \equiv_\mathcal{T} D^\mathcal{T}$, an application of the above lemma would yield the contradiction $C \equiv_\mathcal{T} D^\mathcal{T}$. $\lightning$

According to Lemma 3.4.2, it suffices to check whether $C \sqsubseteq_\mathcal{T} L$ implies $C \equiv_\mathcal{T} L$ for any $L \in \mathsf{Lower}^*(D^\mathcal{T})$. However, this is an immediate consequence of the above lemma: if $C \sqsubseteq_\mathcal{T} L \prec_\emptyset D^\mathcal{T}$, then due to $D \equiv_\emptyset D^\mathcal{T}$ we infer that $C \sqsubseteq_\mathcal{T} L \sqsubsetneq_\emptyset D$, and so it follows that $C \equiv_\mathcal{T} L$. □

## 3.5. ACYCLIC TBOXES

A *concept definition* is an expression of the form $A \equiv C$ where $A \in \Sigma_\mathsf{C}$ is a concept name and where $C \in \mathcal{EL}(\Sigma)$ is a concept description. We then also say that $A$ is a *defined* concept name and $C$ is its *defining* concept description. A concept name that is not defined is called *primitive*. An *acyclic TBox* is a finite set of concept definitions that contains at most one concept definition $A \equiv C$ for each concept name $A \in \Sigma_\mathsf{C}$, and for which the following directed graph, called the *dependency graph* of $\mathcal{T}$, is acyclic.

$$(\Sigma_\mathsf{C}, \{\, (A,B) \mid A,B \in \Sigma_\mathsf{C},\ \exists\, C\colon A \equiv C \in \mathcal{T} \text{ and } B \in \mathsf{Sub}(C)\,\}^+)$$

The *expansion* $C^\mathcal{T}$ of an $\mathcal{EL}$ concept description $C$ with respect to an acyclic TBox $\mathcal{T}$ is obtained from $C$ by exhaustively replacing each defined concept name with its defining concept description. It then holds true that $C$ and its expansion $C^\mathcal{T}$ are equivalent with respect to $\mathcal{T}$, that is, $C \equiv_\mathcal{T} C^\mathcal{T}$. Furthermore, deciding subsumption of two concept descriptions with respect to $\mathcal{T}$ can be reduced to deciding subsumption of the respective expansions with respect to $\emptyset$, that is, it holds true that $C \sqsubseteq_\mathcal{T} D$ if, and only if, $C^\mathcal{T} \sqsubseteq_\emptyset D^\mathcal{T}$.

It is apparent that any acylic TBox is cycle-restricted. Thus, we can simply pull and apply the results from the preceeding section and, in particular, conclude that the subsumption relation $\sqsubseteq_\mathcal{T}$ is neighborhood generated for any acyclic TBox $\mathcal{T}$.

## 3.6. GENERAL TBOXES

A similar situation as for greatest fixed-point semantics arises when considering subsumption with respect to a non-cycle-restricted TBox $\mathcal{T}$.

**Lemma 3.6.1.** *There is an $\mathcal{EL}$ TBox $\mathcal{T}$ over some signature $\Sigma$ such that $\sqsubseteq_\mathcal{T}$ is not bounded and $\sqsupseteq_\mathcal{T}$ is not well-founded.*

*Approbatio.* We demonstrate the claim by giving a counterexample. Define the TBox $\mathcal{T} := \{A \sqsubseteq \exists\, r.\, A\}$ over the signature $\Sigma$ where $\Sigma_\mathsf{C} := \{A\}$ and $\Sigma_\mathsf{R} := \{r\}$. Apparently, then the following infinite chain exists.

$$A \sqsubsetneq_\mathcal{T} \exists\, r.\, A \sqsubsetneq_\mathcal{T} \exists\, r^2.\, A \sqsubsetneq_\mathcal{T} \exists\, r^3.\, A \sqsubsetneq_\mathcal{T} \ldots$$

The following model $\mathcal{I}$ of $\mathcal{T}$ shows that the subsumptions in the chain are indeed strict.

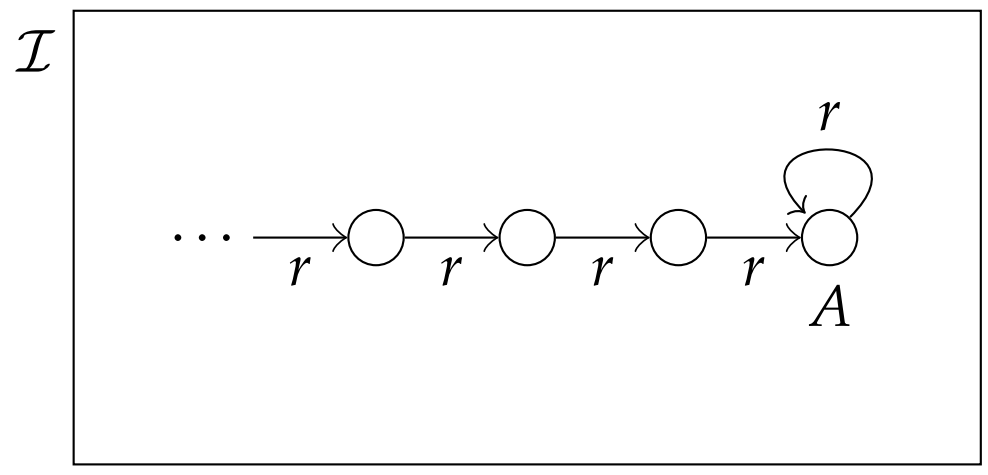

□

**Lemma 3.6.2.** *There is an $\mathcal{EL}$ TBox $\mathcal{T}$ over some signature $\Sigma$ and an $\mathcal{EL}$ concept description $C$ over $\Sigma$ that strictly subsumes some other $\mathcal{EL}$ concept description w.r.t. $\mathcal{T}$, but does not have lower neighbors with respect to $\mathcal{T}$.*

*Approbatio.* We consider a simple signature with exactly one concept name and exactly one role name, i.e., let $\Sigma$ be given by $\Sigma_\mathsf{C} := \{A\}$ and $\Sigma_\mathsf{R} := \{r\}$. We are going to show that $\top$ does not have lower neighbors with respect to the TBox

$$\mathcal{T} := \{\top \sqsubseteq \exists\, r.\, \top,\ A \sqsubseteq \exists\, r.\, A\}.$$

For this purpose, we first prove the validity of the following two statements.

1. If $C$ does not contain the concept name $A$ as a subconcept, then $C \equiv_\mathcal{T} \top$.
2. If in the canonical model of an $\mathcal{EL}(\Sigma)$ concept description $C$ with respect to $\mathcal{T}$ the shortest path from the vertex $C$ to a vertex labelled with $A$ has length $n$, then $C \equiv_\mathcal{T} \exists\, r^n.\, A$.

As a corollary, we then obtain that

$$\mathcal{EL}(\Sigma)/\mathcal{T} = \{\, [\exists\, r^n.\, A]_\mathcal{T} \mid n \in \mathbb{N} \,\} \cup \{[\top]_\mathcal{T}\},$$

and furthermore that the subsumption ordering of these concept descriptions is as follows.

$$A \sqsubsetneq_\mathcal{T} \exists\, r.\, A \sqsubsetneq_\mathcal{T} \exists\, r^2.\, A \sqsubsetneq_\mathcal{T} \exists\, r^3.\, A \sqsubsetneq_\mathcal{T} \ldots \sqsubsetneq_\mathcal{T} \top$$

The following interpretation $\mathcal{I}$ is a model of $\mathcal{T}$ and witnesses the validity of the strictness of the above mentioned subsumptions.

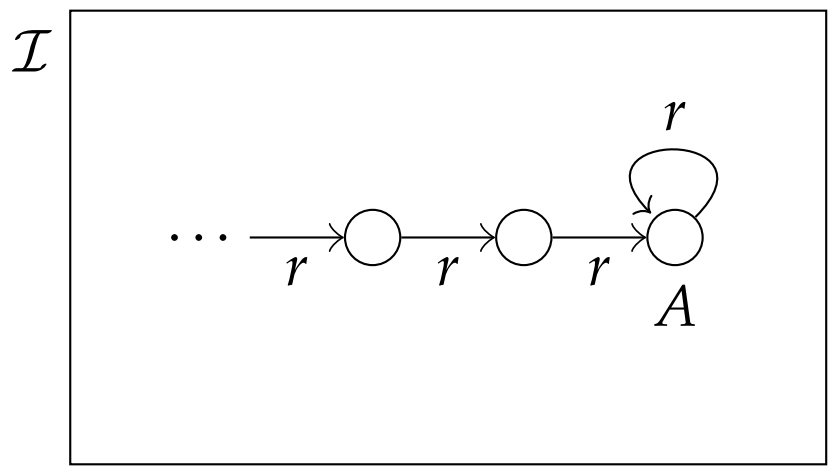

We may now safely conclude that $\top$ indeed does not have any lower neighbors with respect to $\mathcal{T}$.

However, we still have to prove the two statements above, with which we proceed now.

1. Let $C$ be an $\mathcal{EL}(\Sigma)$ concept description which does not contain $A$ as a subconcept. It is easy to verify that in the canonical model $\mathcal{I}_{C,\mathcal{T}}$ there is an $r$-edge from $C$ to $\top$, and the latter has an $r$-loop. Thus, $(\mathcal{I}_{C,\mathcal{T}}, C)$ and $(\mathcal{I}_{\top,\mathcal{T}}, \top)$ are equi-similar, whence $C \equiv_{\mathcal{T}} \top$.

2. As supposed, let $n$ be the length of a shortest path $\vec{p}$ from $C$ to a vertex $D$ with label $A$ within the canonical model $\mathcal{I}_{C,\mathcal{T}}$. In particular, $A$ is an $r$-successor of $D$ and $A$ is an $r$-successor of itself. Henceforth, the other $r$-paths starting with $D$ can already be simulated in the $r$-loop of $A$. All other $r$-paths starting with $C$ may also be simulated by means of $\vec{p}$ and the $r$-loop of $A$. Thus, we conclude that $(\mathcal{I}_{C,\mathcal{T}}, C)$ and the canonical model of $\exists\, r^n.\, A$ w.r.t. $\mathcal{T}$ are equi-similar, that is, $C \equiv_{\mathcal{T}} \exists\, r^n.\, A$. □

**Lemma 3.6.3.** *There is an $\mathcal{EL}$ TBox $\mathcal{T}$ over some signature $\Sigma$ and an $\mathcal{EL}$ concept description $C$ over $\Sigma$ that is strictly subsumed by another $\mathcal{EL}$ concept description w.r.t. $\mathcal{T}$, but does not have upper neighbors with respect to $\mathcal{T}$.*

*Approbatio.* We try to keep things simple, and consider a rather small signature, namely $\Sigma$ defined by $\Sigma_{\mathsf{C}} := \{A, B\}$ and $\Sigma_{\mathsf{R}} := \{r\}$. Furthermore, in order to find a suitable counterexample, we define a TBox by

$$\mathcal{T} := \{\exists\, r.\, A \sqsubseteq A,\ B \sqsubseteq A,\ B \equiv \exists\, r.\, B\}.$$

From the very definition of $\mathcal{T}$ it follows that the following subsumptions hold true.

$$B \equiv_{\mathcal{T}} \exists\, r^n.\, B \sqsubsetneq_{\mathcal{T}} \exists\, r^n.\, A$$
$$\ldots \sqsubsetneq_{\mathcal{T}} \exists\, r^{n+1}.\, A \sqsubsetneq_{\mathcal{T}} \exists\, r^n.\, A \sqsubsetneq_{\mathcal{T}} \ldots \sqsubsetneq_{\mathcal{T}} \exists\, r^2.\, A \sqsubsetneq_{\mathcal{T}} \exists\, r.\, A \sqsubsetneq_{\mathcal{T}} A$$

The following interpretation $\mathcal{I}$ is a model of $\mathcal{T}$ and justifies the strictness of the subsumptions above.

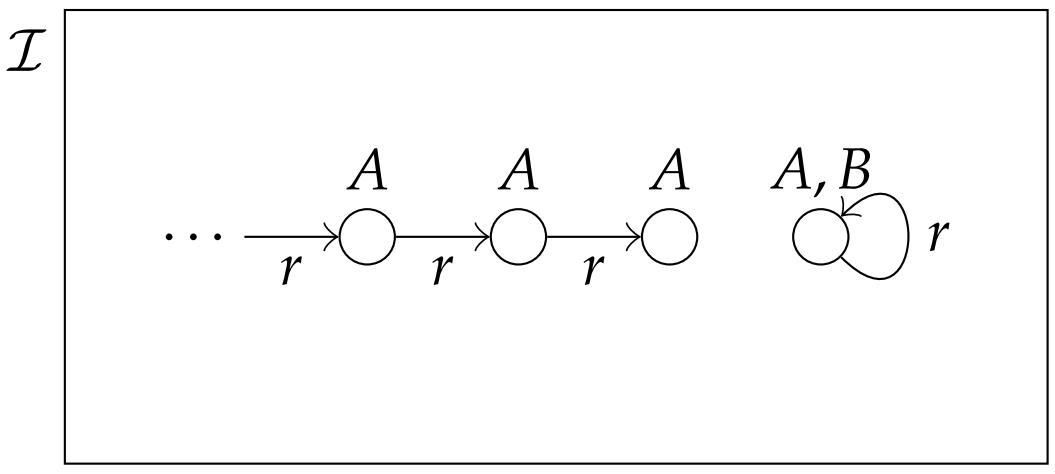

Let $C_n := \exists\, r^n.\, A$ for $n \in \mathbb{N}$. According to the previous observations, the following infinite chain exists.

$$B \sqsubsetneq_{\mathcal{T}} \ldots \sqsubsetneq_{\mathcal{T}} C_{n+1} \sqsubsetneq_{\mathcal{T}} C_n \sqsubsetneq_{\mathcal{T}} \ldots \sqsubsetneq_{\mathcal{T}} C_2 \sqsubsetneq_{\mathcal{T}} C_1 \sqsubsetneq_{\mathcal{T}} C_0 = A$$

The canonical models $\mathcal{I}_{C_n,\mathcal{T}}$ and $\mathcal{I}_{B,\mathcal{T}}$ are depicted below.

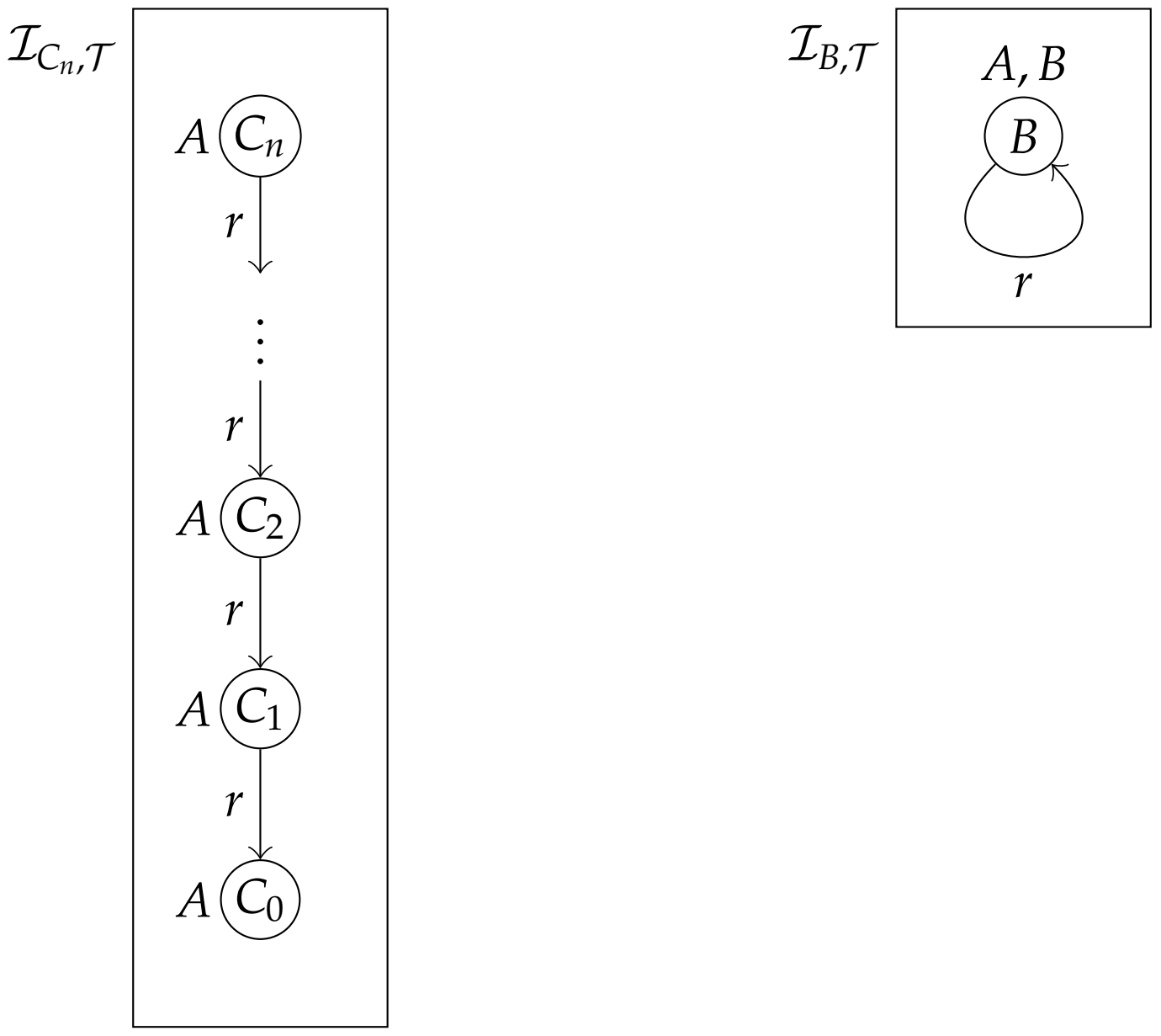

It is readily verified that for each $n \in \mathbb{N}$, there exists a simulation from $(\mathcal{I}_{C_n,\mathcal{T}}, C_n)$ to $(\mathcal{I}_{B,\mathcal{T}}, B)$, but there is no simulation in the converse direction, i.e., it indeed holds true that $B \sqsubsetneq_{\mathcal{T}} C_n$.

Let $C \in \mathcal{EL}(\Sigma)$. We proceed with a case distinction on whether $C$ contains $B$ as a subconcept.

1. Assume that $B$ is a subconcept of $C$. We are going to show that then $C \equiv_{\mathcal{T}} B$. The canonical model $\mathcal{I}_{C,\mathcal{T}}$ contains an $r$-path from the vertex $C$ to some vertex $D$ which has label $B$. Since $B \equiv \exists\, r.\, B \in \mathcal{T}$, the very definition of canonical models yields that each vertex on this path must be labelled with $B$, and hence each of these vertices has $B$ as an $r$-successor. Furthermore, $B$ is an $r$-successor of itself. We conclude that the canonical model $\mathcal{I}_{C,\mathcal{T}}$ has the following structure.

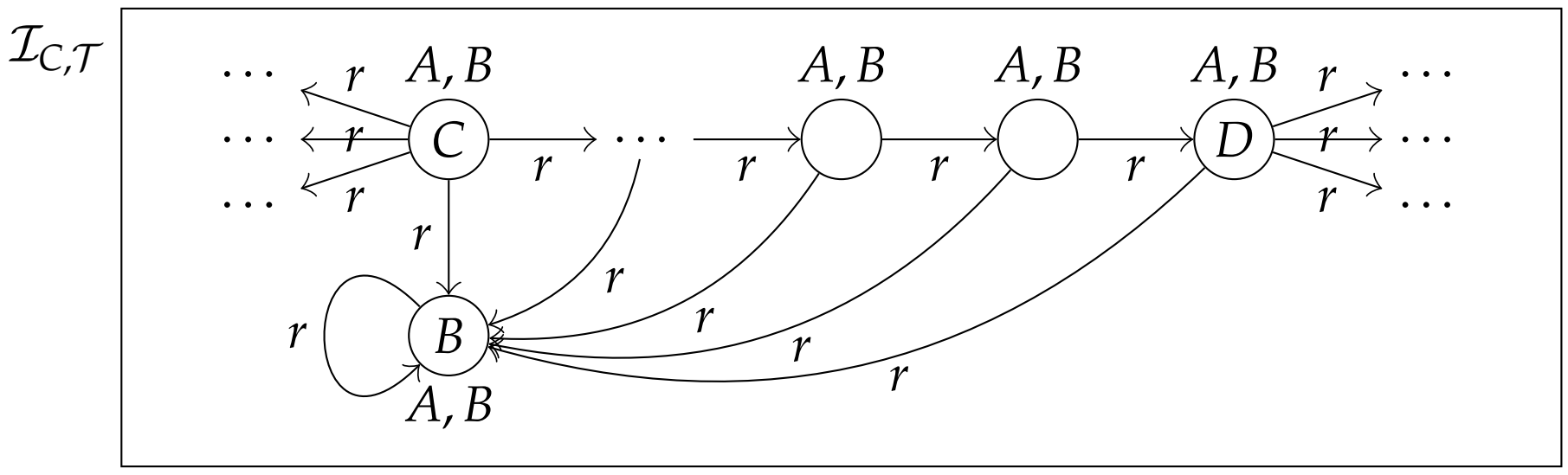

It is not hard to see that $(\mathcal{I}_{B,\mathcal{T}}, B)$ and $(\mathcal{I}_{C,\mathcal{T}}, C)$ are equi-similar, and thus $B \equiv_{\mathcal{T}} C$.

2. Now let $B$ be no subconcept of $C$, and consider only the connected component of the canonical model $\mathcal{I}_{C,\mathcal{T}}$ which contains the vertex $C$. Then this part must be tree-shaped, and each vertex may either have label $A$ or no labels at all. Furthermore, if in a branch of this tree there is a vertex $D$ with label $A$, then all ancestors of $D$ must also have label $A$ due to the presence of the concept inclusion $\exists\, r.\, A \sqsubseteq A$ in $\mathcal{T}$. If we set $n$ to the length of a longest path in this tree, then $(\mathcal{I}_{C,\mathcal{T}}, C)$ can apparently be simulated in $(\mathcal{I}_{C_n,\mathcal{T}}, C_n)$, i.e., $C_n \sqsubseteq_{\mathcal{T}} C$. Furthermore, there exists a simulation from $(\mathcal{I}_{C_m,\mathcal{T}}, C_m)$ to $(\mathcal{I}_{C,\mathcal{T}}, C)$ where within the tree $m$ is the length of a longest path all vertices of which are labelled with $A$. Hence, $C \sqsubseteq_{\mathcal{T}} C_m$.

We conclude that each $\mathcal{EL}(\Sigma)$ concept description $C$ either is equivalent to $B$ w.r.t. $\mathcal{T}$ or there exists an $n \in \mathbb{N}$ such that $B \sqsubsetneq_{\mathcal{T}} C_{n-1} \sqsubsetneq C$, i.e., $B$ does not have upper neighbors with respect to $\mathcal{T}$. □

**Corollarium 3.6.4.** *There is some $\mathcal{EL}$ TBox $\mathcal{T}$ over some signature $\Sigma$ for which the subsumption relation $\sqsubseteq_{\mathcal{T}}$ is not neighborhood generated.* □

## 3.7. RELATIONSHIPS BETWEEN ∅-NEIGHBORS AND $\mathcal{T}$-NEIGHBORS

This section's goal is to explore relationships between neighbors w.r.t. $\emptyset$, neighbors w.r.t. $\mathcal{T}$, and most specific consequences. For this purpose, let $\mathcal{T}$ be some $\mathcal{EL}^{\bot}$ TBox, let $C, D, E$ be $\mathcal{EL}^{\bot}$ concept descriptions, and let $r$ be some role name.

1. We have that $C \prec_{\mathcal{T}} D$ does not imply $\exists\, r.\, C \prec_{\mathcal{T}} \exists\, r.\, D$. As a counterexample define $\mathcal{T} := \{\exists\, r.\, A \equiv \exists\, r.\, \top\}$. Then it holds true that $A \prec_{\mathcal{T}} \top$, but $\exists\, r.\, A \not\prec_{\mathcal{T}} \exists\, r.\, \top$.

2. It does not hold true that $C \prec_{\mathcal{T}} D$ implies $C \sqcap E \prec_{\mathcal{T}} D \sqcap E$. Consider the counterexample $\mathcal{T} := \{A \sqcap B \equiv B\}$: it holds true that $A \prec_{\mathcal{T}} \top$, but $A \sqcap B \not\prec_{\mathcal{T}} B$.

3. $C \prec_{\mathcal{T}} D$ is equivalent to $C^{\mathcal{T}} \prec_{\mathcal{T}} D^{\mathcal{T}}$, since $C \equiv_{\mathcal{T}} C^{\mathcal{T}}$ holds true for all $\mathcal{EL}^{\bot}$ TBoxes $\mathcal{T}$.

4. $C \prec_{\emptyset} D$ does not imply $C \prec_{\mathcal{T}} D$. As a simple counterexample consider $C := A$, $D := \top$, and $\mathcal{T} := \{\top \sqsubseteq A\}$.

5. $C^{\mathcal{T}} \prec_{\emptyset} D^{\mathcal{T}}$ implies $C \prec_{\mathcal{T}} D$.

   *Approbatio.* Assume that $C^{\mathcal{T}}$ is a lower neighbor of $D^{\mathcal{T}}$ with respect to the empty TBox $\emptyset$. We shall immediately conclude that $C^{\mathcal{T}} \sqsubseteq_{\emptyset} D^{\mathcal{T}}$ as well as $C^{\mathcal{T}} \not\sqsupseteq_{\emptyset} D^{\mathcal{T}}$. Applying Section 2.9 yields that $C \sqsubseteq_{\mathcal{T}} D$ and $C \not\sqsupseteq_{\mathcal{T}} D$, i.e., Statement 1 of Definitio 3.1 are satisfied. Now consider an $\mathcal{EL}^{\bot}$ concept description $E$ such that $C \sqsubseteq_{\mathcal{T}} E \sqsubseteq_{\mathcal{T}} D$, i.e., by means of Section 2.9 this is equivalent to $C^{\mathcal{T}} \sqsubseteq_{\emptyset} E^{\mathcal{T}} \sqsubseteq D^{\mathcal{T}}$. By assumption, we may conclude that $E^{\mathcal{T}} \equiv_{\emptyset} C^{\mathcal{T}}$ or $E^{\mathcal{T}} \equiv_{\emptyset} D^{\mathcal{T}}$, i.e., $E \equiv_{\mathcal{T}} C$ or $E \equiv_{\mathcal{T}} D$. Consequently, also Statement 2 of Definitio 3.1 holds true, and thus $C \prec_{\mathcal{T}} D$ as claimed. □

6. We have that $C \prec_{\emptyset} D$ does not always imply $C^{\mathcal{T}} \prec_{\emptyset} D^{\mathcal{T}}$.

7. Furthermore, $C \prec_{\mathcal{T}} D$ does not imply $C \prec_{\emptyset} D$.

   *Approbatio.* Consider the signature $\Sigma$ where $\Sigma_{\mathsf{C}} := \emptyset$ and $\Sigma_{\mathsf{R}} := \{r\}$. Define $\mathcal{T} := \{\exists\, r.\, \top \sqsubseteq \exists\, r.\, \exists\, r.\, \top\}$, and consider the concept descriptions $C := \exists\, r.\, \exists\, r.\, \top$ and $D := \top$. Obviously, modulo equivalence w.r.t. $\mathcal{T}$ there are only two distinct concept descriptions, namely $\exists\, r.\, \top$ and $\top$. In particular, $[\top]_{\mathcal{T}} = \{\top\}$ and $[\exists\, r.\, \top]_{\mathcal{T}} = \mathcal{EL}^{\bot}(\Sigma) \setminus \{\top\}$. We conclude that $C \prec_{\mathcal{T}} D$. However, $C \sqsubsetneq_{\emptyset} \exists\, r.\, \top \sqsubsetneq_{\emptyset} D$, and thus $C \not\prec_{\emptyset} D$. □

8. Eventually, $C \prec_{\mathcal{T}} D$ does not imply $C^{\mathcal{T}} \prec_{\emptyset} D^{\mathcal{T}}$.

   *Approbatio.* Let the signature $\Sigma$ be defined by $\Sigma_{\mathsf{C}} := \{A_1, A_2, B_1, B_2\}$ and $\Sigma_{\mathsf{R}} := \emptyset$, and consider the TBox $\mathcal{T} := \{A_1 \equiv A_2,\ B_1 \equiv B_2\}$. Then, modulo equivalence w.r.t. $\mathcal{T}$ there exists exactly four distinct $\mathcal{EL}(\Sigma)$ concept descriptions, which are $\top$, $A_1$, $B_1$, and $A_1 \sqcap B_1$.

   Now define $C := A_1 \sqcap B_1$ and $D := A_1$. The corresponding most specific consequences satisfy $C^{\mathcal{T}} \equiv_{\emptyset} A_1 \sqcap A_2 \sqcap B_1 \sqcap B_2$ and $D^{\mathcal{T}} \equiv_{\emptyset} A_1 \sqcap A_2$, respectively. It is apparent that $A_1 \sqcap A_2 \sqcap B_1$ is strictly between $C^{\mathcal{T}}$ and $D^{\mathcal{T}}$ with respect to the empty TBox, i.e., $C^{\mathcal{T}} \not\prec_{\emptyset} D^{\mathcal{T}}$.

   Eventually, we can readily verify that $C \prec_{\mathcal{T}} D$. □

# 4. THE DISTRIBUTIVE, GRADED LATTICE OF $\mathcal{EL}$ CONCEPT DESCRIPTIONS

The goal of this section is to explore the properties of the lattice of $\mathcal{EL}$ concept descriptions ordered by subsumption with respect to the empty TBox. In particular, Blyth Blyth, 2005, Chapters 4 and 5 shows that it suffices to investigate whether this lattice is distributive and of locally finite length, such that as an immediate corollary we then obtain that also the Jordan-Dedekind chain condition is satisfied, which states that for each pair $C \sqsubseteq_\emptyset D$, all maximal chains in the intervall $[C, D]$ have the same length. Furthermore, this length can then be utilized to define a distance between $C$ and $D$, and in particular to measure a distance from each concept description $C$ to the top concept description $\top$, which we call the rank of $C$.

## 4.1. DISTRIBUTIVITY

**Lemma 4.1.1.** *For each signature $\Sigma$, the lattice $\boldsymbol{\mathcal{EL}}(\Sigma)$ is distributive, i.e., for all concept descriptions $C, D, E \in \mathcal{EL}(\Sigma)$, it holds true that*

$$\begin{aligned} & C \sqcap (D \vee E) \equiv_\emptyset (C \sqcap D) \vee (C \sqcap E), \\ \textit{and} \quad & C \vee (D \sqcap E) \equiv_\emptyset (C \vee D) \sqcap (C \vee E). \end{aligned}$$

*Approbatio.* We first show that the concept names occuring on the top level are the same for both concept descriptions $C \sqcap (D \vee E)$ and $(C \sqcap D) \vee (C \sqcap E)$. For this purpose we use the fact that the power-set lattice is distributive.

$$\begin{aligned} \mathsf{Conj}(C \sqcap (D \vee E), \Sigma_\mathsf{C}) &= \mathsf{Conj}(C, \Sigma_\mathsf{C}) \cup \mathsf{Conj}(D \vee E, \Sigma_\mathsf{C}) \\ &= \mathsf{Conj}(C, \Sigma_\mathsf{C}) \cup (\mathsf{Conj}(D, \Sigma_\mathsf{C}) \cap \mathsf{Conj}(E, \Sigma_\mathsf{C})) \\ &= (\mathsf{Conj}(C, \Sigma_\mathsf{C}) \cup \mathsf{Conj}(D, \Sigma_\mathsf{C})) \cap (\mathsf{Conj}(C, \Sigma_\mathsf{C}) \cup \mathsf{Conj}(E, \Sigma_\mathsf{C})) \\ &= \mathsf{Conj}(C \sqcap D, \Sigma_\mathsf{C}) \cap \mathsf{Conj}(C \sqcap E, \Sigma_\mathsf{C}) \\ &= \mathsf{Conj}((C \sqcap D) \vee (C \sqcap E), \Sigma_\mathsf{C}) \end{aligned}$$

Now consider an existential restriction $\exists\, r.\, Y \in \mathsf{Conj}((C \sqcap D) \vee (C \sqcap E))$, i.e., there must exist $\exists\, r.\, Y_1 \in \mathsf{Conj}(C \sqcap D)$ and $\exists\, r.\, Y_2 \in \mathsf{Conj}(C \sqcap E)$ such that $Y = Y_1 \vee Y_2$. We need to show that there is some $\exists r. X \in \mathsf{Conj}(C \sqcap (D \vee E))$ with $X \sqsubseteq_\emptyset Y$. If $\exists r. Y_i \in \mathsf{Conj}(C)$ for some $i \in \{1,2\}$, then choose $X := Y_i$. Otherwise it must hold true that $\exists r. Y_1 \in \mathsf{Conj}(D)$ and $\exists r. Y_2 \in \mathsf{Conj}(E)$, which implies $\exists r. (Y_1 \vee Y_2) \in \mathsf{Conj}(D \vee E)$, and hence we may choose $X := Y_1 \vee Y_2$.

Vice versa, let $\exists r. X \in \mathsf{Conj}(C \sqcap (D \vee E))$. If $\exists r. X \in \mathsf{Conj}(C)$, then $\exists r. X \in \mathsf{Conj}((C \sqcap D) \vee (C \sqcap E))$. If otherwise $\exists r. X \in \mathsf{Conj}(D \vee E)$, there exist $\exists r. X_1 \in \mathsf{Conj}(D) \subseteq \mathsf{Conj}(C \sqcap D)$ and $\exists r. X_2 \in \mathsf{Conj}(E) \subseteq \mathsf{Conj}(C \sqcap E)$ such that $X = X_1 \vee X_2$. Thus, it follows that $\exists r. X \in \mathsf{Conj}((C \sqcap D) \vee (C \sqcap E))$ too. □

**Lemma 4.1.2.** *For each signature $\Sigma$, the lattice $\boldsymbol{\mathcal{EL}}(\Sigma)$ is of locally finite length, that is, for all concept descriptions $C, D \in \mathcal{EL}(\Sigma)$ with $C \sqsubseteq_\emptyset D$, every chain in the interval $[C, D]$ has a finite length.*

*Approbatio.* The claim is an immediate consequence of the boundedness of $\sqsubseteq_\emptyset$, which Baader and Morawska showed in Baader and Morawska, 2010, Proof of Proposition 3.5. □

According to Blyth (Blyth, 2005, Chapters 4 and 5), the following statements are obtained as immediate consequences of Lemmata 4.1.1 and 4.1.2.

**Corollarium 4.1.3.** 1. *For each signature $\Sigma$, the lattice $\boldsymbol{\mathcal{EL}}(\Sigma)$ is modular, i.e., for all concept descriptions $C, D, E \in \mathcal{EL}(\Sigma)$, it holds true that*

$$\begin{aligned} (C \sqcap D) \vee (C \sqcap E) &\equiv_\emptyset C \sqcap (D \vee (C \sqcap E)), \\ (C \vee D) \sqcap (C \vee E) &\equiv_\emptyset C \vee (D \sqcap (C \vee E)), \\ C \sqsubseteq_\emptyset D \quad \textit{implies} \quad C \vee (E \sqcap D) &\equiv_\emptyset (C \vee E) \sqcap D, \\ \textit{and} \quad C \sqsupseteq_\emptyset D \quad \textit{implies} \quad C \sqcap (E \vee D) &\equiv_\emptyset (C \sqcap E) \vee D. \end{aligned}$$

2. *For each signature $\Sigma$, the lattice $\boldsymbol{\mathcal{EL}}(\Sigma)$ is both upper and lower semi-modular, i.e., for all concept descriptions $C, D \in \mathcal{EL}(\Sigma)$, it holds true that*

$$C \sqcap D \prec_\emptyset C \quad \textit{if, and only if,} \quad D \prec_\emptyset C \vee D.$$

3. *For each signature $\Sigma$, the lattice $\boldsymbol{\mathcal{EL}}(\Sigma)$ satisfies the* Jordan-Dedekind chain condition, *i.e., for all concept descriptions $C, D \in \boldsymbol{\mathcal{EL}}(\Sigma)$ with $C \sqsubsetneq_\emptyset D$, it holds true that all maximal chains in the interval $[C, D]$ have the same length.* □

## 4.2. RANK FUNCTIONS

The notion of a rank function can be defined for ordered sets. The following definition specifically tailors this notion for the lattice $\boldsymbol{\mathcal{EL}}(\Sigma)$.

**Definitio 4.2.1.** An $\mathcal{EL}$ *rank function* is a mapping $|\cdot| \colon \mathcal{EL}(\Sigma) \to \mathbb{N}$ with the following properties.

1. $|\top| = 0$
2. $C \equiv_\emptyset D$ implies $|C| = |D|$ (equivalence closed)
3. $C \sqsubsetneq_\emptyset D$ implies $|C| \gneq |D|$ (strictly order preserving)
4. $C \prec_\emptyset D$ implies $|C| + 1 = |D|$ (neighborhood preserving)

For an $\mathcal{EL}$ concept description $C$, we say that $|C|$ is the *rank* of $C$. △

**Lemma 4.2.2.** *For each $C \in \mathcal{EL}(\Sigma)$, let $|C| := 0$ if $C \equiv_\emptyset \top$, and otherwise define*

$$|C| := \max\{\, n+1 \mid \exists D_1,\ldots,D_n \in \mathcal{EL}(\Sigma)\colon C \prec_\emptyset D_1 \prec_\emptyset \ldots \prec_\emptyset D_n \prec_\emptyset \top \,\}.$$

*Then, $|\cdot|$ is an $\mathcal{EL}$ rank function.*

*Approbatio.* It is readily verfied that $|\cdot|$ satisfies Statements 1 and 2 of Definitio 4.2.1. We proceed with proving that Statement 4 holds true for $|\cdot|$, which implies the validity of Statement 3 for $|\cdot|$. Consider $\mathcal{EL}$ concept descriptions $C$ and $D$ such that $C \prec_\emptyset D$. Clearly, if we consider a maximal chain from $D$ to $\top$, and add $C$ as prefix, then we have a maximal chain from $C$ to $\top$. It is thus immediate to conclude $|C|+1 = |D|$. □

Since $\boldsymbol{\mathcal{EL}}(\Sigma)$ satisfies the Jordan-Dedekind chain condition, we infer that in order to compute the rank $|C|$ of an $\mathcal{EL}$ concept description $C$ over $\Sigma$ with $C \not\equiv_\emptyset \top$, we simply need to find *one* chain $C \prec_\emptyset D_1 \prec_\emptyset D_2 \prec_\emptyset \ldots \prec_\emptyset D_n \prec_\emptyset \top$, and then it follows that $|C| = n+1$. Furthermore, $|C| = 0$ if $C \equiv_\emptyset \top$.

**Corollarium 4.2.3.** *For each signature $\Sigma$, the lattice $\boldsymbol{\mathcal{EL}}(\Sigma)$ is graded.* □

**Lemma 4.2.4.** *For all $\mathcal{EL}$ concept descriptions $C$ and $D$ over some signature $\Sigma$, the following equation holds true.*

$$|C| + |D| = |C \sqcap D| + |C \vee D|$$

*Approbatio.* follows from Lemma 4.1.2, Corollarium 4.1.3, and (Blyth, 2005, Theorem 4.6). □

**Lemma 4.2.5.** *Let $\mathbf{C}$ be a set of $n$ $\mathcal{EL}$ concept descriptions over $\Sigma$. Then, the following equation holds true.*

$$|\textstyle\bigsqcap \mathbf{C}| = \displaystyle\sum_{i=1}^{n} (-1)^{i+1} \cdot \sum_{\mathbf{D} \in \binom{\mathbf{C}}{i}} |\textstyle\bigvee \mathbf{D}|$$

*Approbatio.* We show the claim by induction on $n$. The induction base where $n \in \{0,1\}$ is trivial, and for $n = 2$ has been shown in Lemma 4.2.4. For the induction step let now $n > 2$. Using the equation from Lemma 4.2.4, we infer the following for each $C \in \mathbf{C}$.

$$|C| + |\textstyle\bigsqcap \mathbf{C} \setminus \{C\}| = |\bigsqcap \mathbf{C}| + |C \vee \bigsqcap \mathbf{C} \setminus \{C\}|$$

By means of the finitely generalized distributivity law and another application of Lemma 4.2.4 we conclude that the following equation holds true for each $C \in \mathbf{C}$.

$$|C| + |\textstyle\bigsqcap \mathbf{C} \setminus \{C\}| = |\bigsqcap \mathbf{C}| + |\bigsqcap\{\, C \vee D \mid D \in \mathbf{C} \setminus \{C\} \,\}|$$

The induction hypothesis allows for replacing the ranks of the $(n-1)$-ary conjunctions, and thus yields the following equation for each $C \in \mathbf{C}$.

$$\begin{aligned}|C| + \sum_{j=1}^{n-1} \sum_{\mathbf{D} \in \binom{\mathbf{C}\setminus\{C\}}{j}} (-1)^{j+1} \cdot |\textstyle\bigvee \mathbf{D}| &= |\textstyle\bigsqcap \mathbf{C}| + \displaystyle\sum_{k=1}^{n-1} \sum_{\mathbf{E} \in \binom{\{\, C \vee D \,\mid\, D \in \mathbf{C}\setminus\{C\} \,\}}{k}} (-1)^{k+1} \cdot |\textstyle\bigvee \mathbf{E}| \\ &= |\textstyle\bigsqcap \mathbf{C}| + \displaystyle\sum_{k=1}^{n-1} \sum_{\mathbf{E} \in \binom{\mathbf{C}\setminus\{C\}}{k}} (-1)^{k+1} \cdot |C \vee \textstyle\bigvee \mathbf{E}|\end{aligned}$$

If we sum up the $n$ equations, then we see that on the left hand side there are exactly $n$ occurences of $|C|$ for each $C \in \mathbf{C}$, and furthermore that for each $j \in \{2, \ldots, n-1\}$ and for each $\mathbf{D} \in \binom{\mathbf{C}}{j}$, there exist exactly $n-j$ occurences of the summand $(-1)^{j+1} \cdot |\bigvee \mathbf{D}|$. On the right hand side, there are, obviously, $n$ occurences of $|\bigsqcap \mathbf{C}|$. Furthermore, for each $k \in \{2, \ldots, n\}$ and for each $\mathbf{E} \in \binom{\mathbf{C}}{k}$, there are exactly $k$ occurences of the summand $(-1)^k \cdot |\bigvee \mathbf{E}|$. Rearranging and then dividing by $n$ eventually yields the induction claim for $n$. □

Let $C = A_1 \sqcap \ldots \sqcap A_m \sqcap \exists\, r_1. C_1 \sqcap \ldots \sqcap \exists\, r_n. C_n$ be a reduced $\mathcal{EL}$ concept description. Then its rank can be computed as follows, cf. Lemma 4.2.5.

$$\begin{aligned} |C| &= |A_1 \sqcap \ldots \sqcap A_m \sqcap \exists\, r_1. C_1 \sqcap \ldots \sqcap \exists\, r_n. C_n| \\ &= |A_1 \sqcap \ldots \sqcap A_m| + |\exists\, r_1. C_1 \sqcap \ldots \sqcap \exists\, r_n. C_n| - |\top| \\ &= m + |\exists\, r_1. C_1 \sqcap \ldots \sqcap \exists\, r_n. C_n| \end{aligned}$$

Furthermore, it holds true that $\exists\, r. C \vee \exists\, s. D \equiv_\emptyset \top$ if $r \neq s$. It follows that we can further simplify the rank computation as follows.

$$\begin{aligned} |\exists\, r_1. C_1 \sqcap \ldots \sqcap \exists\, r_n. C_n| &= |\bigsqcap \{ \bigsqcap \{ \exists\, r_i. C_i \mid i \in \{1, \ldots, n\} \text{ and } r_i = r \} \mid r \in \Sigma_\mathsf{R} \}| \\ &= \sum_{r \in \Sigma_\mathsf{R}} |\bigsqcap \{ \exists\, r_i. C_i \mid i \in \{1, \ldots, n\} \text{ and } r_i = r \}| \end{aligned}$$

The rank of the conjunction of existential restrictions can be computed by means of Lemma 4.2.5, and finally it is readily verified that the rank of one existential restriction $\exists\, r. C$ satisfies the following equation.

$$|\exists\, r. C| = 1 + |\bigsqcap \{ \exists\, r. D \mid C \prec_\emptyset D \}|$$

**Lemma 4.2.6.** *Let $\exists\, r. C$ be an $\mathcal{EL}$ concept description over some signature $\Sigma$. Then the following inequalities hold true.*

$$1 + |C| \leq |\exists\, r. C| \leq 1 + \sum_{i=1}^{|C|} \prod_{j=0}^{i-2} (|C| - j) \leq 1 + |C| \cdot |C|! \leq 1 + |C|^{1+|C|}.$$

*Approbatio.* For each natural number $n$ with $n \leq |C|$, let

$$X_n := \bigsqcap \{ \exists\, r. D \mid C \prec^n D \}.$$

Clearly, it then holds true that $\exists\, r. C \equiv_\emptyset X_0 \sqsubsetneq_\emptyset X_1 \sqsubsetneq_\emptyset X_2 \sqsubsetneq_\emptyset \ldots \sqsubsetneq_\emptyset X_{|C|} \equiv_\emptyset \exists\, r. \top \prec_\emptyset \top$, i.e., $|\exists\, r. C| \geq 1 + |C|$. As a further step, we infer that $|\exists\, r. C| = 1 + \sum_{i=1}^{|C|} \mathsf{d}(X_{i-1}, X_i)$, and the distances $\mathsf{d}(X_{i-1}, X_i)$ can be approximated as follows. Beforehand note that $|\mathsf{Upper}(Y)| \leq |\mathsf{Conj}(Y)| \leq |Y|$ holds true for all $\mathcal{EL}$ concept descriptions $Y$.

Apparently, $\mathsf{d}(X_0, X_1) = 1$ holds true.

In order to construct a chain of neighbors from $X_1$ to $X_2$, we could simply iterate over all top-level conjuncts of $X_1$ and replace each with its *unique* upper neighbor. Of course, the number of top-level conjuncts of $X_1$ is bounded by the number of upper neighbors of $C$, is henceforth bounded by the number of top-level conjuncts of $C$, and thus we obtain that $\mathsf{d}(X_1, X_2) \leq |\mathsf{Conj}(X_1)| \leq |\mathsf{Upper}(C)| \leq |\mathsf{Conj}(C)| \leq |C|$.

The distance between the next two concept descriptions can be approximated as follows.

$$\mathsf{d}(X_2, X_3) \leq |\{\, \exists r.\, E \mid C \prec^2 E \,\}| \leq \sum_{C \prec D} \underbrace{|\{\, \exists r.\, E \mid D \prec E \,\}|}_{\leq |\mathsf{Conj}(D)| \leq |D| \leq |C|-1} \leq |C| \cdot (|C| - 1)$$

Continuing the approach, we infer the following upper bound for the distance between $X_3$ and $X_4$.

$$\begin{aligned}\mathsf{d}(X_3, X_4) &\leq |\{\, \exists r.\, F \mid C \prec^3 F \,\}| \leq \sum_{C \prec^2 E} \underbrace{|\{\, \exists r.\, F \mid E \prec F \,\}|}_{\leq |\mathsf{Conj}(E)| \leq |E| \leq |C|-2} \\ &\leq \sum_{C \prec D} \sum_{D \prec E} (|C| - 2) \leq \sum_{C \prec D} \underbrace{|D|}_{= |C|-1} \cdot (|C| - 2) \leq |C| \cdot (|C| - 1) \cdot (|C| - 2)\end{aligned}$$

In general, for each $i \in \mathbb{N} \cap [1, |C|]$, we observe the following.

$$\begin{aligned} d(X_{i-1}, X_i) &\leq |\{\, \exists r.\, D \mid C \prec^{i-1} D \,\}| \\ &\leq \sum_{C \prec Y_1} \sum_{Y_1 \prec Y_2} \cdots \sum_{Y_{i-2} \prec Y_{i-1}} 1 \\ &\leq |C| \cdot (|C| - 1) \cdot \ldots \cdot (|C| - (i - 2)) \\ &= \prod_{j=0}^{i-2} (|C| - j). \end{aligned}$$

Eventually, we conclude that the following inequalities are satisfied.

$$\begin{aligned} |\exists r.\, C| &= 1 + \sum_{i=1}^{|C|} \mathsf{d}(X_{i-1}, X_i) \\ &\leq 1 + \sum_{i=1}^{|C|} \prod_{j=0}^{i-2} (|C| - j) \\ &\leq 1 + \sum_{i=1}^{|C|} |C|! \\ &= 1 + |C| \cdot |C|! \\ &\leq 1 + |C|^{1+|C|} \end{aligned}$$

□

## 4.3. DISTANCE FUNCTIONS

**Definitio 4.3.1.** An $\mathcal{EL}$ *metric* or $\mathcal{EL}$ *distance function* is a mapping $\mathsf{d} \colon \mathcal{EL}(\Sigma) \times \mathcal{EL}(\Sigma) \to \mathbb{N}$ with the following properties.

1. $\mathsf{d}(C, D) \geq 0$ (non-negative)
2. $\mathsf{d}(C, D) = 0$ if, and only if, $\emptyset \models C \equiv D$ (equivalence closed)
3. $\mathsf{d}(C, D) = \mathsf{d}(D, C)$ (symmetric)
4. $\mathsf{d}(C, E) \leq \mathsf{d}(C, D) + \mathsf{d}(D, E)$ (triangle inequality)

We then also say that $\mathsf{d}(C, D)$ is the *distance* between $C$ and $D$. △

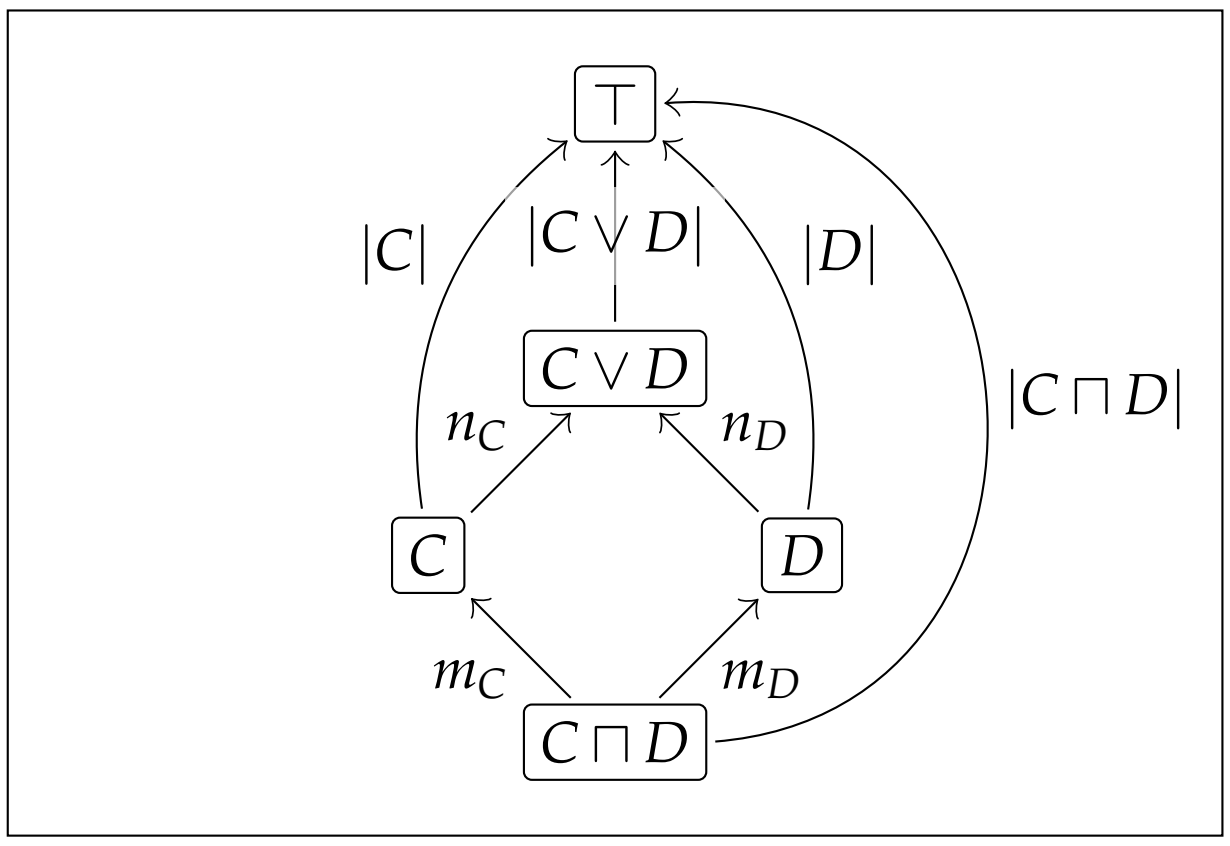

Figure 4.1.: Obtaining a distance function from the rank function

Lemma 4.2.5 for the case $n = 2$ yields that in the rectangle shown in Figure 4.1 opposite edges have the same length, where length means length of a maximal chain between the endpoints. It is easy to see that $|C \sqcap D| = |C| + m_C = |D| + m_D$ and $|C \vee D| = |C| - n_C = |D| - n_D$. Thus, we infer that $m_C = |C \sqcap D| - |C| = |D| - |C \vee D| = n_D$, and similarly that $m_D = n_C$. Consequently, we can define an $\mathcal{EL}$ distance function in the following way.

**Propositio 4.3.2.** *For all $C, D \in \mathcal{EL}(\Sigma)$, define*

$$\mathsf{d}(C, D) := |C \sqcap D| - |C \vee D|.$$

*Then,* d *is an $\mathcal{EL}$ metric.*

*Approbatio.* Statements 1 to 3 are obvious. We proceed with proving the triangle inequality in Statement 4. Fix some $\mathcal{EL}$ concept descriptions $C$, $D$, and $E$. Firstly, we observe that

$$|D| \leq |D \sqcap (C \vee E)| = |(C \sqcap D) \vee (D \sqcap E)|,$$
$$\text{and} \quad |C \sqcap E| \leq |D \sqcap (C \sqcap E)| = |(C \sqcap D) \sqcap (D \sqcap E)|.$$

Since $|X \vee Y| + |X \sqcap Y| = |X| + |Y|$ for all $\mathcal{EL}$ concept descriptions $X, Y, Z$, we infer that

$$|C \sqcap E| + |D| \leq |C \sqcap D| + |D \sqcap E|.$$

Multiplying the inequality with 2, adding some summands, and rearranging now yields

$$|C \sqcap E| - (|C| + |E| - |C \sqcap E|)$$
$$\leq |C \sqcap D| - (|C| + |D| - |C \sqcap D|) + |D \sqcap E| - (|D| + |E| - |D \sqcap E|).$$

Finally, using the identity $|X \vee Y| = |X| + |Y| - |X \sqcap Y|$ we conclude that

$$(|C \sqcap E| - |C \vee E|) \leq (|C \sqcap D| - |C \vee D|) + (|D \sqcap E| - |D \vee E|). \qquad \square$$

The next lemma justifies the name of a distance function. Indeed, if we consider the graph of $\mathcal{EL}$ concept descriptions such that edges exist exactly between neighboring concept descriptions, then the distance $\mathsf{d}(C, D)$ is the length of a shortest path between $C$ and $D$ in this graph.

**Lemma 4.3.3.** *In the graph* $(\mathcal{EL}(\Sigma), \prec_\emptyset \cup \succ_\emptyset)$ *it holds true that* $\mathsf{d}(C, D)$ *is the length of a shortest path from* $C$ *to* $D$ *for all* $C, D \in \mathcal{EL}(\Sigma)$.

*Approbatio.* Set $\sim_\emptyset := \prec_\emptyset \cup \succ_\emptyset$. Firstly, we show by induction over $\ell$ that for all $\mathcal{EL}$ concept descriptions $C$ and $D$ over $\Sigma$ and all paths from $C$ to $D$ of length $\ell$, it holds true that $\mathsf{d}(C, D) \leq \ell$. The induction base where $\ell \in \{0, 1\}$ is obvious. For the induction step now let $C \sim_\emptyset \ldots \sim_\emptyset E \sim_\emptyset D$ be a path of length $\ell > 1$. In particular, the prefix $C \sim_\emptyset \ldots \sim_\emptyset E$ is a path of length $\ell - 1$ from $C$ to $E$, and the induction hypothesis yields that $\mathsf{d}(C, E) \leq \ell - 1$. With the triangle inequality we can infer that

$$\mathsf{d}(C, D) \leq \mathsf{d}(C, E) + \mathsf{d}(E, D) \leq (\ell - 1) + 1 = \ell.$$

It remains to show that for all $C, D \in \mathcal{EL}(\Sigma)$, there exists a path $C \sim_\emptyset \ldots \sim_\emptyset D$ of length $\mathsf{d}(C, D)$. We have already seen that $\mathsf{d}(C, D) = (|C \sqcap D| - |C|) + (|C \sqcap D| - |D|)$, and there exists a path from $C$ to $C \sqcap D$ of length $|C \sqcap D| - |C|$ as well as a path from $C \sqcap D$ to $D$ of length $|C \sqcap D| - |D|$. Conjoining these two paths obviously yields a path from $C$ to $D$ of length $\mathsf{d}(C, D)$. □

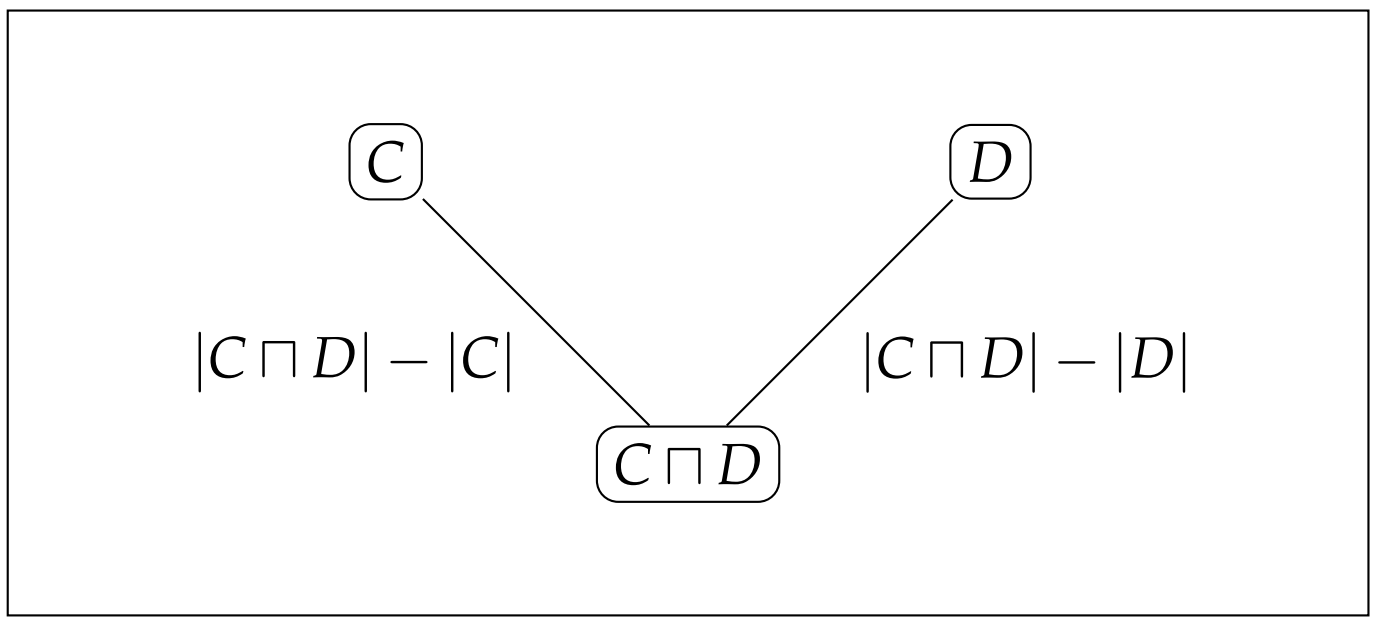

**Corollarium 4.3.4.** $\mathcal{EL}(\Sigma)$ *is a metric lattice, i.e., a lattice which is also a metric space.* □

Furthermore, this metric space is complete, that is, every Cauchy sequence converges modulo equivalence. All subsets of $(\mathcal{EL}(\Sigma)/\emptyset, \mathsf{d})$ are open, since for each $\mathbf{C} \subseteq \mathcal{EL}(\Sigma)/\emptyset$ and each $[C]_\emptyset \in \mathbf{C}$, it holds true that $\mathsf{B}_{\frac{1}{2}}([C]_\emptyset) = \{[C]_\emptyset\} \subseteq \mathbf{C}$. Consequently, all subsets of $(\mathcal{EL}(\Sigma)/\emptyset, \mathsf{d})$ are closed too. It follows that for all metric spaces $(X, \mathsf{d}')$, all mappings $f \colon \mathcal{EL}(\Sigma)/\emptyset \to X$ are continuous.

$(\mathcal{EL}(\Sigma)/\emptyset, \mathsf{d})$ is not bounded, i.e., there is no $\varepsilon \in \mathbb{R}$ such that $\mathsf{d}(C, D) \leq \varepsilon$ for all $C, D \in \mathcal{EL}(\Sigma)$. It is also not precompact or totally bounded, as there do not exist finitely many open balls of radius $\frac{1}{2}$ the union of which covers $\mathcal{EL}(\Sigma)/\emptyset$. Furthermore, this metric space of $\mathcal{EL}$ concept descriptions is not compact; the sequence $(\exists r^n. \top \mid n \in \mathbb{N})$ does contain a converging subsequence. However, it is locally compact, since for each point $[C]_\emptyset$ its neighborhood $\mathsf{B}_{\frac{1}{2}}([C]_\emptyset)$ clearly is compact. If the signature $\Sigma$ is finite, then each closed ball $\{ [D]_\emptyset \mid \mathsf{d}(C, D) \leq \varepsilon \}$ is finite and thus compact; it then follows that $(\mathcal{EL}(\Sigma)/\emptyset, \mathsf{d})$ is proper. We have already shown that all subsets of $\mathcal{EL}(\Sigma)/\emptyset$ are clopen, and hence this metric space is neither connected nor path connected. It is well known that $\mathcal{EL}(\Sigma)/\emptyset$ is countable, and so it is separable.

In a canonical way, the metric space of $\mathcal{EL}$ concept descriptions over some signature $\Sigma$ induces a topological space $\tau_{\mathsf{d}}$ the base of which is the set of open balls $\mathsf{B}_\varepsilon([C]_\emptyset)$ for $\varepsilon \in \mathbb{R}$ and $C \in \mathcal{EL}(\Sigma)$. In particular, $\tau_{\mathsf{d}}$ is the smallest subset of $\wp(\mathcal{EL}(\Sigma)/\emptyset)$ which contains all open balls, and satisfies the following conditions.

1. $\{\emptyset, \mathcal{EL}(\Sigma)/\emptyset\} \subseteq \tau_{\mathsf{d}}$.

2. $\bigcup \mathbf{C} \in \tau_\mathsf{d}$ for all $\mathbf{C} \subseteq \tau_\mathsf{d}$.

3. $\bigcap \mathbf{C} \in \tau_\mathsf{d}$ for all finite $\mathbf{C} \subseteq \tau_\mathsf{d}$.

Since all singletons $\{[C]_\emptyset\}$ are open balls $\mathsf{B}_{\frac{1}{2}}([C]_\emptyset)$, the induced topology $\tau_\mathsf{d}$ contains all these singletons. Due to the $\bigcup$-closedness we conclude that $\tau_\mathsf{d} = \wp(\mathcal{EL}(\Sigma)/\emptyset)$.

It is readily verified that all pairs of distinct points have disjoint neighborhoods, and thus $\tau_\mathsf{d}$ is a Hausdorff space, separated space, or $T_2$ space. Since all topological spaces the base of which are the open balls of some metric space are perfectly normal Hausdorff or $T_6$, we conclude that $\tau_\mathsf{d}$ is even a $T_6$ space. This means that all disjoint (closed) subsets $\mathbf{C}$ and $\mathbf{D}$ of $\mathcal{EL}(\Sigma)/\emptyset$ can be precisely separated by a continuous function $f\colon \mathcal{EL}(\Sigma)/\emptyset \to \mathbb{R}$, i.e., $f^{-1}(\{0\}) = \mathbf{C}$ and $f^{-1}(\{1\}) = \mathbf{D}$.

**Lemma 4.3.5.** *Let $C \in \mathcal{EL}(\Sigma)$, then $\mathsf{d}(C, \bigvee \mathsf{Upper}(C)) = |\mathsf{Upper}(C)|$ modulo equivalence.*

*Approbatio.* We show by induction on $n$ that if $\mathbf{U} \subseteq \mathsf{Upper}(C)$ with $|\mathbf{U}| = n > 0$, then $\mathsf{d}(C, \bigvee \mathbf{U}) = n$. The induction base where $n = 1$ is trivial. For the induction step now assume that $n > 1$, and let $\mathbf{U} = \mathbf{V} \cup \{D\}$ such that $D \notin \mathbf{V}$. Clearly, $\mathsf{d}(C, D) = 1$, and the induction hypothesis yields that $\mathsf{d}(C, \bigvee \mathbf{V}) = n - 1$. Since the conjunction of two non-equivalent upper neighbors of $C$ is equivalent to $C$, it follows that an analogous statement holds true for an arbitrary, but greater than 2, number of upper neighbors, i.e., $C \equiv_\emptyset D \sqcap \bigvee \mathbf{V}$, and hence we can infer by means of Lemma 4.2.4 that opposite sides in the rectangle below have the same length.

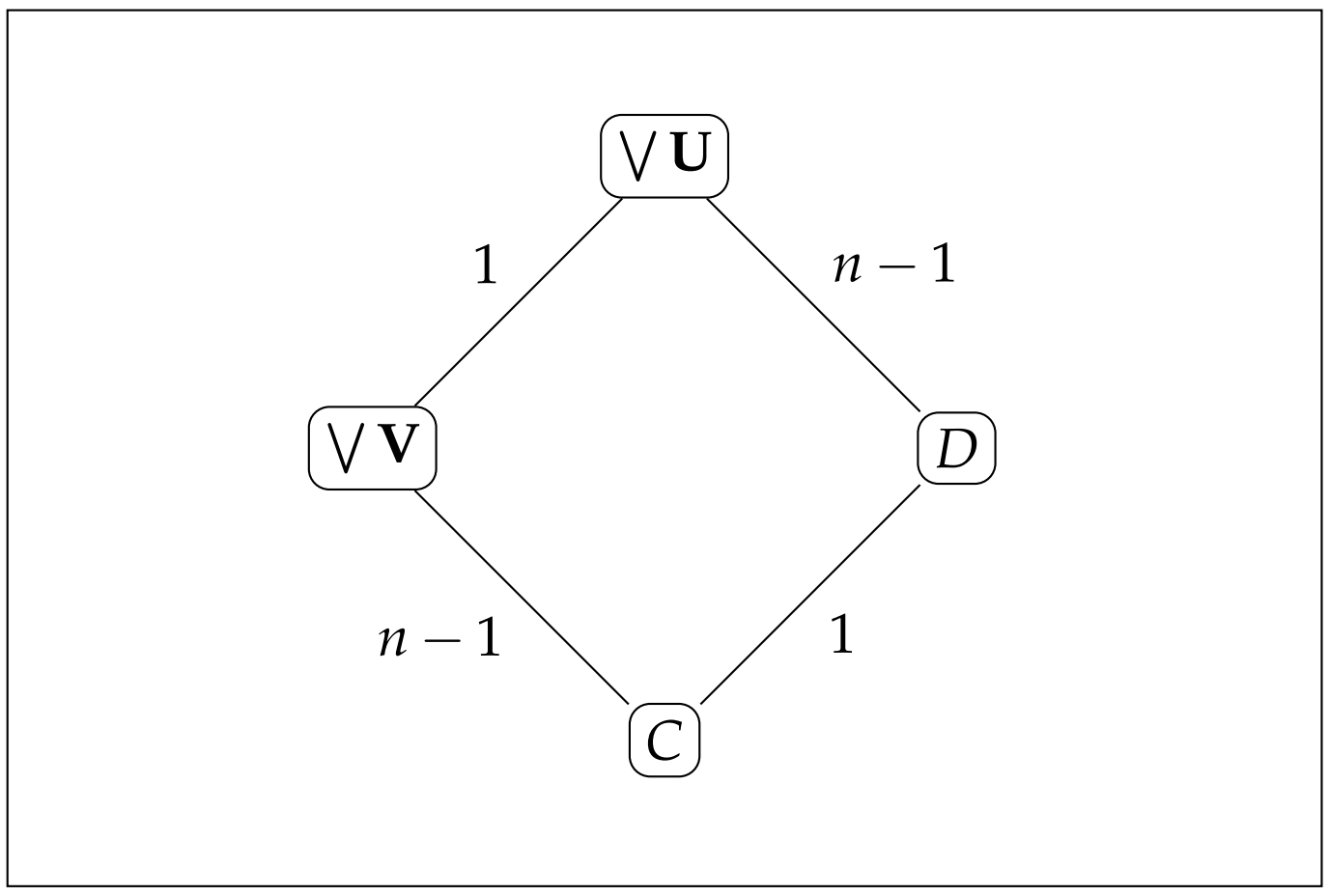

We conclude that $\mathsf{d}(C, \bigvee \mathbf{U}) = n$. □

According the the previous lemma, we can compute the rank of an $\mathcal{EL}$ concept description $C$ as follows.

1. Let $D := C$ and $r := 0$.

2. While $D \not\equiv_\emptyset \top$, compute the set $\mathsf{Upper}(D)$ of upper neighbors of $D$, set $r := r + |\mathsf{Upper}(D)|$ and $D := \bigvee \mathsf{Upper}(D)$.

3. Return $r$.

## 4.4. SIMILARITY FUNCTIONS

In Ecke, Peñaloza, and Turhan, 2015 Ecke, Peñaloza, and Turhan defined the notion of a concept similarity measure as a function of type $\mathcal{EL}(\Sigma) \times \mathcal{EL}(\Sigma) \to [0,1]$, and then considered so-called *relaxed instances* of concept descriptions with respect to ontologies. Simply speaking, $a$ is a relaxed instance of $C$ if there is a concept that is similar enough to $C$ and has $a$ as an instance. It is straight-forward to consider these relaxed instances also with respect to the distance function we have just introduced. More formally, we define them as follows.

**Definitio 4.4.1.** Consider an interpretation $\mathcal{I}$ over some signature $\Sigma$ and a concept description $C \in \mathcal{EL}(\Sigma)$, and let $n \in \mathbb{N}$. Then, the expression $Ↄ \leq n.C$ is called a *relaxed concept description*, and its extension is defined by

$$(Ↄ \leq n.C)^{\mathcal{I}} := \bigcup \{ D^{\mathcal{I}} \mid D \in \mathcal{EL}(\Sigma) \text{ and } \mathsf{d}(C,D) \leq n \}.$$

Suppose that $\mathcal{O}$ is an ontology over some signature $\Sigma$, and further let $a \in \Sigma_{\mathsf{I}}$ be an individual name, $C \in \mathcal{EL}(\Sigma)$ a concept description, and $n \in \mathbb{N}$. We then say that $a$ is a *relaxed instance* of $C$ with respect to $\mathcal{O}$ and distance threshold $n$, denoted as $\mathcal{O} \models a \sqsubseteq Ↄ \leq n.C$, if it holds true that $a^{\mathcal{I}} \in (Ↄ \leq n.C)^{\mathcal{I}}$ for each model $\mathcal{I}$ of $\mathcal{O}$. △

For transforming our distance function $\mathsf{d}$ into a similarity function $\mathsf{s}\colon \mathcal{EL}(\Sigma) \times \mathcal{EL}(\Sigma) \to [0,1]$ we can proceed as follows. We begin with transforming $\mathsf{d}$ into a metric with range $[0,1)$. For that purpose, we choose an order-preserving, sub-additive function $f\colon [0,\infty) \to [0,1)$ with $\ker(f) = \{0\}$. Note that a function $f\colon [0,\infty) \to \mathbb{R}$ is sub-additive if $f'' < 0$ and $f(0) = 0$. Then $f \circ \mathsf{d}$ is such a metric with range $[0,1)$. Suitable functions are the following.

- $f\colon x \mapsto \frac{x}{1+x}$ or more generally $f\colon x \mapsto (\frac{x}{1+x})^y$ for $y > 0$
- $f\colon x \mapsto 1 - \frac{1}{2^x}$ or more generally $f\colon x \mapsto 1 - y^x$ for $y \in (0,1)$

Then, $\mathsf{s} := 1 - f \circ \mathsf{d}$ is a similarity function on $\mathcal{EL}(\Sigma)$. It is easy to verify that then $\mathsf{s}$ satifies the following properties which have been defined by Lehmann and Turhan in Lehmann and Turhan, 2012, for all $\mathcal{EL}$ concept descriptions $C, D, E$ over $\Sigma$.

1. $\mathsf{s}(C,D) = \mathsf{s}(D,C)$ (symmetric)
2. $1 + \mathsf{s}(C,D) \geq \mathsf{s}(C,E) + \mathsf{s}(E,D)$ (triangle inequality)
3. $\emptyset \models C \equiv D$ implies $\mathsf{s}(C,E) = \mathsf{s}(D,E)$ (equivalence invariant)
4. $\emptyset \models C \equiv D$ if, and only if, $\mathsf{s}(C,D) = 1$ (equivalence closed)
5. $\emptyset \models C \sqsubseteq D \sqsubseteq E$ implies $\mathsf{s}(C,D) \geq \mathsf{s}(C,E)$ (subsumption preserving)
6. $\emptyset \models C \sqsubseteq D \sqsubseteq E$ implies $\mathsf{s}(C,E) \leq \mathsf{s}(D,E)$ (reverse subsumption preserving)

However, as it turns out such a similarity measure $1 - f \circ \mathsf{d}$ does not satisfy the property of *structural dependance*. For instance, consider a signature $\Sigma$ without role names and such that $\Sigma_C := \{A\} \cup \{ B_n \mid n \in \mathbb{N} \}$. It is now readily verified that

$$(1 - f \circ \mathsf{d})(A \sqcap \bigsqcap \{ B_\ell \mid \ell \leq n \}, \bigsqcap \{ B_\ell \mid \ell \leq n \}) = 1 - f(1)$$

for all $n \in \mathbb{N}$, and since $f(1) > 0$ we conclude that the sequence does not converge to $1$ for $n \to \infty$.

For extending our rank function $|\cdot|$ and our distance function $\mathsf{d}$ to $\mathcal{EL}^{\bot}$, we can simply define $|\bot| := \infty$, $\mathsf{d}(\bot, \bot) := 0$, and $\mathsf{d}(\bot, C) := \mathsf{d}(C, \bot) := \infty$ for $C \not\equiv_{\emptyset} \bot$. When transforming the extended metric into a similarity measure then two concept descriptions have a similarity of $0$ if, and only if, exactly one of them is unsatisfiable. In $\mathcal{EL}$ without the bottom concept description $\bot$, a similarity of $0$ can never occur when utilizing the above construction.

## 4.5. COMPUTATIONAL COMPLEXITY

We close this section with some first investigations on the complexities of decision problems and computation problems related to the introduced rank function.

### 4.5.1. COMPUTING THE RANK FUNCTION

As it turns out, the rank of $\mathcal{EL}$ concept descriptions can be asymptotically $\mathsf{rd}(C)$-exponential in the size of $C$. This vast growth makes it practically impossible to compute the rank function. It is easy to prove that in the case where $\mathsf{rd}(C) = 1$, the rank of $C$ can be at least exponential with respect to the size of $C$. To see this, consider the concept description $C_n := \exists\, r. \bigsqcap\{A_1, \ldots, A_n\}$ for each $n \in \mathbb{N}$. It is well-known that there are exponentially many subsets of $\{A_1, \ldots, A_n\}$ with $\lfloor \frac{n}{2} \rfloor$ elements; let $X_1, \ldots, X_\ell$ be an enumeration of these, and define $D_m := \bigsqcap\{\, \exists\, r. \bigsqcap X_i \mid i \in \{m, \ldots, \ell\}\,\}$. Clearly, then $C_n \sqsubsetneq_{\emptyset} D_1 \sqsubsetneq_{\emptyset} D_2 \sqsubsetneq_{\emptyset} \ldots \sqsubsetneq_{\emptyset} D_\ell \sqsubsetneq_{\emptyset} \top$ is an exponentially long chain of strict subsumptions. We conclude that $|C_n|$ is at least exponential in $n$.

In the sequel of this section, we show that there is a sequence of $\mathcal{EL}$ concept descriptions $C_n$ such that $\mathsf{rd}(C_n) = n$ and further that $|C_n|$ is asymptotically bounded below by $(2, ||C_n||) \uparrow\uparrow n$, i.e., the following holds true.

$$|C_n| \succeq \underbrace{2^{2^{\cdot^{\cdot^{\cdot^{2^{||C_n||}}}}}}}_{n \text{ times}}$$

We start with defining the signature $\Sigma_k$ by $(\Sigma_k)_{\mathsf{C}} := \{A_1, \ldots, A_k\}$ for some $k \in \mathbb{N}$ with $k \geq 3$ and $(\Sigma_k)_{\mathsf{R}} := \{r\}$. Note that the precondition $k \geq 3$ is essential, as we will discuss later. Our aim now is to construct a suitable sequence of concept descriptions $C_n \in \mathcal{EL}(\Sigma)$. However, we shall not do this directly, but rather translate this problem into a similar problem within order theory. For this purpose, it is necessary to introduce some notions.

As usual, a *partially ordered set* (abbrv. poset) $\mathbb{P}$ is a pair $(P, \leq)$ consisting of a set $P$ and binary relation $\leq$ on $P$ that is reflexive, antisymmetric, and transitive. An *ideal* in $\mathbb{P}$ is some subset of $P$ that is closed under $\leq$, that is, a subset $I \subseteq P$ such that $p \leq i$ implies $p \in I$ for each $i \in I$ and for any $p \in P$. The *prime ideal* of some element $p \in P$ is the ideal

$$\downarrow p := \{\, q \mid q \in P \text{ and } q \leq p \,\}.$$

Apparently, any ideal is a union of prime ideals. As further abbreviation, let $\downarrow Q := \bigcup\{\, \downarrow q \mid q \in Q \,\}$. We denote the set of all ideals in $\mathbb{P}$ by $\mathsf{Ideals}(\mathbb{P})$.

Two elements $p, q \in P$ are *comparable* in $\mathbb{P}$ if either $p \leq q$ or $q \leq p$ holds true. A *chain* in $\mathbb{P}$ is a subset $C \subseteq P$ such that any two elements in $C$ are comparable, while an *antichain* in $\mathbb{P}$ is a subset $A \subseteq P$ such that no two (distinct) elements in $A$ are comparable. The *height* of $\mathbb{P}$ is defined as the supremum over all cardinalities of chains in $\mathbb{P}$, denoted by $\mathsf{height}(\mathbb{P})$. The *width* of $\mathbb{P}$ is defined as the supremum over all cardinalities of antichains in $\mathbb{P}$, denoted by $\mathsf{width}(\mathbb{P})$. Dilworth's theorem, cf. (Dilworth, 1950), states that there is always a partition of $P$ into $n$ disjoint chains if $\mathbb{P}$ has width $n$. In particular, we have that $|P| \leq \mathsf{height}(\mathbb{P}) \cdot \mathsf{width}(\mathbb{P})$. A further important theorem that connects the aforementioned notions is the following. According to Steiner (1993), for each partially ordered set $\mathbb{P} := (P, \leq)$, it holds true that

$$2^{\mathsf{width}(\mathbb{P})} + |P| - \mathsf{width}(\mathbb{P}) \leq |\mathsf{Ideals}(\mathbb{P})| \leq \left(\frac{|P| + \mathsf{width}(\mathbb{P})}{\mathsf{width}(\mathbb{P})}\right)^{\mathsf{width}(\mathbb{P})}.$$

Let $\mathbb{P} := (P, \leq)$ and $\mathbb{Q} := (Q, \sqsubseteq)$ be partially ordered sets. We call some mapping $f\colon P \to Q$ *order-preserving* if $x \leq y$ implies $f(x) \sqsubseteq f(y)$ for any elements $x, y \in P$. Furthermore, $f$ is *order-reflecting* if $f(x) \sqsubseteq f(y)$ implies $x \leq y$ for any elements $x, y \in P$. It is easy to verify that $f$ and $g$ are both order-reflecting, if $g \circ f = \mathrm{id}_P$ and $f \circ g = \mathrm{id}_Q$ holds true, and further both $f$ and $g$ are order-preserving. Another immediate corollary is that any order-reflecting mapping is injective. We call some such mapping $f\colon P \to Q$ an *order-isomorphism* from $\mathbb{P}$ to $\mathbb{Q}$, denoted as $f\colon \mathbb{P} \simeq \mathbb{Q}$, if it is bijective, order-preserving, and order-reflecting. We conclude that in order to prove that two partially ordered sets $\mathbb{P}$ and $\mathbb{Q}$ are isomorphic, it suffices to find two mutually inverse mappings between $P$ and $Q$ which are both order-preserving.

Let now $P_0^k := (\Sigma_k)_\mathsf{C}$ and inductively define the posets $\mathbb{P}_n^k := (P_n^k, \subseteq)$ as follows.

$$\begin{aligned} P_1^k &:= \wp(P_0^k) \\ P_{n+1}^k &:= \mathsf{Ideals}(\mathbb{P}_n^k) \quad \text{for any } n \in \mathbb{N}_+ \end{aligned}$$

Note that, if we set $\mathbb{P}_0^k := (P_0^k, \emptyset)$, then $P_1^k = \mathsf{Ideals}(\mathbb{P}_0^k)$ is satisfied as well. Furthermore, we define the following posets $\mathbb{E}_n^k := (E_n^k, \sqsupseteq_\emptyset)$.

$$\begin{aligned} E_1^k &:= \{ \textstyle\bigsqcap \mathbf{A} \mid \mathbf{A} \subseteq (\Sigma_k)_\mathsf{C} \} \\ E_{n+1}^k &:= \{ \textstyle\bigsqcap \{ \exists r.C \mid C \in \mathbf{C} \} \mid \mathbf{C} \subseteq E_n^k \} \quad \text{for any } n \in \mathbb{N}_+ \end{aligned}$$

Figure 4.2 displays the poset $\mathbb{P}_2^3$ and Figure 4.3 shows the poset $\mathbb{E}_2^3$; apparently, both are isomorphic to $\mathsf{FCD}(3)$, the free distributive lattice on three generators.

**Lemma 4.5.1.1.** *$\mathbb{P}_n^k$ and $\mathbb{E}_n^k/\emptyset$ are isomorphic for any $n \in \mathbb{N}_+$.*

*Approbatio.* To ease readability, we shall not distinguish between equivalence classes of $\mathbb{E}_n^k$ w.r.t. $\emptyset$ and their representatives. Furthermore, we are going to prove the claim by induction on $n$.

For the induction base let $n = 1$. It is readily verified that

$$\begin{aligned} \iota_1^k\colon \mathbb{P}_1^k &\simeq \mathbb{E}_1^k \\ \mathbf{A} &\mapsto \textstyle\bigsqcap \mathbf{A} \end{aligned}$$

is an isomorphism from $\mathbb{P}_1^k$ to $\mathbb{E}_1^k$, and has the following inverse isomorphism.

$$\begin{aligned} \kappa_1^k\colon \quad \mathbb{E}_1^k &\simeq \mathbb{P}_1^k \\ \textstyle\bigsqcap \mathbf{A} &\mapsto \mathbf{A} \end{aligned}$$

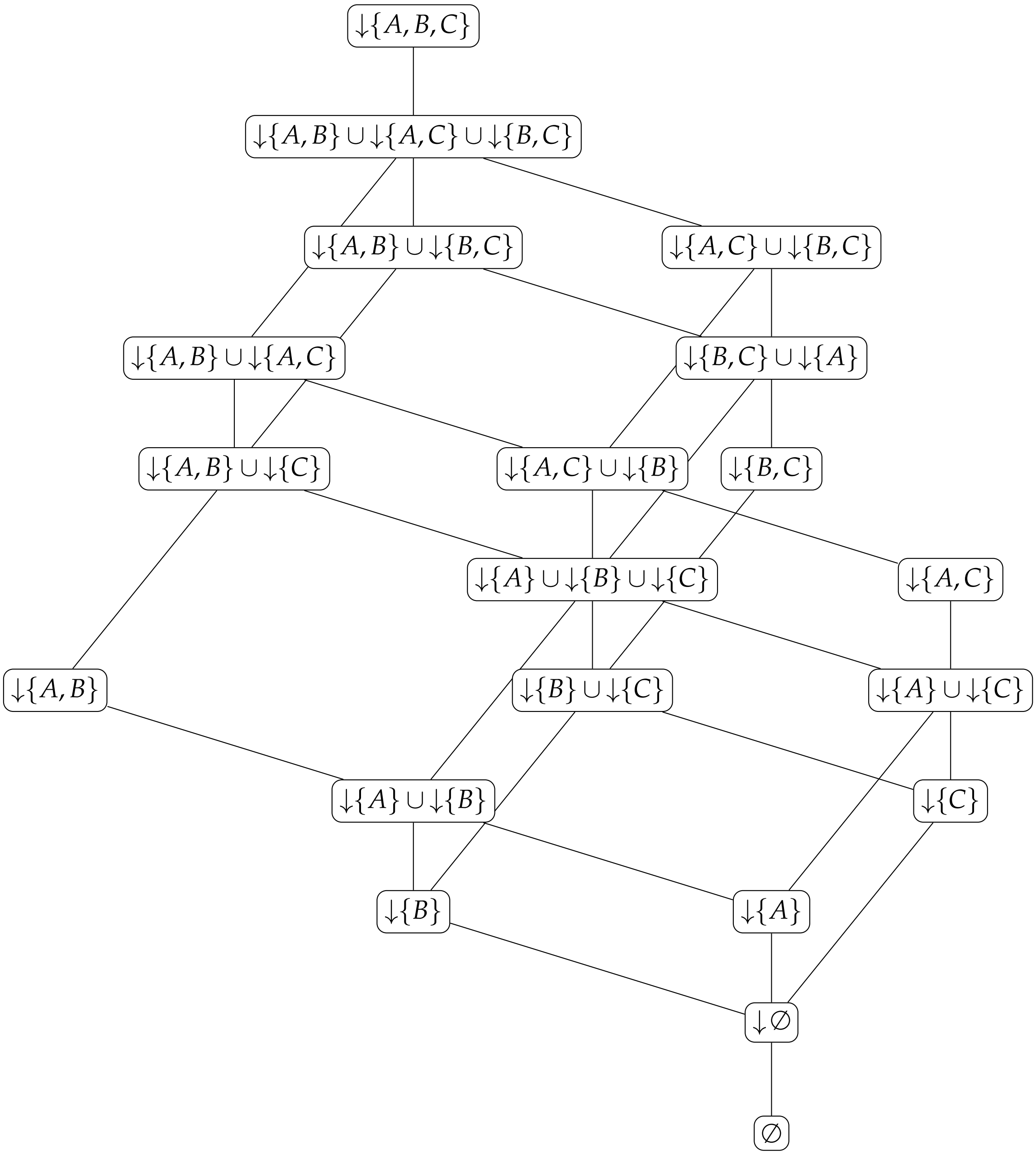

Figure 4.2.: The ordered set $\mathbb{P}_2^3$

Regarding the induction step assume that $n > 1$. We now show that

$$\begin{aligned} \iota_{n+1}^k \colon \qquad & \mathbb{P}_{n+1}^k \xrightarrow{\sim} \mathbb{E}_{n+1}^k \\ & \{p_1, \ldots, p_m\} \mapsto \exists\, r.\, \iota_n^k(p_1) \sqcap \cdots \sqcap \exists\, r.\, \iota_n^k(p_m) \end{aligned}$$

is an isomorphism from $\mathbb{P}_{n+1}^k$ to $\mathbb{E}_{n+1}^k$, and that its inverse isomorphism is as follows.

$$\begin{aligned} \kappa_{n+1}^k \colon \qquad & \mathbb{E}_{n+1}^k \xrightarrow{\sim} \mathbb{P}_{n+1}^k \\ & \exists\, r.\, C_1 \sqcap \cdots \sqcap \exists\, r.\, C_m \mapsto \downarrow\kappa_n^k(C_1) \cup \cdots \cup \downarrow\kappa_n^k(C_m) \end{aligned}$$

- Consider $\{p_1, \ldots, p_\ell\}$, $\{p_1, \ldots, p_m\} \in P_{n+1}^k$ where it holds true that $\ell \leq m$, that is, $\{p_1, \ldots, p_\ell\} \subseteq$ $\{p_1, \ldots, p_m\}$. In particular, both $\{p_1, \ldots, p_\ell\}$ and $\{p_1, \ldots, p_m\}$ are ideals in $\mathbb{P}_n^k$. It is obvious that

$$\iota_{n+1}^k(\{p_1, \ldots, p_\ell\}) \sqsupseteq_\emptyset \iota_{n+1}^k(\{p_1, \ldots, p_m\})$$

holds true.

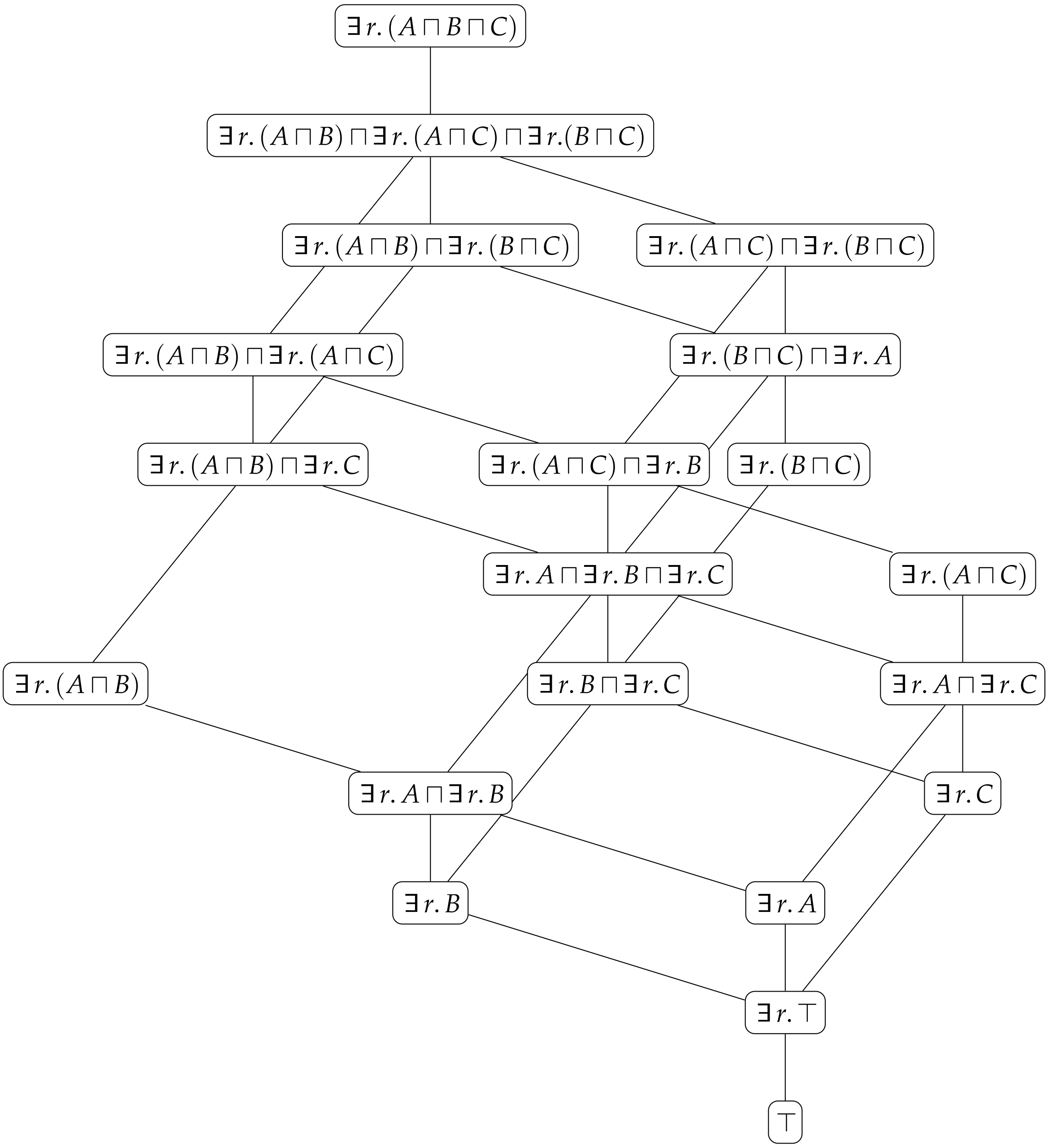

Figure 4.3.: The ordered set $\mathbb{E}_2^3$

- Furthermore, we have the following.

$$
\begin{aligned}
&(\kappa^k_{n+1} \circ \iota^k_{n+1})(\{p_1,\dots,p_m\})\\
&= \kappa^k_{n+1}(\exists\, r.\, \iota^k_n(p_1) \sqcap \dots \sqcap \exists\, r.\, \iota^k_n(p_m))\\
&= {\downarrow}\, \kappa^k_n(\iota^k_n(p_1)) \cup \dots \cup {\downarrow}\, \kappa^k_n(\iota^k_n(p_m))\\
&= {\downarrow}\, p_1 \cup \dots \cup {\downarrow}\, p_m\\
&= \{p_1,\dots,p_m\}
\end{aligned}
$$

  The penultimate equation follows from the induction hypothesis, while the last equation is true, since $\{p_1,\dots,p_m\}$ is already an ideal.

- Assume that $\exists\, r.\, C_1 \sqcap \dots \sqcap \exists\, r.\, C_\ell \sqsupseteq_\emptyset \exists\, r.\, D_1 \sqcap \dots \sqcap \exists\, r.\, D_m$. We shall now prove that ${\downarrow}\, \kappa^k_n(C_1) \cup \dots \cup {\downarrow}\, \kappa^k_n(C_\ell)$ is a subset of ${\downarrow}\, \kappa^k_n(D_1) \cup \dots \cup {\downarrow}\, \kappa^k_n(D_m)$. So, consider some element $p$ of the former, i.e., $p \subseteq \kappa^k_n(C_i)$ for some index $i \in \{1,\dots,\ell\}$, which is equivalent to $\iota^k_n(p) \sqsupseteq_\emptyset C_i$. We now have that there exists some index $j \in \{1,\dots,m\}$ such that $C_i \sqsupseteq_\emptyset D_j$, and we infer that $p \subseteq \kappa^k_n(D_j)$.

- Eventually, we prove that $\iota^k_{n+1} \circ \kappa^k_{n+1} = \mathrm{id}$.

$$
\begin{aligned}
&(\iota^k_{n+1} \circ \kappa^k_{n+1})(\exists\, r.\, C_1 \sqcap \dots \sqcap \exists\, r.\, C_m)\\
=\ & \iota^k_{n+1}({\downarrow}\, \kappa^k_n(C_1) \cup \dots \cup {\downarrow}\, \kappa^k_n(C_m))\\
=\ & \bigsqcap \{\, \exists\, r.\, \iota^k_n(p) \mid p \in {\downarrow}\, \kappa^k_n(C_1) \cup \dots \cup {\downarrow}\, \kappa^k_n(C_m) \,\}\\
\equiv_\emptyset\ & \bigsqcap \{\, \exists\, r.\, \iota^k_n(p) \mid p \in \{\kappa^k_n(C_1),\dots,\kappa^k_n(C_m)\} \,\}\\
=\ & \exists\, r.\, \iota^k_n(\kappa^k_n(C_1)) \sqcap \dots \sqcap \exists\, r.\, \iota^k_n(\kappa^k_n(C_m))\\
\equiv_\emptyset\ & \exists\, r.\, C_1 \sqcap \dots \sqcap \exists\, r.\, C_m
\end{aligned}
$$

  The first equivalence follows from the fact that $p \subseteq q$ implies $\iota^k_n(p) \sqsupseteq_\emptyset \iota^k_n(q)$ and, consequently, $\exists\, r.\, \iota^k_n(p) \sqsupseteq_\emptyset \exists\, r.\, \iota^k_n(q)$. The second equivalence is an immediate consequence of our induction hypothesis. □

**Lemma 4.5.1.2.** *$|\exists\, r^n.\, \bigsqcap(\Sigma_k)_\mathsf{C}| \geq \mathrm{height}(\mathbb{P}^k_{n+1})$ holds true for each $n \in \mathbb{N}$.*

*Approbatio.* Fix some $n \in \mathbb{N}$. In the previous lemma we have shown that $\mathbb{P}^k_{n+1}$ and $\mathbb{E}^k_{n+1}$ are isomorphic, and so it follows that $\mathrm{height}(\mathbb{P}^k_{n+1}) = \mathrm{height}(\mathbb{E}^k_{n+1})$. Apparently, $\emptyset$ is the smallest element of $\mathbb{P}^k_{n+1}$ and $P^k_n$ is the greatest element of $\mathbb{P}^k_{n+1}$. Since $\iota^k_{n+1}$ is an order-isomorphism, it follows that (the equivalence class with representative) $\iota^k_{n+1}(\emptyset)$ is the smallest element of $\mathbb{E}^k_{n+1}$ and that (the equivalence class with representative) $\iota^k_{n+1}(P^k_n)$ is the greatest element of $\mathbb{E}^k_{n+1}$. It obviously holds true that $\iota^k_{n+1}(\emptyset) \equiv_\emptyset \top$. Furthermore, we show by induction on $n$ that $\iota^k_{n+1}(P^k_n) \equiv_\emptyset \exists\, r^n.\, \bigsqcap(\Sigma_k)_\mathsf{C}$ is satisfied. If $n = 0$, then $\iota^k_1(P^k_0) = \iota^k_1((\Sigma_k)_\mathsf{C}) = \bigsqcap(\Sigma_k)_\mathsf{C} = \exists\, r^0.\, \bigsqcap(\Sigma_k)_\mathsf{C}$ holds true. Now let $n > 0$. Then we have the following.

$$
\begin{aligned}
&\iota^k_{n+2}(P^k_{n+1})\\
&= \bigsqcap \{\, \exists\, r.\, \iota^k_{n+1}(p) \mid p \in P^k_{n+1} \,\}\\
&= \exists\, r.\, \iota^k_{n+1}(P^k_n)\\
&= \exists\, r.\, \exists\, r^n.\, \bigsqcap(\Sigma_k)_\mathsf{C}\\
&= \exists\, r^{n+1}.\, \bigsqcap(\Sigma_k)_\mathsf{C}
\end{aligned}
$$

The second equality follows from the fact that $P_n^k$ is the greatest element within $P_{n+1}^k$, which means that $\iota_{n+1}^k(p) \sqsupseteq_\emptyset \iota_{n+1}^k(P_n^k)$ is satisfied for any $p \in P_{n+1}^k$. The penultimate equation is a consequence of the induction hypothesis. Eventually, we conclude that the rank of $\exists\, r^n. \bigsqcap(\Sigma_k)_C$ must be greater than or equal to the height of $\mathbb{P}_{n+1}^k$. □

The next lemma shows that each antichain $A$ in $\mathbb{P}_n^k$ induces an antichain in $\mathbb{P}_{n+1}^k$ the cardinality of which is exponential in $|A|$. In particular, this implies that the width of $\mathbb{P}_{n+1}^k$ is exponential in the width of $\mathbb{P}_n^k$. Since the width of $\mathbb{P}_1^k$, the power-set lattice of $\{A_1, \ldots, A_k\}$, obviously is asymptotically bounded above and below by $2^k$, it now follows by induction that the width of $\mathbb{P}_n^k$ is asymptotically bounded above and below by $(2, k) \uparrow\uparrow n$, that is,

$$\mathsf{width}(\mathbb{P}_n^k) \asymp (2, k) \uparrow\uparrow n.$$

**Lemma 4.5.1.3.** *Let $n > 0$ and consider some antichain $A$ in $\mathbb{P}_n^k$ such that $|A| = 2 \cdot \ell$. Then, the following $A'$ is an antichain in $\mathbb{P}_{n+1}^k$.*

$$A' := \{\, \downarrow a_1 \cup \cdots \cup \downarrow a_\ell \mid \{a_1, \ldots, a_\ell\} \in \tbinom{A}{\ell} \,\}$$

*Approbatio.* Consider two mutually distinct $\{a_1, \ldots, a_\ell\}$ and $\{b_1, \ldots, b_\ell\}$ in $\binom{A}{\ell}$. It is readily verified that $\downarrow a_1 \cup \cdots \cup \downarrow a_\ell$ and $\downarrow b_1 \cup \cdots \cup \downarrow b_\ell$ are elements of $\mathbb{P}_{n+1}^k$, and we shall now show that these are incomparable with respect to $\subseteq$.

From the assumption $\{a_1, \ldots, a_\ell\} \neq \{b_1, \ldots, b_\ell\}$ it follows that, without loss of generality, $a_1 \notin \{b_1, \ldots, b_\ell\}$, that is, $a_1 \neq b_i$ for any index $i \in \{1, \ldots, \ell\}$. Now the precondition that $A$ is an antichain yields that $a_1 \not\subseteq b_i$ for each index $i \in \{1, \ldots, \ell\}$, and we infer that $a_1 \notin \downarrow b_i$ for each $i$, which means that

$$a_1 \notin \downarrow b_1 \cup \cdots \cup \downarrow b_\ell.$$

Furthermore, it is trivial that $a_1 \in \downarrow a_1$, and consequently we have that

$$a_1 \in \downarrow a_1 \cup \cdots \cup \downarrow a_\ell.$$

We conclude that $\downarrow a_1 \cup \cdots \cup \downarrow a_\ell \not\subseteq \downarrow b_1 \cup \cdots \cup \downarrow b_\ell$. The converse direction follows analogously, and so we have that $\downarrow a_1 \cup \cdots \cup \downarrow a_\ell$ and $\downarrow b_1 \cup \cdots \cup \downarrow b_\ell$ are indeed not comparable. □

**Propositio 4.5.1.4.** $\mathsf{width}(\mathbb{P}_n^k)$ *is asympotically bounded above and below by $(2, k) \uparrow\uparrow n$ for any $n \in \mathbb{N}_+$.*

*Approbatio.* We prove the statement by induction on $n$. For the induction base where $n = 1$, Sperner's theorem, cf. (Sperner, 1928), yields that

$$\mathsf{width}(\mathbb{P}_1^k) = \tbinom{k}{\lfloor \frac{k}{2} \rfloor}.$$

Furthermore, for the *central binomial coefficients* it is well-known that $\binom{2 \cdot k}{k} \sim \frac{4^k}{\sqrt{\pi \cdot k}}$ is satisfied. We infer that

$$\mathsf{width}(\mathbb{P}_1^k) \sim \frac{2^k}{\sqrt{\frac{\pi}{2} \cdot k}}$$

or more simplified that $\mathsf{width}(\mathbb{P}_1^k) \asymp 2^k$.

Now for the induction step let $n > 1$. The induction hypothesis states that there exists some antichain in $\mathbb{P}_n^k$ that has $n$-exponential cardinality in $k$. An application of Lemma 4.5.1.3 yields an antichain in $\mathbb{P}^k_{n+1}$ that has $(n+1)$-exponential cardinality in $k$, which implies that $\mathrm{width}(\mathbb{P}^k_{n+1})$ is at least $(n+1)$-exponential in $k$. Since $|P^k_{n+1}| \leq (2,k) \uparrow\uparrow (n+1)$ holds true, we conclude that $\mathrm{width}(\mathbb{P}^k_{n+1})$ is at most $(n+1)$-exponential in $k$ as well. □

**Lemma 4.5.1.5.** $\mathrm{height}(\mathbb{P}^k_{n+1}) \geq \mathrm{width}(\mathbb{P}^k_n)$ *for each* $n \in \mathbb{N}_+$.

*Approbatio.* We show that any antichain $A$ in $\mathbb{P}^k_n$ induces a chain $C$ in $\mathbb{P}^k_{n+1}$ such that $|A| = |C|$, which obviously implies our claim. Thus, consider some such antichain $A = \{a_1, \ldots, a_\ell\}$ in $\mathbb{P}^k_n$. We define $C$ as follows.

$$C := \{c_1, \ldots, c_\ell\} \quad \text{where} \quad c_i := {\downarrow} a_1 \cup \cdots \cup {\downarrow} a_i$$

It is apparent that $C$ consists of ideals of $\mathbb{P}^k_n$, that is, $C \subseteq P^k_{n+1}$ is satisfied. Furthermore, we can readily verify that any two elements in $C$ are comparable with respect to $\subseteq$. We conclude that $C$ is a chain. It remains to prove that $|A| = |C|$. Of course, $|C| \leq |A|$ follows from the very definition of $C$. We show the converse inequality by demonstrating that no two elements of $C$ are equal. Let $1 \leq i_1 < i_2 \leq \ell$. Of course, it then holds true that $c_{i_1} \subseteq c_{i_2}$. Now consider some $a_j$ where $j \in \{i_1+1, \ldots, i_2\}$. Since $A$ is an antichain, it follows that $a_j$ is $\subseteq$-incomparable to each $a_h$ for $h \in \{1, \ldots, i_1\}$, which implies that $a_j \notin {\downarrow} a_h$ although $a_j \in {\downarrow} a_j$. Consequently, the set inclusion is strict. □

**Corollarium 4.5.1.6.** $\mathrm{height}(\mathbb{P}^k_{n+1})$ *is asympotically bounded above and below by* $(2,k) \uparrow\uparrow n$ *for any* $n \in \mathbb{N}_+$*, that is,*

$$\mathrm{height}(\mathbb{P}^k_{n+1}) \asymp (2,k) \uparrow\uparrow n. \qquad \square$$

**Corollarium 4.5.1.7.** $|\exists\, r^n. \bigsqcap(\Sigma_k)_{\mathsf{C}}|$ *is asympotically bounded above and below by* $(2,k) \uparrow\uparrow n$ *for any* $n \in \mathbb{N}_+$*, that is,*

$$|\exists\, r^n. \bigsqcap(\Sigma_k)_{\mathsf{C}}| \asymp (2,k) \uparrow\uparrow n. \qquad \square$$

We close this section with a justification that the precondition $k \geq 3$ is crucial for the non-elementary growth of the ranks of the concept descriptions $\exists\, r^n. \bigsqcap(\Sigma_k)_{\mathsf{C}}$. The proof of this asymptotic behaviour heavily relies on the fact that the width of $\mathbb{P}^k_{n+1}$ is exponential in $\mathbb{P}^k_n$. However, this does not hold true in case $k < 3$. To see this, reconsider the proof of Lemma 4.5.1.3. For $k < 3$ any antichain in $\mathbb{P}^k_1$ has at most two elements. Since for the central binomial coefficient it holds true that $\binom{2}{1} = 2$, we can only infer that there must exist some antichain in $\mathbb{P}^k_2$ with two elements and so on and so forth. In fact, we can easily verify that $\mathrm{width}(\mathbb{P}^k_n) \leq 2$ is always satisfied. In particular, we have the following.

- Each $\mathbb{P}^0_n$ is a chain of height $n$.
- Each $\mathbb{P}^1_n$ is a chain of height $n+1$.
- Each $\mathbb{P}^2_n$ consists of two chains of height $n$ that are connected by two incomparable elements, i.e., these are of the following form.

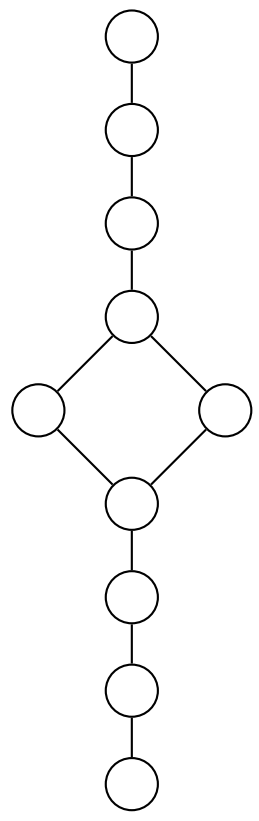

In fact, the upcoming Lemma 4.5.1.9 shows that the ranks of the concept descriptions $\exists\, r^n. \top$, $\exists\, r^n. A$, and $\exists\, r^n. (A \sqcap B)$ are all linear in $n$.

**Lemma 4.5.1.8.** *For each $n \in \mathbb{N}$ and each reduced concept description C, it holds true that*

$$\mathsf{Upper}(\exists\, r^{n+1}. C) = \{\exists\, r^n. \bigsqcap\{\exists\, r. D \mid C \prec_\emptyset D\}\}.$$

*Approbatio.* We show the claim by induction over $n$. The base case where $n = 0$ has been proven in Propositio 3.1.2.1. For the inductive case let $n > 0$. According to Propositio 3.1.2.1 and the induction hypothesis, it holds true that

$$\begin{aligned} \mathsf{Upper}(\exists\, r^{n+2}. C) &= \{\exists\, r. \exists\, r^{n+1}. C\} \\ &= \{\exists\, r. E \mid E \in \mathsf{Upper}(\exists\, r^{n+1}. C)\} \\ &= \{\exists\, r. \exists\, r^n. \bigsqcap\{\exists\, r. D \mid C \prec_\emptyset D\}\} \\ &= \{\exists\, r^{n+1}. \bigsqcap\{\exists\, r. D \mid C \prec_\emptyset D\}\}. \end{aligned}$$ □

**Lemma 4.5.1.9.** *Let $n \in \mathbb{N}$. Then the following equalities hold true.*

1. $|\exists\, r^n. \top| = n$
2. $|\exists\, r^n. A| = n + 1$
3. $|\exists\, r^n. (A \sqcap B)| = 2 \cdot (n + 1)$

*Approbatio.*

1. We show the claim by induction on $n$. If $n = 0$, then $\exists\, r^n. \top = \top$ and we can thus immediately conclude that $|\exists\, r^n. \top| = 0$. Let $n > 0$. Since it holds true that $\exists\, r^{n+1}. C \prec_\emptyset \exists\, r^n. \bigsqcap\{\exists\, r. D \mid C \prec_\emptyset D\}$, we infer that $|\exists\, r^{n+1}. \top| = 1 + |\exists\, r^n. \top| = 1 + n$.
2. We know that $\exists\, r^n. A \prec_\emptyset \exists\, r^n. \top$, and so we infer that $|\exists\, r^n. A| = 1 + |\exists\, r^n. \top| = 1 + n$.

| $k \backslash n$ | 0 | 1 | 2 | 3 | 4 | 5 | 6 |
|---|---|---|---|---|---|---|---|
| 0 | 0 | 1 | 2 | 3 | 4 | 5 | 6 |
| 1 | 1 | 2 | 3 | 4 | 5 | 6 | 7 |
| 2 | 2 | 4 | 6 | 8 | 10 | 12 | 14 |
| 3 | 3 | $8 \geq \binom{3}{1}$ | $20 \geq 4$ | $84 \geq \binom{4}{2}$ | $8573 \geq \binom{6}{3}$ | $? \geq \binom{20}{10}$ | $? \geq \binom{184756}{92378}$ |
| 4 | 4 | $16 \geq \binom{4}{2}$ | $168 \geq \binom{6}{3}$ | $? \geq \binom{20}{10}$ | $? \geq \binom{184756}{92378}$ | $? \gtrapprox \binom{2.33 \cdot 10^{55614}}{1.16 \cdot 10^{55614}}$ | ? |
| 5 | 5 | $32 \geq \binom{5}{2}$ | $7581 \geq \binom{10}{5}$ | $? \geq \binom{252}{126}$ | $? \gtrapprox \binom{3.63 \cdot 10^{74}}{1.82 \cdot 10^{74}}$ | ? | ? |
| 6 | 6 | $64 \geq \binom{6}{3}$ | $? \geq \binom{20}{10}$ | $? \geq \binom{184756}{92378}$ | $? \gtrapprox \binom{2.33 \cdot 10^{55614}}{1.16 \cdot 10^{55614}}$ | ? | ? |
| 7 | 7 | $128 \geq \binom{7}{3}$ | $? \geq \binom{35}{17}$ | $? \gtrapprox \binom{4.54 \cdot 10^{9}}{2.27 \cdot 10^{9}}$ | ? | ? | ? |
| 8 | 8 | $256 \geq \binom{8}{4}$ | $? \geq \binom{70}{35}$ | $? \gtrapprox \binom{1.12 \cdot 10^{20}}{5.61 \cdot 10^{19}}$ | ? | ? | ? |
| 9 | 9 | $512 \geq \binom{9}{4}$ | $? \geq \binom{126}{63}$ | $? \gtrapprox \binom{6.03 \cdot 10^{36}}{3.02 \cdot 10^{36}}$ | ? | ? | ? |
| 10 | 10 | $1024 \geq \binom{10}{5}$ | $? \geq \binom{252}{126}$ | $? \gtrapprox \binom{3.63 \cdot 10^{74}}{1.82 \cdot 10^{74}}$ | ? | ? | ? |

Table 4.1.: Some ranks of $\exists\, r^n.\, (A_1 \sqcap \cdots \sqcap A_k)$ and corresponding lower bounds of widths of $\mathbb{P}_n^k$.

3. We first observe that the following neighboring subsumptions hold true.

$$\begin{aligned}
\exists\, r^n.\, (A \sqcap B) &\prec_\emptyset \exists\, r^{n-1}.\, (\exists\, r.\, A \sqcap \exists\, r.\, B) \\
&\prec_\emptyset \exists\, r^{n-2}.\, (\exists\, r^2.\, A \sqcap \exists\, r^2.\, B) \\
&\prec_\emptyset \dots \\
&\prec_\emptyset \exists\, r^{n-j}.\, (\exists\, r^j.\, A \sqcap \exists\, r^j.\, B) \\
&\prec_\emptyset \dots \\
&\prec_\emptyset \exists\, r^n.\, A \sqcap \exists\, r^n.\, B \\
&\prec_\emptyset \exists\, r^n.\, A \sqcap \exists\, r^n.\, \top \\
&\equiv_\emptyset \exists\, r^n.\, A
\end{aligned}$$

We conclude that $|\exists\, r^n.\, (A \sqcap B)| = (n+1) + |\exists\, r^n.\, A| = 2 \cdot (n+1)$. □

Of course, for the border case where $k = 3$ Lemma 4.5.1.3 does not immediately show a start of the non-elementary growth either. This is due to the fact that then $\mathbb{P}_1^3$ has width 3, and the central binomial coefficient $\binom{3}{1} = \binom{3}{2}$ evaluates to 3, i.e., an application of Lemma 4.5.1.3 does not induce a bigger antichain in $\mathbb{P}_2^3$. However, we have seen in Figure 4.2 that $\mathbb{P}_2^3$ has an antichain of cardinality 4. Now the sequence $(\, x_n \mid n \in \mathbb{N} \text{ and } n \geq 2\,)$ where $x_2 := 4$ and $x_{n+1} := \binom{x_n}{\lfloor \frac{x_n}{2} \rfloor}$ grows non-elementarily and each $\mathbb{P}_n^3$ contains an antichain of cardinality $x_n$.

| $n$ | 2 | 3 | 4 | 5 | 6 | 7 | 8 |
|---|---|---|---|---|---|---|---|
| $x_n$ | 4 | $\binom{4}{2} = 6$ | $\binom{6}{3} = 20$ | $\binom{20}{10} = 184756$ | $\binom{184756}{92378} \approx 2.33 \cdot 10^{55614}$ | $\approx \binom{2.33 \cdot 10^{55614}}{1.16 \cdot 10^{55614}}$ | ? |

Eventually, we have run some experiments in which we tried to compute ranks of concept descriptions of the form $\exists\, r^n.\, (A_1 \sqcap \cdots \sqcap A_k)$. The result are listed in Table 4.1. As expected, this only works for sufficiently small values of $n$ and $k$ and henceforth we have provided lower bounds, namely lower bounds of the widths of $\mathbb{P}_n^k$, for these ranks if possible. Please note the anomaly in Table 4.1 for the case where $k = 3$ and $n = 2$ as explained above.

### 4.5.2. DECISION PROBLEMS RELATED TO THE RANK FUNCTION

Given a concept description $C$ and a natural number $n$ (in binary encoding), then we can decide in triple exponential time whether the rank of $C$ is equal to $n$, at most $n$, or at least $n$. A procedure can construct a chain of $n$ neighbors and then check whether $\top$ is reached. If $n$ is fixed, then this requires only deterministic polynomial time.

There are three decision problems tightly related with the rank function on $\mathcal{EL}$ concept descriptions: for some given $\mathcal{EL}$ concept description $C$ and a number $n \in \mathbb{N}$, the first one asks if the rank of $C$ equals $n$, the second one asks whether $n$ is an upper bound for $|C|$, and the third one asks if $|C| \geq n$. In particular, we define these three decision problems as follows.

$$\begin{aligned}
\mathbf{P}_{\mathcal{EL}\text{-Rank}} &:= \{ (C,n) \mid C \in \mathcal{EL}(\Sigma),\ n \in \mathbb{N},\ \text{and } |C| = n \} \\
\mathbf{P}_{\mathcal{EL}\text{-Rank-Upper-Bound}} &:= \{ (C,n) \mid C \in \mathcal{EL}(\Sigma),\ n \in \mathbb{N},\ \text{and } |C| \leq n \} \\
\mathbf{P}_{\mathcal{EL}\text{-Rank-Lower-Bound}} &:= \{ (C,n) \mid C \in \mathcal{EL}(\Sigma),\ n \in \mathbb{N},\ \text{and } |C| \geq n \}
\end{aligned}$$

In the following, we shall investigate the relationships between these problems in terms of reducibility and we shall provide bounds for their complexities.

**Lemma 4.5.2.1.** *The following Turing reductions exist.*

1. $\mathbf{P}_{\mathcal{EL}\text{-Rank}} \leq_T^{\mathbf{P}} \mathbf{P}_{\mathcal{EL}\text{-Rank-Upper-Bound}}$
2. $\mathbf{P}_{\mathcal{EL}\text{-Rank}} \leq_T^{\mathbf{P}} \mathbf{P}_{\mathcal{EL}\text{-Rank-Lower-Bound}}$
3. $\mathbf{P}_{\mathcal{EL}\text{-Upper-Bound-Rank}} \leq_T^{\mathbf{P}} \mathbf{P}_{\mathcal{EL}\text{-Rank}}$ *if numbers are unarily encoded.*
4. $\mathbf{P}_{\mathcal{EL}\text{-Upper-Bound-Rank}} \leq_T \mathbf{P}_{\mathcal{EL}\text{-Rank}}$

*Approbatio.*

1. It is easy to see that $(C,n) \in \mathbf{P}_{\mathcal{EL}\text{-Rank}}$ holds true if, and only if, $(C,n) \in \mathbf{P}_{\mathcal{EL}\text{-Rank-Upper-Bound}}$ as well as $(C,n-1) \notin \mathbf{P}_{\mathcal{EL}\text{-Rank-Upper-Bound}}$ are satisfied. This shows that in order to construct some Turing machine that decides $\mathbf{P}_{\mathcal{EL}\text{-Rank}}$, we could query an oracle Turing machine for $\mathbf{P}_{\mathcal{EL}\text{-Rank-Upper-Bound}}$ twice. Indeed, this Turing machine runs in polynomial time for the following reason. If $n$ is unarily encoded, then its predecessor $n-1$ can be computed in constant time. Otherwise if $n$ is efficiently encoded, i.e. without loss of generality, if $n$ is binarily encoded, then its predecessor $n-1$ can be computed in linear time (with respect to the size of $n$, i.e., the length of an encoding of $n$) as follows.
   - Find the lowest bit in $n$ that is $1$. (In particular, all lower bits in $n$ are then $0$.)
   - Flip this bit and all lower ones.
2. Analogously as for Statement 1, since $(C,n) \in \mathbf{P}_{\mathcal{EL}\text{-Rank}}$ holds true if, and only if, $(C,n) \in \mathbf{P}_{\mathcal{EL}\text{-Rank-Lower-Bound}}$ as well as $(C,n+1) \notin \mathbf{P}_{\mathcal{EL}\text{-Rank-Lower-Bound}}$ are satisfied.
3. It is apparent that $(C,n) \in \mathbf{P}_{\mathcal{EL}\text{-Upper-Bound-Rank}}$ is satisfied if, and only if, $(C,m) \in \mathbf{P}_{\mathcal{EL}\text{-Rank}}$ holds true for some $m \leq n$. It follows that we can construct a Turing machine deciding $\mathbf{P}_{\mathcal{EL}\text{-Upper-Bound-Rank}}$ which uses an oracle for $\mathbf{P}_{\mathcal{EL}\text{-Rank}}$. Obviously, if $n$ is unarily encoded, then the number of queries to an oracle for $\mathbf{P}_{\mathcal{EL}\text{-Rank}}$ is linear in $n$.

4. In case of efficient encodings, i.e., if $n$ is binarily encoded, then it might be the case that we need to pose an exponential number of queries to the oracle for $\mathbf{P}_{\mathcal{EL}\text{-RANK}}$, which implies that the oracle Turing machine constructed for Statement 3 now has an exponential time complexity. Thus, we can now only infer that there must exist some Turing reduction, but not necessarily a polynomial time Turing reduction. □

**Lemma 4.5.2.2.** *The following statements are satisfied for each complexity class* **C**.

1. $\mathbf{P}_{\mathcal{EL}\text{-RANK-UPPER-BOUND}} \in \mathbf{C}$ *implies* $\mathbf{P}_{\mathcal{EL}\text{-RANK}} \in \mathbf{P}^{\mathbf{C}}$
2. $\mathbf{P}_{\mathcal{EL}\text{-RANK-LOWER-BOUND}} \in \mathbf{C}$ *implies* $\mathbf{P}_{\mathcal{EL}\text{-RANK}} \in \mathbf{P}^{\mathbf{C}}$
3. *If* $\mathbf{P}_{\mathcal{EL}\text{-RANK}} \in \mathbf{C}$ *and numbers are unarily encoded, then* $\mathbf{P}_{\mathcal{EL}\text{-RANK-UPPER-BOUND}} \in \mathbf{P}^{\mathbf{C}}$.
4. *If* $\mathbf{P}_{\mathcal{EL}\text{-RANK}} \in \mathbf{C}$, *then* $\mathbf{P}_{\mathcal{EL}\text{-RANK-UPPER-BOUND}} \in \mathbf{PSpace}^{\mathbf{C}}$.

*Approbatio.* 1. Lemma 4.5.2.1 shows that $\mathbf{P}_{\mathcal{EL}\text{-RANK-UPPER-BOUND}} \in \mathbf{C}$ implies $\mathbf{P}_{\mathcal{EL}\text{-RANK}} \in \mathbf{P}^{\mathbf{C}}$.

2. The proof is very similar as for Statement 1.
3. We already know from Lemma 4.5.2.1 that $\mathbf{P}_{\mathcal{EL}\text{-RANK-UPPER-BOUND}} \in \mathbf{P}^{\mathbf{C}}$ if $\mathbf{P}_{\mathcal{EL}\text{-RANK}} \in \mathbf{C}$ and numbers are unarily encoded.
4. Again, we use arguments from Lemma 4.5.2.1 and its proof. It is readily verified that the size of $m$ is bounded by the size of $n$ whenever $m \leq n$ holds true. Thus, we only need polynomial space to successively enumerate all such numbers $m$ that are smaller than $n$. We conclude that $\mathbf{P}_{\mathcal{EL}\text{-RANK-UPPER-BOUND}} \in \mathbf{PSpace}^{\mathbf{C}}$ if $\mathbf{P}_{\mathcal{EL}\text{-RANK}} \in \mathbf{C}$ is satisfied. □

**Lemma 4.5.2.3.** *1. If numbers are unarily encoded, then* $\mathbf{P}_{\mathcal{EL}\text{-RANK-UPPER-BOUND}} \in \mathbf{2EXP}$, *otherwise* $\mathbf{P}_{\mathcal{EL}\text{-RANK-UPPER-BOUND}} \in \mathbf{3EXP}$.

2. *If $n$ is fixed, then we can decide whether* $(C, n) \in \mathbf{P}_{\mathcal{EL}\text{-RANK-UPPER-BOUND}}$ *in deterministic polynomial time w.r.t.* $||C||$.
3. *The same upper complexity bounds hold true for* $\mathbf{P}_{\mathcal{EL}\text{-RANK-LOWER-BOUND}}$ *and* $\mathbf{P}_{\mathcal{EL}\text{-RANK}}$.

*Approbatio.* 1. In order to decide whether $(C, n) \in \mathbf{P}_{\mathcal{EL}\text{-RANK-UPPER-BOUND}}$, we can use the following procedure.

   a) Set $D := C$ and $i := 0$.
   b) While $i < n$, replace $D$ with an upper neighbor of $D$ if it exists, and increment $i$.
   c) If $D \equiv_\emptyset \top$, then accept $(C, n)$, otherwise reject.

   Each computation of an upper neighbor of $D$ requires quadratic time w.r.t. $||D||$, and the size of such an upper neighbor is also quadratic in $||D||$. Since the loop is executed $n$ times, we conclude that it needs time in $\mathcal{O}(||C||^2 + (||C||^2)^2 + \ldots + ||C||^{2^n})$. If $n$ is unarily encoded, this is obviously double exponential. Otherwise if $n$ is binarily encoded, then the length of the encoding of $n$ is $\log_2(n)$, and thus we infer that the procedure needs triple exponential time w.r.t. the size of the encoding of $(C, n)$.

2. In case of $n$ being fixed, the above procedure is apparently polynomial in $||C||$ with exponent $2 + 2^2 + \cdots + 2^{2^n}$.

3. There are obvious variations of the above algorithm with similar complexities for the other decision problems $\mathbf{P}_{\mathcal{EL}\text{-Rank-Lower-Bound}}$ and $\mathbf{P}_{\mathcal{EL}\text{-Rank}}$. □

# 5. CONCLUSION

We have investigated the *neighborhood problem* for the description logic $\mathcal{EL}$ and some of its variants. We found that existence of neighbors can in general only be guaranteed for the case of $\mathcal{EL}$ without a TBox, without the bottom concept description, and without greatest fixed-point semantics. The presence of a TBox, the bottom concept description, or greatest fixpoint semantics allow for the construction of concept descriptions that do not have neighbors in certain directions. For the case of $\mathcal{EL}$ we proposed sound and complete procedures for deciding neighborhood as well as for computing all upper neighbors and all lower neighbors, respectively. Furthermore, we have shown that deciding neighborhood and computing all upper neighbors requires only deterministic polynomial time. More specifically, the number of (equivalence classes of) upper neighbors is linear, and any upper neighbor has a quadratic size. All lower neighbors of an $\mathcal{EL}$ concept description can be enumerated in deterministic exponential time—in particular, the number of (equivalence classes of) lower neighbors is always exponential, and each lower neighbor has a quadratic size. There is a non-deterministic polynomial time procedure such that, for any lower neighbor $L$ of the input, it has one (successful) computation path that returns a concept description equivalent to $L$.

As further results, we have proven that the lattice of $\mathcal{EL}$ concept descriptions is distributive, modular, graded, and metric. In particular, this means that there exists a rank function as well as a distance function on this lattice. Unfortunately, the rank function shows a non-elementary growth, which directly prohibits its practical usage. We have seen that even a supposedly simple concept description can have a multi-exponential rank: the rank of $\exists\, r^n. (A_1 \sqcap \cdots \sqcap \ldots A_k)$ is asympotically bounded above and below by $(2, k) \uparrow\uparrow n$. Furthermore, some first complexity results on decision problems related to these rank and distance functions were found. However, the exact complexities are currently unknown; we do not know whether the presented upper bounds are sharp, and lower bounds are also not available.

As an important consequence we infer that the algorithm *NextClosures* (Kriegel, 2016b) can be utilized for enumerating canonical bases of closure operators in $\mathcal{EL}$, cf. (Kriegel, 2018a, Section 8).

Other possible future research could consider extensions to more expressive description logics. Of course, these logics should be considered without any TBox or with cycle-restricted TBoxes for deciding existence of neighbors in general. Eventually, a further direction for future research is a more fine-grained characterization of existence of neighbors. That is, given a description logic where neighbors need not exist in general, how can we decide whether a concept description has neighbors and how can we enumerate these?

## ACKNOWLEDGEMENTS

The author gratefully thanks both Franz Baader and Bernhard Ganter for fruitful discussions on neighborhood generated orders and further thanks Franz Baader again for proof-reading as well as for providing comments that helped to improve the quality of this document. Furthermore, the author thanks the anonymous reviewers for their constructive hints and helpful remarks on those parts of this report that have already been published in (Kriegel, 2018b).

# A. APPENDIX

## A.1. ASYMPTOTIC NOTATIONS

We use the following notions the origins of which have been described by Knuth (1976). Let $g\colon \mathbb{N} \to \mathbb{N}$ be a function. Then, the sets $\mathcal{O}(g)$, $\Omega(g)$, and $\Theta(g)$ are defined as follows.

$$\mathcal{O}(g) := \{\, f \mid f\colon \mathbb{N} \to \mathbb{N} \text{ and } \exists\, c \in \mathbb{R}_+ \;\exists\, n_0 \in \mathbb{N} \;\forall\, n \geq n_0\colon f(n) \leq c \cdot g(n) \,\}$$
$$\Omega(g) := \{\, f \mid f\colon \mathbb{N} \to \mathbb{N} \text{ and } \exists\, c \in \mathbb{R}_+ \;\exists\, n_0 \in \mathbb{N} \;\forall\, n \geq n_0\colon f(n) \geq c \cdot g(n) \,\}$$
$$\Theta(g) := \{\, f \mid f\colon \mathbb{N} \to \mathbb{N} \text{ and } \exists\, c, d \in \mathbb{R}_+ \;\exists\, n_0 \in \mathbb{N} \;\forall\, n \geq n_0\colon c \cdot g(n) \leq f(n) \leq d \cdot g(n) \,\}$$

Obviously, we have that $g \in \mathcal{O}(f)$ is equivalent to $f \in \Omega(g)$, and further that $\Theta(f) = \mathcal{O}(f) \cap \Omega(f)$ holds true. It is not hard to verify that $f \in \mathcal{O}(g)$ is equivalent to $\limsup_{n\to\infty} \frac{f(n)}{g(n)} < \infty$, and dually $f \in \Omega(g)$ is equivalent to $\liminf_{n\to\infty} \frac{f(n)}{g(n)} > 0$. Furthermore, we write $f \preceq g$ and say that $f$ is *asymptotically bounded above* by $g$ if $f \in \mathcal{O}(g)$, we write $f \succeq g$ and say that $f$ is *asymptotically bounded below* by $g$ if $f \in \Omega(g)$, and we write $f \asymp g$ and say that $f$ is *asymptotically bounded above and below* by $g$ if $f \in \Theta(g)$. Another notation that we use within this document is the following. We write $f \sim g$ and say that $f$ *asymptotically equals* $g$ if $\lim_{n\to\infty} \frac{f(n)}{g(n)} = 1$. Clearly, $f \sim g$ implies $f \asymp g$.

## A.2. KNUTH'S UP-ARROW NOTATION

For better readability, we use Knuth's *up-arrow notation*, that is, we set

$$x \uparrow\uparrow n := \underbrace{x^{x^{\cdot^{\cdot^{\cdot^{x^x}}}}}}_{n \text{ times}}$$

and further we define the following syntactic sugar as another abbreviation.

$$(x, y) \uparrow\uparrow n := \underbrace{x^{x^{\cdot^{\cdot^{\cdot^{x^{x^y}}}}}}}_{n \text{ times}}$$

## A.3. COMPLEXITY CLASSES

The following standard complexity classes are used within this document.

$$\begin{aligned}
\mathbf{P} &:= \{\, L \mid L \text{ is decided by a det. TM } \mathcal{M} \text{ in time } \mathcal{O}(x \mapsto x^n) \text{ for some } n \in \mathbb{N} \,\} \\
\mathbf{NP} &:= \{\, L \mid L \text{ is decided by a non-det. TM } \mathcal{M} \text{ in time } \mathcal{O}(x \mapsto x^n) \text{ for some } n \in \mathbb{N} \,\} \\
\mathbf{PSpace} &:= \{\, L \mid L \text{ is decided by a TM } \mathcal{M} \text{ in space } \mathcal{O}(x \mapsto x^n) \text{ for some } n \in \mathbb{N} \,\} \\
\mathbf{EXP} &:= \{\, L \mid L \text{ is decided by a det. TM } \mathcal{M} \text{ in time } \mathcal{O}(x \mapsto 2^{x^n}) \text{ for some } n \in \mathbb{N} \,\} \\
\mathbf{2EXP} &:= \{\, L \mid L \text{ is decided by a det. TM } \mathcal{M} \text{ in time } \mathcal{O}(x \mapsto 2^{2^{x^n}}) \text{ for some } n \in \mathbb{N} \,\} \\
\mathbf{3EXP} &:= \{\, L \mid L \text{ is decided by a det. TM } \mathcal{M} \text{ in time } \mathcal{O}(x \mapsto 2^{2^{2^{x^n}}}) \text{ for some } n \in \mathbb{N} \,\} \\
\mathbf{nEXP} &:= \{\, L \mid L \text{ is decided by a det. TM } \mathcal{M} \text{ in time } \mathcal{O}(x \mapsto (2, x^m) \uparrow\uparrow n) \text{ for some } m \in \mathbb{N} \,\}
\end{aligned}$$

Furthermore, we define the *polynomial hierarchy* as usual.

$$\begin{aligned}
\Delta_0^{\mathbf{P}} := \Sigma_0^{\mathbf{P}} := \Pi_0^{\mathbf{P}} &:= \mathbf{P} \\
\Delta_{n+1}^{\mathbf{P}} &:= \mathbf{P}^{\Sigma_n^{\mathbf{P}}} \\
\Sigma_{n+1}^{\mathbf{P}} &:= \mathbf{NP}^{\Sigma_n^{\mathbf{P}}} \\
\Pi_{n+1}^{\mathbf{P}} &:= (\mathbf{coNP})^{\Sigma_n^{\mathbf{P}}} \\
\mathbf{PH} &:= \bigcup \{\, \Delta_n^{\mathbf{P}} \mid n \in \mathbb{N} \,\}
\end{aligned}$$

In particular, it holds true that $\Delta_1^{\mathbf{P}} = \mathbf{P}$, $\Sigma_1^{\mathbf{P}} = \mathbf{NP}$, and $\Pi_1^{\mathbf{P}} = \mathbf{coNP}$.